\documentclass[aps,prd,preprintnumbers,showkeys,nofootinbib,
superscriptaddress,fleqn,floatfix,tightenlines,10pt]{revtex4-2}
\usepackage{amsmath,amsfonts,amssymb,amscd,amsxtra,amsthm}
\usepackage{graphicx}  
\usepackage{epstopdf}
\usepackage{dcolumn}  
\usepackage{bm}          
\usepackage{slashed}
\usepackage{cancel}
\usepackage{float}
\usepackage{mathtools}
\usepackage{amsbsy}
\usepackage{amstext}

\usepackage[utf8]{inputenc} 
\usepackage{booktabs} 
\usepackage[normalem]{ulem} 
\usepackage[dvipsnames]{xcolor} 
\usepackage{tabularx}
\usepackage{enumitem}  
\usepackage{array} 
\usepackage{slashed}
\usepackage{tikz}
\usepackage{float}
\usepackage{multirow}
\usepackage{cancel}
\renewcommand\sout{\bgroup \color{red} \ULdepth=-.5ex \ULset}

\def\ep{\epsilon}
\makeatletter

\begin{document}  
 \preprint{INHA-NTG-06/2022}
\title{Mechanical structure of a spin-1 particle} 
\author{June-Young Kim}
\email[E-mail: ]{Jun-Young.Kim@ruhr-uni-bochum.de}
\affiliation{Institut f\"ur Theoretische Physik II, Ruhr-Universit\"at
  Bochum, D-44780 Bochum, Germany}

\author{Bao-Dong Sun}
\email[E-mail: ]{bao-dong.sun@m.scnu.edu.cn}
\affiliation{Guangdong Provincial Key Laboratory of Nuclear Science,
Institute of Quantum Matter, \\ South China Normal University,
Guangzhou 510006, China}  
\affiliation{Guangdong-Hong Kong Joint Laboratory of Quantum Matter,\\
Southern Nuclear Science Computing Center, \\ South China Normal
University, Guangzhou 510006, China 
}
\affiliation{Helmholtz Institut f\"ur Strahlen- und Kernphysik and Bethe
   Center for Theoretical Physics, Universit\"at Bonn, D-53115 Bonn, Germany}

\author{Dongyan Fu}
\email[E-mail: ]{fudongyan@ihep.ac.cn}
\affiliation{Institute of High Energy Physics, Chinese Academy of
  Sciences, Beijing 100049, China}
\affiliation{School of Physical Sciences, University of Chinese
  Academy of Sciences, Beijing 101408, China}

\author{Hyun-Chul Kim}
\email[E-mail: ]{hchkim@inha.ac.kr}
\affiliation{Department of Physics, Inha University, Incheon 22212,
South Korea}
\affiliation{School of Physics, Korea Institute for Advanced Study
(KIAS), Seoul 02455, South Korea}

\date{\today}
\begin{abstract}
 We investigate the mechanical structure of a spin-1
 particle. Introducing three different frameworks, i.e., the
 three-dimensional (3D) Breit frame, the two-dimensional (2D) Breit
 frame, and the 2D infinite momentum frame (equivalently the
 two-dimensional Drell-Yan frame), we scrutinize the 2D and 3D
 energy-momentum tensor (EMT) distributions in these frames. We first
 derive the EMT distributions in the 2D Breit frame by performing the
 Abel transformation. The mass distribution in the 2D Breit frame
 contains an additional monopole contribution induced
 geometrically. The pressure distribution in the 2D Breit frame also
 gets an induced monopole structure. When the Lorentz boost is 
 carried out, the mass distribution in the 2D infinite-momentum frame
 acquires the induced dipole term. Similarly, we also have the induced
 dipole contributions to the pressure and shear-force densities.   
 We visualize the 2D mass distributions when the spin-1 particle
 is polarized along the $x$- and $z$-axes. We observe that the 2D mass
 distribution in the infinite momentum frame exhibit clearly the
 induced dipole structure when the spin-1 particle is polarized along
 the $x$-axis. We also discuss the strong force fields inside a
 polarized spin-1 particle.  
\end{abstract}
\maketitle

\section{Introduction}
The gravitational form factors (GFFs) or energy-momentum tensor (EMT)
form factors of a hadron show its mechanical structure (see a recent 
review~\cite{Polyakov:2018zvc}). They carry essential information on
how its mass and spin are distributed inside, how its stability is
acquired, and how the internal force fields are laid out. It implies
that the gravitational form factors are equally important as the
electromagnetic (EM) form factors. In the modern understanding of
hadronic form factors, the gravitational form factors can be
understood as the second Mellin moments of the generalized parton  
distributions (GPDs)~\cite{Muller:1994ses, Ji:1996ek,
  Radyushkin:1996nd, Radyushkin:1996ru, Ji:1996nm}. 
The GFFs can also be defined from the matrix element of the EMT 
operator that comes from the response of the hadron to a variation of
the external space-time metric tensor~\cite{Kobzarev:1962wt,
  Pagels:1966zza}. The two-dimensional
(2D) inverse Fourier transforms of the GFFs provide the mass and
angular momentum probability distributions in the transverse    
plane~\cite{Burkardt:2000za, Ralston:2001xs, Burkardt:2002hr}. These
probability distributions in the transverse plane are called the
transverse densities~\cite{Miller:2007uy}. However, the most
intriguing form factor among the GFFs is the $D$-term (Druck-term)
form factor, which furnishes the pressure and shear-force
densities. They describe the stability mechanism and mechanics of the
hadron. In contrast with spin-0 (pion) and spin-1/2 particles
(nucleon), higher multipole form factors emerge for the spin-1 particle,
as was shown in the case of the EM form factors of the deuteron and
spin-1 particle~\cite{Carlson:2009ovh, Sun:2018tmk, Lorce:2022jyi,
  Kim:2022bia, Pefkou:2021fni, Polyakov:2018rew}. A similar situation can also be found for the EMT
form factors of the $\Delta$ isobar~\cite{Kim:2020lrs}.

Traditionally, the EM form factors of the nucleon have been used for
exhibiting how the charge and magnetization distributions are
spatially distributed inside a nucleon in the Breit
frame (BF)~\cite{Yennie:1957}. Similarly, the nucleon EMT form factors 
yield the three-dimensional (3D) mass, angular momentum, pressure, and 
shear-force densities for the nucleon~\cite{Polyakov:2002yz,
  Goeke:2007fp}. However, there have been serious criticisms on the 3D 
densities of the nucleon~\cite{Yennie:1957, Burkardt:2000za,
  Burkardt:2002hr, Miller:2007uy, Jaffe:2020ebz}, because the 3D
densities can only be probabilistically meaningful for atoms and
nuclei, not for the nucleon. The reasoning of the criticism lies in
the fact that the intrinsic size of the nucleon ($R_N$) is comparable
to the corresponding Compton wavelength~\cite{Belitsky:2005qn},
$\lambdabar=\hbar/mc$ ($R_N\sim 4\lambdabar$), which may bring about
relativistic corrections up to $20~\%$. As expected, the relativistic
corrections to the distributions for the deuteron are known to be
negligible~\cite{Kim:2022bia, Lorce:2022jyi}, whereas those to the
$\rho$ meson should be rather strong. Thus, it is necessary to resolve 
this problem for the $\rho$ meson. Choosing the infinite-momentum 
frame (IMF), one can avoid the relativistic corrections but obtain the 
2D transverse densities, which suffer from a deficiency of losing
information on the longitudinal direction. As proposed in
Refs.~\cite{Lorce:2020onh, Lorce:2022jyi}, however, one can view the 3D
densities as the quasi-probabilistic ones defined in phase space,
taking them fully relativistically. From this perspective, the 3D
distributions in the BF provide the intuitive and physical meaning. Then
the Abel transformation allows one to project the 3D densities in the
BF to the 2D transverse densities~\cite{Panteleeva:2021iip,
  Kim:2021jjf, Kim:2021kum}, which was already introduced in
describing deeply virtual Compton scattering~\cite{Polyakov:2007rv,
  Moiseeva:2008qd}.   The 2D BF EM distributions for the 
  $\rho$ meson have been studied~\cite{Sun:2018tmk} where a wave packet
  is applied to the localized $\rho$-meson state. Note that,
  very recently, a novel prescription of defining the 3D distribution
  was studied in Refs.~\cite{Epelbaum:2022fjc, Panteleeva:2022khw}. 

In Refs.~\cite{Kim:2021jjf, Panteleeva:2021iip}, it was shown that the 
Abel transformation projects the 3D BF distributions onto the 2D 
IMF densities. The Abel images of the 3D BF distributions
without any non-relativistic approximation are exactly
identified as the 2D IMF densities that can be obtained from the 2D
Fourier transform of the corresponding form factors. To find the Abel
images of the 3D distributions in the BF, one should know the hadronic
matrix element in the IMF or the expressions for the 2D distributions
in the IMF. In Ref.~\cite{Kim:2021jjf}, the four different 3D BF EMT
distributions are related to the 2D IMF ones and a complete set of the
2D IMF distributions were determined. The 2D IMF distributions for the
transversely polarized nucleon can be found by using the obtained 
Abel images from the 3D BF distributions.   

In the meanwhile, Freese and Miller~\cite{Freese:2021mzg} criticized
the Abel transformation from the 3D BF distributions to the 2D IMF 
ones. They argued that the Abel transformation is merely defined as
the integral of a spherical symmetric 3D distribution over $z$-axis.
Having performed the integral of the nucleon EMT distributions
$T^{00}(\bm{r})$ or $T^{ij}(\bm{r})$ over $z$, they obtained the 
Abel images in the BF. However, this integral merely projects the 3D
BF distribution onto the 2D BF one without any relativistic
effects. So, their Abel images are different from those defined in
Refs.~\cite{Kim:2021jjf, Panteleeva:2021iip}. Knowing this difference
is essential to understand the physical implications of the Abel
images.  Though we have technical difficulties to construct the Abel 
images in the IMF from the 3D distributions in the BF for higher
spin particles such as the $\rho$ meson and $\Delta$ isobar, we can 
formally establish the Abel images in the IMF for these
particles~\cite{Kim:2021jjf, Panteleeva:2021iip}. The only trouble
arises from the fact that relativistic contributions 
bring about a number of kernels composed of the Laplacians into the Abel
transformation~\cite{Kim:2022bia}. Depending on the order of the
differential equations, we need to provide additional boundary
conditions on the distributions so that we can perform the
integrals. An easy way to circumvent such complexities is to project
3D BF distributions to 2D BF ones as done in 
Ref.~\cite{Freese:2021mzg}. Then, we can proceed to the IMF by
increasing the $z$-component of the momentum ($P_z$) to infinity. In
the case of the spin-1 particle, however, the procedure from $P_z=0$
to $P_z=\infty$ is very complicated as shown in
Ref.~\cite{Kim:2022bia}. Nevertheless, we want to emphasize  
the following point: in Ref.~\cite{Kim:2021kum} it was explicitly shown
that by introducing the kernel for relativistic effects, the 2D
IMF transverse charge densities of both the proton and neutron were 
derived from the nonrelativistic charge and magnetization
distributions in 3D space. The results for the transverse charge
densities are exactly the same as those obtained in
Ref.~\cite{Miller:2007uy}. Thus, the IMF Abel images directly obtained
from the 3D distributions~\cite{Kim:2021jjf, Panteleeva:2021iip,
  Kim:2021kum} are immune to the criticism raised in
Ref.~\cite{Freese:2021mzg}. 

In the present work, we want to investigate the EMT densities of a
spin-1 particle, focusing on the mechanical structure of
the GFFs and the corresponding densities of the spin-1 particle
instead of describing the GFFs of the spin-1
particle~\cite{Sun:2020wfo, Lorce:2022jyi, Kim:2022bia}
quantitatively. The outline of the current work is given as follows:  
In the next Section, we define the GFFs of a spin-1 particle in three
different frames such as the 3D BF, 2D BF, and 2D IMF. In
Section~\ref{sec:3}, we show how to obtain the densities in the 
three different frames. In Section~\ref{sec:4}, we formulate the Abel
transformation for the EMT densities. In Section~\ref{sec:5}, we
introduce a toy model, where the GFFs are parametrized as simple
quadrupole types, and derive the 2D EMT densities. We discuss the
physical implications of the numerical results and 2D force fields
inside a spin-1 particle. In the final section, we summarize the
current work and draw conclusions.

\section{Gravitational form factors of a spin-1 particle}
\label{sec:2}

The EMT operator of quantum chromodynamics (QCD) can be derived either
by varying the action of QCD under the Poincar{\'e} transformation
with the symmetrization imposed~\cite{Belinfante:1939, Pauli:1940dq,
  Callan:1970ze} or by taking variation of the QCD action, 
which is coupled to the weak classical torsionless gravitational
background field, with respect to the metric tensor of this curved
background field~\cite{Parker:2009uva, Polyakov:2018zvc}. The total
EMT operator consisting of the quark $(q)$ and gluon $(g)$ parts are
then written by  
\begin{align}
\hat{T}^{\mu \nu} = \sum_{q} \hat{T}^{\mu \nu}_{q} + \hat{T}^{\mu
  \nu}_{g},  
\end{align} 
where the quark and gluon parts are expressed respectively as
\begin{align}
\hat{T}^{\mu \nu}_{q}  &= \frac{1}{4} \bar{\psi}_{q} \left( -
  i\overleftarrow{\mathcal{D}}^{\mu}
  \gamma^{\nu} -
  i\overleftarrow{\mathcal{D}}^{\nu}
  \gamma^{\mu}+
  i\overrightarrow{\mathcal{D}}^{\mu}
  \gamma^{\nu}+
  i\overrightarrow{\mathcal{D}}^{\nu}
  \gamma^{\mu}\right) \psi_{q} - g^{\mu \nu}
  \bar{\psi}_{q} \left(
  -\frac{i}{2}\overleftarrow{\slashed{\mathcal{D}}}
+\frac{i}{2}\overrightarrow{\slashed{\mathcal{D}}} 
  - m_{q} \right) \psi_{q}, \cr 
\hat{T}^{\mu \nu}_{g} &= F^{a,\nu \eta} F^{a,  \ \nu}_{ \ \ \eta} + 
\frac{1}{4} g^{\mu \nu} F^{a,\kappa \eta}F^{a,}_{\ \ \kappa \eta}.
\end{align}
Here, $\overrightarrow{\mathcal{D}}_{\mu} =
\overrightarrow{\partial}_{\mu} + i g t^{a} A^{a}_{\mu}$ and
$\overleftarrow{\mathcal{D}}_{\mu} = \overleftarrow{\partial}_{\mu} -
i g t^{a} A^{a}_{\mu}$ are covariant derivatives. $t^a$ stand for the
SU(3) color group generators that satisfy the commutation relations
$[t^{a},t^{b}]=if^{abc} t^{c}$ and are normalized as  
$\mathrm{tr}(t^{a} t^{b}) = \frac{1}{2}\delta^{ab}$.
$\psi_q$ is the quark operator with flavor index $q$ and $m_q$
represents the corresponding current quark mass.  
$F^{a,\mu\eta}$ denotes the gluon field strength expressed as
$F^{a}_{\mu \nu} = \partial_{\mu} A^{a}_{\nu} - \partial_{\nu}
A^{a}_{\nu} - g f^{abc} A^{b}_{\mu} A^{c}_{\nu}$. 
The total EMT operator is a conserved current: $\partial^{\mu}
\hat{T}_{\mu \nu}=0$. 

The matrix element of the total EMT current operator for a spin-1
particle is parametrized in terms of the six different form
factors~\cite{Cotogno:2019vjb, Cosyn:2019aio, Polyakov:2019lbq}  
\begin{align}
\langle p^\prime,\lambda^\prime| \hat T_{\mu\nu}(0) |p,\lambda\rangle
&= \biggl[
2P_\mu P_\nu  \Bigl(
- {\ep^{\prime*}\cdot \ep} \, A_0 (t)
 + \frac{{\ep^{\prime*}\cdot P} \, {\ep \cdot P}}{m^2}
 \, A_1(t) \Bigl) \cr
 &+2\left[P_\mu(\ep^{\prime*}_{\nu} \,\ep\cdot P+\ep_{\nu}\,
\ep^{\prime*}\cdot P)
+P_\nu(\ep^{\prime*}_{\mu}\, \ep\cdot
P+\ep_{\mu} \,\ep^{\prime*}\cdot
P) \right] \, J (t) \cr
 &+\frac12(\Delta_\mu \Delta_\nu-\eta_{\mu\nu}\Delta^2)
 \Bigl(
{\ep^{\prime*}\cdot \ep} \, D_0 (t)
+\frac{{\ep^{\prime*}\cdot P} \, {\ep \cdot P}}{m^2}  \, D_1(t)\Bigl) \cr 
&+\Bigl[\frac12(\ep_{\mu}
\ep^{\prime*}_{\nu}+\ep^{\prime*}_{\mu}\ep_{\nu})\Delta^2
-(\ep^{\prime*}_{\mu}\Delta_\nu+\ep^{\prime*}_{\nu} \Delta_\mu)\,\ep\cdot P
\cr
&+(\ep_{\mu} \Delta_\nu+\ep_{\nu}
\Delta_\mu)\,\ep^{\prime*}\cdot P
-4\eta_{\mu\nu} \, {\ep^{\prime*}\cdot P} \, {\ep\cdot P} \Bigl] \,
  E(t) \biggr], 
\label{eq:3}
\end{align}
where the one-particle state for the spin-1 particle is normalized as
$\langle p', \lambda'| p, \lambda \rangle = 2p^{0}
(2\pi)^{3}\delta_{\lambda'  \lambda} \delta^{(3)}(\bm{p}'-\bm{p})$
with the spin polarizations of the initial $\lambda$ and final
$\lambda'$ states. $P=(p+p')/2$ designates the average of the
four-momenta $p$ and $p'$, and $\Delta=p'-p$ represents the
four-momentum transfer with $\Delta^{2} = t$. $P$ and $\Delta$ are
orthogonal each other, i.e., $P\cdot \Delta =0$. The polarization
vectors are defined as $\epsilon'_{\mu}=\epsilon'_{\mu}(p',\lambda')$,
$\epsilon_{\mu}=\epsilon_{\mu}(p,\lambda)$ for simplicity. In this
work, we choose canonical spin states (see relevant
discussions~\cite{Polyzou:2012ut, Lorce:2019sbq, Lorce:2020onh}). It
can be obtained by applying the rotationless boost operator to the
spin-one polarization vector 
$\epsilon^{\mu}(0,\lambda)=(0,\hat{\bm{\epsilon}}_{\lambda})$ in the
rest frame. The explicit expression of the spin-one vector
$\epsilon^{\mu}$ in any frame is defined by 
\begin{align}
\epsilon^{\mu}(p, \lambda) = \left( \frac{\bm{p}\cdot
  \hat{\bm{\epsilon}}_{\lambda}}{ m},\hat{{\epsilon}}_{\lambda} +
  \frac{\bm{p}\cdot \hat{\bm{\epsilon}}_{\lambda}}{ m(m+p_{0})}\bm{p}
  \right) 
\end{align}
with 
\begin{align}
\hat{\bm{\epsilon}}_{x}=(1,0,0), \ \ \
  \hat{\bm{\epsilon}}_{y}=(0,1,0), \ \ \
  \hat{\bm{\epsilon}}_{z}=(0,0,1), 
\end{align}
in the Cartesian basis. $\eta_{\mu\nu}$ is the metric tensor
$\eta_{\mu\nu}=\mathrm{diag}(1,-1,-1,-1)$. $A_0(t)$, $A_1(t)$, $J(t)$,
$D_0(t)$, $D_1(t)$, and $E(t)$ are the six independent GFFs of the
spin-1 particle. Since the total~(quark$+$gluon) current is conserved,
we do not consider the non-conserving terms arising from the quark and
gluon parts separately (see Ref.~\cite{Polyakov:2019lbq} for the
separate quark and gluon GFFs). 

Since the GFFs of a spin-1 particle reveal higher multipole
structures, it is useful to define the quadrupole operator
$\hat{Q}^{ij}$. It is defined in terms of the spin operator
$\hat{S}^{i}$ as   
\begin{align}
\hat{Q}^{ij} &= \frac{1}{2}\left( \hat{S}^{i}\hat{S}^{j}
               +\hat{S}^{j}\hat{S}^{i}
               -\frac{2}{3}S(S+1)\delta^{ij}\right), 
\end{align}
where the indices $i$ and $j$ run over $1$, $2$, and
$3$. $\hat{Q}^{ij}$ is a symmetric irreducible tensor operator. The
components of the spin operators can be expressed 
in  terms of the SU(2) Clebsch-Gordan coefficients $C^{S \sigma'}_{S
  \sigma 1 a}$ in the spherical basis  
\begin{align}
\hat{S}^{a}_{\sigma'\sigma} = \sqrt{S(S+1)}C^{S \sigma'}_{S \sigma 1
  a} \ \ \ \mathrm{with} \ \ \ (a=0,\pm1. \  \  \sigma,\sigma'=0,
  \cdot\cdot\cdot,\pm S). 
\end{align}

\subsection{Gravitational form factors of the spin-1 particle in the
  3D Breite frame}
We first examine the matrix element of the EMT
operators~\eqref{eq:3} in the 3D BF. 
To examine the multipole structure of the GFFs, we need to define the
$n$-rank irreducible Cartesian tensors, which are expressed
respectively in coordinate and momentum spaces
\begin{align}
Y^{i_{1}i_{2}...i_{n}}_{n}(\Omega_{r}) =
  \frac{(-1)^{n}}{(2n-1)!!}r^{n+1}\partial^{i_{1}} 
\partial^{i_{2}}\cdots\partial^{i_{n}}\frac{1}{r},  
  \ \ \ \ \ Y^{i_{1}i_{2}...i_{n}}_{n}(\Omega_{\Delta}) =
  \frac{(-1)^{n}}{(2n-1)!!}\Delta^{n+1}\partial^{i_{1}}
\partial^{i_{2}}\cdots\partial^{i_{n}}\frac{1}{\Delta}, 
\end{align}
where the first three of $Y^{i_{1}i_{2}...i_{n}}_{n}(\Omega_{r})$ are
explcitly written as 
\begin{align}
Y_{0}(\Omega_{r})=1, \ \  Y^{i}_{1}(\Omega_{r})=\frac{r^{i}}{r},  \ \ 
Y^{ij}_{2}(\Omega_{r})=\frac{r^{i}r^{j}}{r^{2}}-\frac{1}{3}\delta^{ij}.
\label{eq:tensor}
\end{align}

In the BF, the four-momenta of the initial and final states are
respectively expressed as $p=(p^0,\,\bm{p})$ and $p'=(p^0,\,-\bm{p})$,
which yield $\Delta^0=0$ and $\bm{P}=\bm{0}$. Then we consider the
matrix element of each component of the EMT current:
\begin{align}
\langle p', \lambda' | \hat{T}^{00} (0) | p, \lambda \rangle&= 2m^{2}
  \mathcal{E}_{0}(t)  \delta_{\lambda'  \lambda} +  4m^{2}\tau
  \mathcal{E}_{2}(t)  \hat{Q}^{kl}  Y^{kl}_{2}(\Omega_{\Delta}), \cr 
\langle p', \lambda' | \hat{T}^{0i} (0) | p, \lambda \rangle&= 2m^{2}
   \sqrt{\tau} i   \epsilon^{ijk}   S^{k}
   Y^{k}_{1}(\Omega_{\Delta})   \mathcal{J}_{1}(t), \cr 
\langle p', \lambda' | \hat{T}^{ij} (0) | p, \lambda \rangle&= 
2m^{2}\tau \left(Y^{ij}_2(\Omega_{\Delta}) - \frac{2}{3} 
\delta^{ij}\right) \mathcal{D}_{0}(t) \delta_{\lambda \lambda'}
+ 8\tau^{2}m^{2} \left(Y^{ij}_{2}(\Omega_{\Delta}) - 
\frac{2}{3}\delta^{ij} \right)Y^{kl}_{2}(\Omega_{\Delta}) 
\hat{Q}^{kl} \mathcal{D}_{3}(t). \cr
& +4m^{2} \tau \left(Y_{2}^{jk}(\Omega_{\Delta})
\hat{Q}^{ik}+Y^{ik}_{2}(\Omega_{\Delta})\hat{Q}^{jk}-
\frac{1}{3}\hat{Q}^{ij}-\delta^{ij} Y^{kl}_{2}(\Omega_{\Delta})
\hat{Q}^{kl}\right)\mathcal{D}_{2}(t),
\end{align}
where $\tau= -\frac{t}{4m^{2}}$ and $\hat{Q}^{ij}_{\lambda \lambda'}=
\langle \lambda'  | \hat{Q}^{ij} |  \lambda \rangle$. For brevity, we
define $\hat{Q}^{ij}:=\hat{Q}^{ij}_{\lambda \lambda'}$ from now on. 
Examining these expressions for each component of the matrix element,
one can easily relate the multipole form factors $\mathcal{E}_0(t)$,
$\mathcal{E}_2(t)$, $\mathcal{J}_1(t)$, $\mathcal{D}_0(t)$,
$\mathcal{D}_2(t)$, and $\mathcal{D}_3(t) $ to the GFFs in
Eq.~\eqref{eq:3}  
\begin{align}
\mathcal{E}_{0}(t)
&=
 A_0(t) -\frac{1}{3}\tau \Bigl[ -{5} A_0(t)  +3 D_0(t)+ 4J(t) -2E(t) +A_1(t) 
 \Bigr]
 \cr
&-\frac{2}{3}\tau^{2} \Bigl[ -A_0(t)  + D_0(t) + {2}J(t) - 2E(t) +
  A_1(t) +  \frac12D_1(t) \Bigr] 
-\frac{1}{3}\tau^{3}  \Bigl[ A_1(t) +  D_1(t)  \Bigl], \cr
\mathcal{E}_{2}(t)
&=-A_0(t) + {2}J(t)  -E(t)  + \frac12A_1(t) 
 \cr
&
+ \tau \Bigl[ -A_0(t)  + D_0(t) + 2J(t) - 2E(t) + A_1(t) +
  {1\over2}D_1(t)  \Bigr] 
 \cr
& + \frac{1}{2} \tau^{2}  \Bigl[ A_1(t) +  D_1(t)  \Bigl], \cr
\mathcal{J}_{1}(t)&=
 J(t)  + \tau \Bigl( J(t)   -  E(t) \Bigl), \cr
\mathcal{D}_0(t)&=
-D_0(t)+{4\over3} E(t) -\frac{1}{3}\tau \Bigl[ 2D_0(t)-2 E(t)+ D_1(t)
                  \Bigl] - \frac{1}{3} \tau^{2}   D_1(t), \cr 
{\mathcal D}_2(t)&= -E(t), \cr
{\mathcal D}_3(t)&=  \frac14 \Bigl[ 2D_0(t)-2 E(t)+ D_1(t) \Bigl] +
                   \frac{1}{4}\tau   D_1(t). 
\label{eq:EMTMFF}
\end{align}
\subsection{Gravitational form factors of the spin-1 particle in the
  2D Breite frame} 
In~Refs.~\cite{Lorce:2018egm, Lorce:2020onh}, the elastic frame~(EF)
was introduced to study how the hadronic matrix element undergoes
changes under the Lorentz boost. It was shown that this frame
naturally interpolates between the 2D BF and 2D IMF for both the 
nucleon~\cite{Lorce:2020onh} and the deuteron~\cite{Lorce:2022jyi}. 
The EF also allows one to define a quasi-probabilistic
distribution for a moving hadron in the Wigner sense. To trace down
the origin of both the geometrical and relativistic effects, we   
scrutinize how the multipole structure of the EMT matrix element
emerges in 2D space. If we restrict ourselves to 2D space, we have
to define the 2D $n$-rank irreducible tensors respectively in
2D coordinate and momentum spaces as follows: 
\begin{align}
X^{i_{1}\ldots i_{n}}_{n}(\theta_{x_{\perp}}) =
  \frac{(-1)^{n+1}}{(2n-2)!!} x_{\perp}^{n} \partial^{i_{1}}
  \ldots \partial^{i_{n}} \ln{x_{\perp}}, \ \ X^{i_{1}\ldots
  i_{n}}_{n}(\theta_{\Delta_{\perp}}) = \frac{(-1)^{n+1}}{(2n-2)!!}
  \Delta_{\perp}^{n} \partial^{i_{1}} \ldots \partial^{i_{n}}
  \ln{\Delta_{\perp}}  
\end{align}
with $n>0$ and $i_{n}=1,\,2$. $x_\perp$ is the radial distance from
the center of the 2D plane. The first three of $X_n^{i_1\cdots i_n}
(\theta_{x_{\perp}})$ are given as 
\begin{align}
X_{0}(\theta_{x_{\perp}}):=1, \ \
  X^{i}_{1}(\theta_{x_{\perp}})=\frac{x_{\perp}^{i}}{x_{\perp}},  \ \
  X^{ij}_{2}(\theta_{x_{\perp}})=\frac{x_{\perp}^{i}x_{\perp}^{j}}
{x_{\perp}^{2}}-\frac{1}{2}\delta^{ij}. 
\label{eq:tensor2}
\end{align}

In the EF, the spacelike momentum transfer
$\bm{\Delta}=(\bm{\Delta}_{\perp},0)$ lies in the 2D transverse 
plane. The frame satisfies the following conditions: 
$\bm{P}=(\bm{0},P_{z})$ and $\Delta^0 = 0$. If we take the 2D BF,
i.e., $P_{z}=0$, the matrix element for each component of 
$\hat{T}^{\mu\nu}$ is derived as   
\begin{align}
\langle p', \lambda' | \hat{T}^{00} (0) | p, \lambda \rangle &= 2m^{2} 
  \mathcal{E}_{(0,1)}(t)  \delta_{\sigma'  \sigma} +
  2m^{2}  \mathcal{E}_{(0,0)}(t)  \delta_{\lambda'  3}
  \delta_{\lambda  3} + 4m^{2}  \tau  \mathcal{E}_{2}(t)
  \hat{Q}^{kl}  X^{kl}_{2}(\theta_{\Delta_{\perp}}),  \cr 
\langle p', \lambda' | \hat{T}^{0i}(0) | p, \lambda \rangle &= 2m^{2} 
  \sqrt{\tau} i  \epsilon^{3li}  \hat{S}^{3}_{\lambda' \lambda}  
X^{l}_{1}(\theta_{\Delta_{\perp}})  \mathcal{J}_{1}(t),  \cr 
\langle p', \lambda' | \hat{T}^{ij} (0) | p, \lambda \rangle &=
  2m^{2}\tau   \left[  \left(\frac{1}{3}\mathcal{D}_{2}(t)  -
  \frac{1}{2}\mathcal{D}_{0}(t)\right)  \delta_{\sigma'
  \sigma}+  \left(-\frac{2}{3}\mathcal{D}_{2}(t)  -
  \frac{1}{2}\mathcal{D}_{0}(t)\right)  \delta_{\lambda'
  3}  \delta_{\lambda  3}\right]  \delta^{ij}  \cr 
& +2m^{2}\tau X^{ij}_{2}(\theta_{\Delta_{\perp}})  \delta_{\lambda'
  \lambda} \mathcal{D}_{0}(t) + 4m^{2}\tau\bigg{[}\hat{Q}^{ik}
  X^{jk}_{2}(\theta_{\Delta_{\perp}}) + \hat{Q}^{jk}
  X^{ik}(\theta_{\Delta_{\perp}}) - \hat{Q}^{lm}
  X^{lm}_{2}(\theta_{\Delta_{\perp}}) \delta^{ij}\bigg{]}
  \mathcal{D}_{2}(t)  \cr 
&+  8m^{2}\tau^{2} \hat{Q}^{lm}
  \left(X^{lm}_{2}(\theta_{\Delta_{\perp}})
  +\frac{1}{2}\delta^{lm}\right)
  \left(X^{ij}_{2}(\theta_{\Delta_{\perp}}) -
  \frac{1}{2}\delta^{ij}\right) \mathcal{D}_{3}(t)  
\label{eq:2DBF_FF}
\end{align}
with
\begin{align}
\mathcal{E}_{(0,0)}(t)=\mathcal{E}_{0}(t) +\frac{2}{3}\tau
  \mathcal{E}_{2}(t), \ \ \ \mathcal{E}_{(0,1)}(t)=\mathcal{E}_{0}(t)
  -\frac{1}{3}\tau \mathcal{E}_{2}(t), 
\label{eq:15}
\end{align}
where the definition of the multipole form factors is the same as 
Eq.~\eqref{eq:EMTMFF} and the indices run over $i,j=1,2$. We use the
following short-handed notation 
$\delta_{\sigma'\sigma}= \delta_{\sigma'\sigma} \delta_{\lambda'
  \sigma'} \delta_{\lambda \sigma}$ with $\sigma',\sigma=1,2$. We want
to mention that the 2D multipole form factors exhibit a distinctive
feature in contrast to those in the 3D BF. As shown in
Eq.~\eqref{eq:15}, the mass form factors in the 2D BF,
$\mathcal{E}_{(0,0)}$ and $\mathcal{E}_{(0,1)}$, acquire respectively
the positive and negative contributions from the quadrupole mass form
factor $\mathcal{E}_{2}$ defined in the 3D BF. 
It originates from the presence of the quadrupole structure and 
the geometrical difference between the 2D and 3D
spaces. This feature has been already observed in the charge
distributions of a higher spin particle~$(S\geq1)$ such as the $\rho$
meson~\cite{Kim:2022bia}. We want to emphasize that they do not
come from relativistic effects.  

\subsection{Gravitational form factors of the spin-1 particle in the
  infinite-momentum frame} 
If we take the value of the $P_{z}$ as infinity in the EF, we arrive
naturally at the IMF. Thus, by taking $P_z\to \infty$, we obtain the
matrix element in the IMF:
\begin{align}
\langle p', \lambda' | \hat{T}^{00} (0) | p, \lambda \rangle &= 2P^{2}_{z}
  \mathcal{E}^{\mathrm{IMF}}_{(0,0)}(t)  \delta_{\lambda'  3}\delta_{3
  \lambda} +  2P^{2}_{z}  \mathcal{E}^{\mathrm{IMF}}_{(0,1)}(t)
  \delta_{\sigma'  \sigma} \cr 
 &+  2P^{2}_{z} \sqrt{\tau}\mathcal{E}^{\mathrm{IMF}}_{1}(t) i
   \epsilon^{3jk} \hat{S}^{j}_{\lambda' \lambda}
   X^{k}_{1}(\theta_{\Delta_{\perp}})   +4P^{2}_{z} \tau
   \mathcal{E}^{\mathrm{IMF}}_{2}(t) \hat{Q}^{kl}
   X^{kl}_{2}(\theta_{\Delta_{\perp}}),  \cr 
\langle p', \lambda' | \hat{T}^{0i}_{a} (0) | p, \lambda \rangle &=  2 m
  P_{z}  \sqrt{\tau}  i  \epsilon^{3li}  \hat{S}^{3}_{\lambda'
  \lambda}  X^{l}_{1}(\theta_{\Delta_{\perp}})  \mathcal{J}^{\mathrm{IMF}}_{1}(t)
  + 4m  P_{z}  \tau  \left(X^{ik}_{2}(\theta_{\Delta_{\perp}}) -
\frac{1}{2}\delta^{ik}\right)
 \mathcal{J}^{\mathrm{IMF}}_{2}(t)  \hat{Q}^{3k},  \cr 
\langle p', \lambda' | \hat{T}^{ij} (0) | p, \lambda \rangle &= 
 2m^{2}\tau \left[ \left( \frac{1}{3} \mathcal{D}^{\mathrm{IMF}}_{2}
   -\frac{1}{2}\mathcal{D}^{\mathrm{IMF}}_{(0,1)}   \right)
   \delta_{\sigma'   \sigma}  +   \left(   -\frac{2}{3}
    \mathcal{D}^{\mathrm{IMF}}_{2}   
-\frac{1}{2}\mathcal{D}^{\mathrm{IMF}}_{(0,0)}  \right)
   \delta_{\lambda'   3}   \delta_{\lambda 3} \right]
   \delta^{ij}   \cr 
& +  2m^{2}\tau X^{ij}_{2}(\theta_{\Delta_{\perp}})
  \left[\delta_{\sigma' \sigma} \mathcal{D}^{\mathrm{IMF}}_{(0,1)}(t)
  +  \delta_{\lambda' 3}\delta_{\lambda 3}
  \mathcal{D}^{\mathrm{IMF}}_{(0,0)}(t) \right] \cr 
&+ 4m^{2}\tau \bigg{[} \hat{Q}^{ik}
  X^{jk}_{2}(\theta_{\Delta_{\perp}}) + \hat{Q}^{jk}
  X^{ik}_{2}(\theta_{\Delta_{\perp}}) - \hat{Q}^{lm}
  X^{lm}_2(\theta_{\Delta_{\perp}}) \delta^{ij}\bigg{]}
  \mathcal{D}^{\mathrm{IMF}}_{2}(t)   \cr 
& +8m^{2} \tau^{3/2} i \epsilon^{lm3} \hat{S}^{l}
  X^{m}_{1}(\theta_{\Delta_{\perp}})
  \left(X^{ij}_{2}(\theta_{\Delta_{\perp}}) -
  \frac{1}{2}\delta^{ij}\right) \mathcal{D}^{\mathrm{IMF}}_{1}(t) \cr 
&+  8 m^{2} \tau^{2} \hat{Q}^{lm}
  \left(X^{lm}_{2}(\theta_{\Delta_{\perp}}) +
  \frac{1}{2}\delta^{lm}\right)
  \left(X^{ij}_{2}(\theta_{\Delta_{\perp}}) -
  \frac{1}{2}\delta^{ij}\right) \mathcal{D}^{\mathrm{IMF}}_{3}(t), 
\label{eq:16}
\end{align}
where the multipole form factors are related to the GFFs in
Eq.~\eqref{eq:EMTMFF}
\begin{align}
&\mathcal{E}^{\mathrm{IMF}}_{(0,0)}(t) = \frac{1}{3(1+\tau)^{2}}
  \left[12 \tau \mathcal{J}_{1} - 3(\tau-1)\mathcal{E}_{0} +  \tau
  (2+4\tau)\mathcal{E}_{2}+  \tau
  (\tau-1)(3\mathcal{D}_{0}-2\mathcal{D}_{2}) - 4\tau^{2}
  (1+2\tau)\mathcal{D}_{3}  \right], \cr 
&\mathcal{E}^{\mathrm{IMF}}_{(0,1)}(t) = \frac{1}{3(1+\tau)^{2}}
  \left[6 \tau \mathcal{J}_{1} + 3\mathcal{E}_{0}  - \tau
  \mathcal{E}_{2}- 3 \tau \mathcal{D}_{0} - \tau \mathcal{D}_{2}-3
  \tau^{2} \mathcal{D}_{2} +  2\tau^{2} \mathcal{D}_{3}  \right],  \cr 
&\mathcal{E}^{\mathrm{IMF}}_{1}(t) = \frac{1}{3(1+\tau)^{2}}
  \left[-6(\tau-1)  \mathcal{J}_{1} - 6\mathcal{E}_{0}  +2 \tau
  \mathcal{E}_{2}+ 6 \tau \mathcal{D}_{0} - 4\tau ( \mathcal{D}_{2} +
  \tau\mathcal{D}_{3} ) \right] ,  \cr 
&\mathcal{E}^{\mathrm{IMF}}_{2}(t) =- \frac{1}{3(1+\tau)^{2}} \left[6
  \mathcal{J}_{1} - 3\mathcal{E}_{0}  -  (3+2\tau) \mathcal{E}_{2}+ 3
  \tau \mathcal{D}_{0} +(\tau +3 ) \mathcal{D}_{2} + 2 \tau
  (3+2\tau)\mathcal{D}_{3}  \right] ,  \cr 
&\mathcal{J}^{\mathrm{IMF}}_{1}(t) =
  \frac{\mathcal{J}_{1}-\tau\mathcal{D}_{2}}{1+\tau}, \ \
  \mathcal{J}^{\mathrm{IMF}}_{2}(t) = \frac{\mathcal{J}_{1}+
  \mathcal{D}_{2}}{1+\tau}, \cr 
&\mathcal{D}^{\mathrm{IMF}}_{(0,1)}(t)
  =\mathcal{D}_{0}+\frac{\tau}{3}G_{W}, \
  \mathcal{D}^{\mathrm{IMF}}_{(0,0)}(t) =
  \mathcal{D}_{0}+\frac{4\tau}{3}G_{W}, \cr 
&\mathcal{D}^{\mathrm{IMF}}_{1}(t) = \frac{1}{4} G_{W}, \
  \mathcal{D}^{\mathrm{IMF}}_{2}(t) = \mathcal{D}_{2}, \
  \mathcal{D}^{\mathrm{IMF}}_{3}(t) =\mathcal{D}_{3} -
  \frac{1}{4}G_{W}, 
\label{eq:IMF_FF}
\end{align}
with
\begin{align}
&G_{W}(t)= -\frac{2(3\mathcal{D}_{0}+\mathcal{D}_{2}-
2\tau \mathcal{D}_{3})}{3(1+\tau)}.
\end{align}
It is straightforward to understand the meaning of each term in
Eq.~\eqref{eq:IMF_FF}. In the case of the $00$ and $0k$-components,
they are subjected to both the Wigner spin rotation and get mixed with
the other components of the EMT current under the Lorentz boost, so
that they vary as shown in Eq.~\eqref{eq:IMF_FF}. It is remarkable to
see that the dipole and quadrupole contributions are respectively
induced by the Lorentz boost as shown in Eq.~\eqref{eq:16}. When it
comes to the matrix element of $\hat{T}^{ij}$, 
the $D$-term form factors undergo changes by the Wigner spin rotation
under the Lorentz boost except for
$\mathcal{D}_2^{\mathrm{IMF}}(t)$. Interestingly, it was found  
that this effect of the Wigner spin rotation can be parametrized in
terms of one combination of the form factors, $G_W$, as done for
the EM form factors in Ref.~\cite{Lorce:2022jyi}. In addition to that,
the monopole form factors acquire the geometrical contribution in the
presence of the quadrupole form factors. As shown in
Eq.~\eqref{eq:2DBF_FF} there was no term such as
$\mathcal{D}^{\mathrm{IMF}}_{1}$ in the $ij$-component of the EMT
current in the BF. It implies that the form factor $G_W(t)$ only
appears by the Wigner spin rotation under the Lorentz boost. Thus, the
induced dipole form factor $\mathcal{D}^{\mathrm{IMF}}_{1}$ is solely
due to the relativistic effect. 

The multipole form factors at $t=0$ in the IMF can be given as
follows:   
\begin{align}
&\mathcal{E}^{\mathrm{IMF}}_{(0,1)}(0)
  =\mathcal{E}^{\mathrm{IMF}}_{(0,0)}(0)=\mathcal{E}_{0}(0), \ \ \
  \mathcal{E}^{\mathrm{IMF}}_{1}(0) = 2 \mathcal{J}_{1}(0) -2
  \mathcal{E}_{0}(0), \ \ \ \mathcal{E}^{\mathrm{IMF}}_{2}(0) = -2
  \mathcal{J}_{1}(0) + \mathcal{E}_{0}(0)+\mathcal{E}_{2}(0)  -
  \mathcal{D}_{2}(0), \cr 
&\mathcal{J}^{\mathrm{IMF}}_{1}(0) = \mathcal{J}_{1}(0), \ \
  \mathcal{J}^{\mathrm{IMF}}_{2}(0) = \mathcal{J}_{1}(0) +
  \mathcal{D}_{2}(0), \cr 
&\mathcal{D}^{\mathrm{IMF}}_{(0,1)}(0)
  =\mathcal{D}^{\mathrm{IMF}}_{(0,0)}(0) = \mathcal{D}_{0}(0), \
  \mathcal{D}^{\mathrm{IMF}}_{1}(0) = -\frac{1}{2}\mathcal{D}_{0}(0) -
  \frac{1}{6} \mathcal{D}_{2}(0), \cr 
&\mathcal{D}^{\mathrm{IMF}}_{2}(0) = \mathcal{D}_{2}(0), \
  \mathcal{D}^{\mathrm{IMF}}_{3}(0) = \mathcal{D}_{3}(0) +
  \left[\frac{1}{2}\mathcal{D}_{0}(0) + \frac{1}{6}
  \mathcal{D}_{2}(0)\right]. 
  \label{eq:IMF_nor}
\end{align}
The results from the light-front formalism should coincide with  
those from the IMF. As a cross-check, we have carried out the
same calculation in the light-front formalism (See
Appendix~\ref{Appendix:A}).

\section{Definition of the EMT distributions
\label{sec:3}}

While the 3D EMT distributions of a spin-one particle can not be
interpreted as a probability densities because of the ambiguous
relativistic corrections, we can understand it as a
quasi-probabilistic distribution by the Wigner distribution. This
quasi-probabilistic distribution conveys 
information on the internal structure of a hadron in a fully
relativistic picture. The matrix element of the EMT current for a
physical state $|\psi \rangle$ can be expressed in terms of the Wigner
distribution as~\cite{Lorce:2020onh}  
\begin{align}
\langle \hat{T}^{\mu \nu}(\bm{r}) \rangle= \int
  \frac{d^{3} \bm{P}}{(2\pi)^{3}}\int d^{3} \bm{R} \, W(\bm{R},\bm{P})
  \langle \hat{T}^{\mu \nu}(\bm{r}) \rangle_{\bm{R},\bm{P}}, 
\label{eq:20}
\end{align}
where $W(\bm{R},\bm{P})$ represents the Wigner distribution 
given by 
\begin{align}
W(\bm{R},\bm{P}) &=\int \frac{d^{3} \bm{\Delta}}{(2\pi)^{3}}
   e^{-i\bm{\Delta}\cdot \bm{R}}   \tilde{\psi}^{*}\left(\bm{P} +
   \frac{\bm{\Delta}}{2}\right)   \tilde{\psi}\left(\bm{P} -
   \frac{\bm{\Delta}}{2}\right) \cr 
&=\int d^{3}\bm{z} \,e^{-i\bm{z}\cdot \bm{P}} {\psi}^{*}\left(\bm{R} -
  \frac{\bm{z}}{2}\right) {\psi}\left(\bm{R} +
  \frac{\bm{z}}{2}\right). 
\label{eq:Wig}
\end{align}
The average position $\bm{R}$ and momentum $\bm{P}$ are defined as
$\bm{R}=(\bm{r}'+\bm{r})/2$ and $\bm{P}=(\bm{p}'+\bm{p})/2$, respectively. 
$\bm{\Delta}=\bm{p}'-\bm{p}$ denotes the momentum transfer, which
enables us to get access to the internal structure of a particle. The
variable $\bm{z}=\bm{r}'-\bm{r}$ stands for the position separation
between the initial and final particles. 
The Wigner distribution contains information on the wave packet of a 
particle 
\begin{align}
 \psi(\bm{r}) = \langle \bm{r} |\psi \rangle  = \int \frac{d^3
  \bm{p}}{(2\pi^3)} e^{i\bm{p}\cdot \bm{r}} 
  \tilde{\psi}(\bm{p}), \ \ \  \tilde{\psi}(\bm{p}) =
  \frac{1}{\sqrt{2p^{0}}}\langle p| \psi \rangle, 
\end{align}
with the plane-wave states $|p\rangle$ and
  $|\bm{r}\rangle$ respectively normalized as  
 $\langle p'|p\rangle = 2 p^{0}(2\pi)^3
\delta^{(3)}(\bm{p}'-\bm{p})$ and $\langle \bm{r}'|\bm{r}\rangle
=  \delta^{(3)}(\bm{r}'-\bm{r})$. The position state
$|\bm{r}\rangle$ localized at $\bm{r}$ at time $t=0$ is defined
as a Fourier transform of the momentum eigenstate $|p\rangle$
\begin{align}
|\bm{r} \rangle = \int \frac{d^{3}\bm{p}}{(2\pi)^{3}\sqrt{2p^{0}}}
       e^{-i\bm{p}\cdot \bm{r}} | p \rangle. 
\end{align}
If we integrate
over the average position and momentum, then the probabilistic density
in either position or momentum space is recovered to be 
\begin{align}
  \int   \frac{d^{3}\bm{P}}{(2\pi)^{3}}\,W_{N}(\bm{R},\bm{P}) =|
  \psi_N\left(\bm{R} \right) |^{2},\;\;\;
\int d^{3}\bm{R}\,W_{N}(\bm{R},\bm{P}) =| \tilde{\psi}_N\left(\bm{P} \right)
  |^{2}.
\end{align}

Given $\bm{P}$ and $\bm{R}$, the matrix element $\langle
\hat{T}^{\mu \nu} (\bm{r})\rangle_{\bm{R},\bm{P}}$ conveys information on
the internal structure of the particle localized around the average
position $\bm{R}$ and average momentum $\bm{P}$. 
This can be expressed as the 3D Fourier transform of the matrix
element $\langle  p', \lambda' | \hat{T}^{\mu \nu}(0) | p, \lambda
\rangle$:   
\begin{align}
\langle \hat{T}^{\mu \nu}(\bm{r}) \rangle_{\bm{R},\bm{P}}= \langle
  \hat{T}^{\mu \nu}(0) \rangle_{-\bm{x},\bm{P}} = \int
  \frac{d^{3}\bm{\Delta}}{(2\pi)^{3}} e^{-i\bm{x} \cdot \bm{\Delta} }
  \frac{1}{\sqrt{2p^{0}}\sqrt{2p'^{0}}}\langle  p', \lambda' |
  \hat{T}^{\mu \nu}(0) | p, \lambda \rangle, 
  \label{eq:25}
\end{align} 
with the shifted position vector $\bm{x}=\bm{r}-\bm{R}$. 
The matrix element $\langle  p', \lambda' |
  \hat{T}^{\mu \nu}(0) | p, \lambda \rangle$ was discussed in the
  previous Section.

\subsection{EMT distributions in the 3D Breit frame}
Having integrated over $\bm{P}$ of Eq.~\eqref{eq:20}, we find that the
part of the wave packet can be factorized. Thus, the target in the BF
is understood as a localized state around $\bm{R}$ from the Wigner
perspective. In this frame, Eq.~\eqref{eq:25} is reduced to   
\begin{align}
T^{\mu \nu}_{\mathrm{BF}}(\bm{x},\lambda', \lambda) = \langle
  \hat{T}^{\mu \nu}(0) \rangle_{-\bm{x},\bm{0}}  
=  \int \frac{d^{3}\bm{\Delta}}{2P_{0}(2\pi)^{3}} e^{-i\bm{x} \cdot
  \bm{\Delta} }   \langle  p', \lambda' |
  \hat{T}^{\mu \nu}(0) | p, \lambda \rangle. 
\label{eq:26}
\end{align} 
From now on we use $\bm{r}$ instead of $\bm{x}$, i.e.,
$\bm{x}=\bm{r}-\bm{R} \to \bm{r}$. We expand the temporal component
of the EMT distributions in terms of the multipole distributions as
follows:  
\begin{align}
T^{00}_{\mathrm{BF}}(\bm{r},\lambda', \lambda) = \varepsilon_{0}(r)
  \delta_{\lambda'\lambda} + \varepsilon_{2}(r) \hat{Q}^{ij}
  Y^{ij}_{2}(\Omega_{r}), 
\label{eq:27}
\end{align}
where
\begin{align}
&\varepsilon_0(r) =  m {\tilde {\cal E}}_0(r), \ \varepsilon_2(r) =
  -{1\over 2m} r{d\over dr} {1\over r}{d\over dr} \, {\tilde {\mathcal
  E}}_2(r),  
\end{align}
with
\begin{align}
{\tilde {\mathcal E}}_{0,2}(r) = 2m \int{d^3\Delta\over
  2P_{0}(2\pi)^3} e^{-i{\bm\Delta}\cdot{\bm r}} {\mathcal E}_{0,2}(t)   .
\end{align}
Note that since the angle integration of $Y_2^{ij}$ 
vanishes, i.e., $\int d\Omega_{r}Y^{ij}_{2}(\Omega_{r})
= 0$, we have the normalization of the mass distribution
$\mathcal{E}_0(0)$ as follows: 
\begin{align}
\int d^{3}r \, T^{00}_{\mathrm{BF}}(\bm{r},\lambda',\lambda)= \int 
  d^{3}r  \, \varepsilon_{0}(r) \delta_{\lambda' \lambda}= m
  \mathcal{E}_{0}(0) \delta_{\lambda' \lambda}, 
\end{align} 
which gives $\mathcal{E}_{0}(0)=1$.

We can get the spin distribution by contracting $T_{\mathrm{BF}}^{0k}$
by $\epsilon^{ijk} r^j$
\begin{align}
J^{i}(\bm{r},\lambda', \lambda)&=\epsilon^{ijk} r^{j}T^{0
  k}_{\mathrm{BF}}(\bm{r},\lambda',
  \lambda) \cr 
&= \hat{S}^{j}_{\lambda' \lambda}   \int
  \frac{d^{3}\bm{\Delta}}{(2\pi)^{3}} e^{-i\bm{r} \cdot \bm{\Delta} } 
  \left[\bigg{(}\bar{\mathcal{J}}_{1}(t) + \frac{2}{3}t
  \frac{d\bar{\mathcal{J}}_{1}(t)}{dt}\bigg{)} \delta^{ij} +
  \bigg{(}\Delta^{i}\Delta^{j}-\frac{1}{3} \bm{\Delta}^{2}
  \delta^{ij}\bigg{)} \frac{d\bar{\mathcal{J}}_{1}(t)}{dt}\right] 
\label{eq:5s}
\end{align} 
with $\bar{\mathcal{J}}_{1}:=m\mathcal{J}_{1}/P_{0}$.
Integrating the angular-momentum distribution $J^{i}$ over $\bm{r}$
  yields $\int d^{3}r \, J^{i}(\bm{r},\lambda', \lambda) = 
  \hat{S}^{i}_{\lambda' \lambda} \mathcal{J}_{1}(0)$, so that we have
  the obvious normalization for spin, i.e., $\mathcal{J}_{1}(0)=1$. 

The spatial component of the EMT distributions can be parametrized in
terms of the elastic pressure and shear forces, given
by\cite{Polyakov:2019lbq}  
\begin{align}
T^{ij}_{\mathrm{BF}}({\bm r}, \lambda', \lambda ) &=p_0(r)
   \delta^{ij}+s_0(r)Y_2^{ij}(\Omega_{r})  + p_2(r)  \hat{Q}^{ij} +
  s_2(r)  
2 \left[\hat{Q}^{ip}Y_{2}^{pj}(\Omega_{r})+\hat
   Q^{jp}Y_{2}^{pi}(\Omega_{r}) -\delta^{ij} 
\hat{Q}^{pq}Y_{2}^{pq}(\Omega_{r})    \right] \cr 
&- \frac{1}{m^{2}} \hat Q^{pq} \partial^{p}  
\partial^{q} \left[ p_3(r) \delta^{ij} 
+ s_3(r) Y_2^{ij}(\Omega_{r})\right],
\end{align}
where
\begin{align}
&p_n(r) =  {1\over 6m} \ \frac{1}{r^2}\frac{d}{dr} r^2\frac{d}{dr}\
  {\tilde {\cal D}}_n(r), \ s_n(r) = -{1\over 4m} r{d\over dr} {1\over
  r}{d\over dr} \, {\tilde {\mathcal D}}_n(r)  
  \label{eq:pressure_3D}
\end{align}
with
\begin{align}
{\tilde {\mathcal D}}_{n}(r) = 2m \int{d^3\Delta\over 2P_{0}(2\pi)^3}
  e^{-i{\bm\Delta}\cdot{\bm r}} {\mathcal D}_{n}(t).  
\label{eq:D-termden}
\end{align}
The conservation of the EMT current furnishes the equilibrium
conditions as 
\begin{align}
p_{n}^{\prime}(r) + \frac{2}{3} s_{n}^{\prime}(r) +
  \frac{2}{r}s_{n}(r) = 0,  \ \ \ (n=0,2,3). 
\end{align}
The pressures given in Eq.~\eqref{eq:pressure_3D}
satisfy the von Laue conditions: 
\begin{align}
\int d^{3}r \, p_{n}(r) = 0.
\end{align} 
Note that all the monopole and quadrupole distributions are related to
the dimensionless constant $D$-terms 
\begin{align}
\mathcal{D}_{n}(0) = m \int d^{3}r \, r^2 p_{n}(r) = -\frac{4}{15}m
  \int d^{3} r \, r^{2} s_{n}(r), \ \ \ (n=0,2,3). 
\end{align}

\subsection{EMT distributions in the 2D Breit frame}
The EF distributions depend on the impact parameter $x_{\perp}$
($\bm{r}=(\bm{x}_{\perp}, \, x_{z})$) and momentum
$\bm{P}=(\bm{0},P_{z})$ where a spin-one particle moves along the
$z$-direction without loss of generality. In this frame,
Eq.~\eqref{eq:25} is reduced to   
\begin{align}
T^{\mu \nu}_{\mathrm{EF}}(\bm{x}_{\perp},P_{z},\lambda', \lambda):=
  \int dx_{z} \langle \hat{T}^{\mu \nu}(0) \rangle_{-\bm{r},\bm{0}}  
=  \int
  \frac{d^{2}\bm{\Delta}_{\perp}}{2P_{0}(2\pi)^{2}}
  e^{-i\bm{x}_{\perp} \cdot \bm{\Delta}_{\perp} } 
  \langle  p', \lambda' |
  \hat{T}^{\mu \nu}(0) | p, \lambda \rangle \bigg{|}_{\Delta_{z}=0}.
\label{eq:6}
\end{align} 
Before investigating the EMT distributions in the IMF, one should
first separate the geometrical contributions from the relativistic
ones. By doing that, we can explicitly show that the relativistic
corrections are different from the geometrical ones. 

In this subsection, we examine the distributions in the 2D BF by taking 
$P_{z}\to 0$. The temporal component of the EMT current in the 2D EF
is given by  
\begin{align}
 T^{00}_{\mathrm{EF}}(\bm{x}_{\perp},0,\lambda',
  \lambda)&=  \delta_{3\lambda} \delta_{\lambda'3}
  \varepsilon^{\mathrm{(2D)}}_{(0,0)}(x_{\perp})+
  \delta_{\sigma'\sigma}
   \varepsilon^{\mathrm{(2D)}}_{(0,1)}(x_{\perp})
+  \hat{Q}^{ij} X^{ij}_{2}(\theta_{x_{\perp}}) 
  \varepsilon^{\mathrm{(2D)}}_{2}(x_{\perp}). 
\label{eq:j0}
\end{align}
The mass distribution is defined by 
\begin{align}
\varepsilon^{\mathrm{(2D)}}_{(0,l)}(x_{\perp}) =  m {\tilde {\cal
  E}}_{(0,l)}(x_{\perp}), \ \varepsilon^{\mathrm{(2D)}}_2(x_{\perp}) =
  -{1\over 2m} x_{\perp}{d\over dx_{\perp}} {1\over x_{\perp}}{d\over
  dx_{\perp}} \,  {\tilde {\mathcal E}}_2(x_{\perp}), 
\end{align}
where
\begin{align}
{\tilde {\mathcal E}}_{(0,l),\, 2}(x_{\perp}) = 2m
  \int{d^2\Delta\over 2P_{0}(2\pi)^2}
  e^{-i{\bm\Delta}_{\perp}\cdot{\bm x}_{\perp}} {\mathcal E}_{(0,l),\,
  2}(t), \ \ \ (l=0,1). 
\end{align}
At the zero momentum transfer $t=0$, the normalizations of the form
factors in the 2D BF are found to be 
\begin{align}
\mathcal{E}_{(0,l)}(0)= \frac{1}{m} \int d^{2} x_{\perp}
  \varepsilon^{\mathrm{(2D)}}_{(0,l)}(x_{\perp}), \ \
  \mathcal{E}_{2}(0)= -\frac{m}{4}\int d^{2} x_{\perp} \,
  x^{2}_{\perp} \varepsilon^{\mathrm{(2D)}}_{2}(x_{\perp}), 
  \mbox{ with } (l=0,1). 
\end{align}
One can see that the newly introduced form factors
$\mathcal{E}_{(0,l)}$ are normalized to be 
$\mathcal{E}_{(0,0)}(0)=\mathcal{E}_{(0,1)}(0)=\mathcal{E}(0) =1$
from Eq.~\eqref{eq:15}. These features will be revisited in 
Secion~\ref{sec:4}. 

We obtain the spin distribution as 
\begin{align}
J^{3}_{\mathrm{(2D)}}(\bm{x}_{\perp},\lambda', \lambda)&=\epsilon^{3jk}
  x_{\perp}^{j}T^{0 k}_{\mathrm{EF}}(\bm{x}_{\perp},0
  ,\lambda', \lambda) =  \hat{S}^{3}_{\lambda'
  \lambda}   \int   \frac{d^{2}\bm{\Delta}}{(2\pi)^{2}} 
e^{-i\bm{x}_{\perp} \cdot \bm{\Delta}_{\perp} }
  \left[ \bar{\mathcal{J}}_{1}(t) + 
t \frac{d\bar{\mathcal{J}}_{1}(t)}{dt}\right]
\label{eq:500}
\end{align} 
with $\bar{\mathcal{J}}_{1}(t):= m\mathcal{J}_{1}(t)/P_0$. 
Integrating the 2D angular-momentum distribution
$J^{3}_{\mathrm{(2D)}}$ over $\bm{x}_{\perp}$ yields $\int d^{2}x_{\perp} \,
J^{3}_{\mathrm{(2D)}}(\bm{x}_{\perp},\lambda', \lambda) =
\hat{S}^{3}_{\lambda' \lambda} \mathcal{J}_{1}(0)$, so that we get the 
normalization for spin, i.e., $\mathcal{J}_1(0)=1$. It gives
exactly the same constraint given in Eq.~\eqref{eq:5s}.

As done in the 3D case, the spatial component of the EMT distributions
in the EF can be constructed as follows 
\begin{align}
T^{ij}_{\mathrm{EF}}({\bm x}_\perp,0, \lambda', \lambda )
&= \int {d^2 \Delta \over 2P_{0}(2\pi)^2 } e^{-i \bm{\Delta}_{\perp}
  \cdot \bm{ x}_{\perp}} 
\langle p^\prime, \lambda^\prime \, |{\hat T}^{ij}(0)|p,\lambda
  \rangle  \cr
&=\bigg{(}p^{\mathrm{(2D)}}_0(x_{\perp}) -
  \frac{2}{3}p^{\mathrm{(2D)}}_2(x_{\perp})\bigg{)} \delta^{ij}
  \delta_{\sigma' \sigma}+s^{\mathrm{(2D)}}_0(x_{\perp})
  X_2^{ij}(\theta_{x_{\perp}}) \delta_{\sigma' \sigma}   \cr 
&+ \bigg{(}p^{\mathrm{(2D)}}_0(x_{\perp}) +
  \frac{4}{3}p^{\mathrm{(2D)}}_2(x_{\perp})\bigg{)} \delta^{ij}
  \delta_{\lambda' 3}\delta_{\lambda 3}+s^{\mathrm{(2D)}}_0(x_{\perp})
  X_2^{ij}(\theta_{x_{\perp}}) \delta_{\lambda' 3}\delta_{\lambda 3}
  \cr 
&+  2s^{\mathrm{(2D)}}_2(x_{\perp}) 
 \left[\hat{Q}^{ip}X_{2}^{pj}(\theta_{x_{\perp}})+\hat
  Q^{jp}X_{2}^{pi}(\theta_{x_{\perp}}) - \delta^{ij} \hat
  Q^{pq}X_{2}^{pq}(\theta_{x_{\perp}})  \right] \cr 
 & -\frac{1}{m^{2}}\hat{Q}^{pq} \partial^{p}\partial^{q}
   \bigg{(}s^{\mathrm{(2D)}}_{3}(x_{\perp}) X^{ij}(\theta_{x_{\perp}})
   + p^{\mathrm{(2D)}}_{3}(x_{\perp}) \delta^{ij}\bigg{)}, 
   \label{eq:Tij_2D}
\end{align}
where $p_n^{(\mathrm{2D})}(x_\perp)$ and
$s_n^{\mathrm{(2D)}}(x_{\perp})$ are expressed as 
\begin{align}
\label{eq:pressure_force_0}
p^{\mathrm{(2D)}}_n(x_{\perp})  =  {1\over 8m} \
   \frac{1}{x_{\perp}}\frac{d}{dx_{\perp}}
   x_{\perp}\frac{d}{dx_{\perp}}\ {\tilde {\cal D}}_n(x_{\perp}), \ \
   s^{\mathrm{(2D)}}_n(x_{\perp}) = -{1\over 4m} x_{\perp}{d\over
   dx_{\perp}} {1\over x_{\perp}}{d\over dx_{\perp}} \, {\tilde
   {\mathcal D}}_n(x_{\perp})
\end{align}
with
\begin{align}
{\tilde {\mathcal D}}_{n}(x_{\perp}) = 2m \int{d^2\Delta\over
  2E(2\pi)^2} e^{-i\bm{\Delta}_{\perp}\cdot\bm{x}_{\perp}} {\mathcal
  D}_{n}(t).  
\end{align}
A remarkable difference between 2D BF and 3D BF
distributions is that the $p_{2}^{\mathrm{(2D)}}$ distribution is
induced to the monopole structure as shown in the mass distribution,
which should be distinguished from the relativistic effects that will
be discussed later. One can find the monopole term of
$T_{\mathrm{EF}}^{ij}$ in Eq.~\eqref{eq:Tij_2D}. As previously shown,
we obtain the equilibrium conditions by the conservation of the EMT
current 
\begin{align}
p_{n}^{\mathrm{(2D)}\prime}(x_{\perp}) + \frac{1}{2}
  s_{n}^{\mathrm{(2D)}\prime}(x_{\perp}) +
  \frac{1}{r}s_{n}^{\mathrm{(2D)}}(x_{\perp}) = 0, \ \ \ (n=0,2,3) ,
\end{align}
which result in the von Laue stability conditions for all the monopole 
and quadrupole pressures: 
\begin{align}
\int d^{2}x_{\perp} p^{\mathrm{(2D)}}_{n}(x_{\perp})=0.
\end{align}
They are related to the dimensionless constant $D$-terms
\begin{align}
\mathcal{D}_{n}(0) = 2m \int d^{2}x_{\perp} \, x_{\perp}^2
  p^{\mathrm{(2D)}}_{n}(x_{\perp}) = -\frac{1}{2}m \int d^{2}
  x_{\perp} \, x_{\perp}^{2} s^{\mathrm{(2D)}}_{n}(x_{\perp}), \ \ \
  (n=0,2,3). 
\end{align}

Once we have the spatial component of the EMT distributions, we can
look into how the 2D stress tensor provides information on the 
internal forces that make a hadron stable.  It would be interesting to 
investigate the strong force fields and visualize them in 2D 
space. The internal local force fields are given by  
\begin{align}
\hat{\bm{x}}_{\perp}^{i}T^{ij}_{\mathrm{EF}}
  (\bm{x}_{\perp},0,\lambda',\lambda)=  
  \frac{dF_{r}}{dS_{r}}\hat{\bm{x}}^{j}_{\perp} +
  \frac{dF_{\theta}}{dS_{r}}\hat{\bm{\theta}}^{j}_{\perp}, \ \
  \hat{\bm{\theta}}_{\perp}^{i}T^{ij}_{\mathrm{EF}}
(\bm{x}_{\perp},0,\lambda',\lambda)=
  \frac{dF_{r}}{dS_{\theta}}\hat{\bm{x}}^{j}_{\perp} +
  \frac{dF_{\theta}}{dS_{\theta}}\hat{\bm{\theta}}^{j}_{\perp}. 
\end{align}
Each force field acting on the infinitesiaml area is derived as 
\begin{align}
\frac{dF_{r}}{dS_{r}}&= \bigg{[}
  \left(p^{\mathrm{(2D)}}_{0} +
   \frac{1}{2}s^{\mathrm{(2D)}}_{0}\right) 
  - \frac{1}{3} \left(2p^{\mathrm{(2D)}}_{2}+
  s^{\mathrm{(2D)}}_{2}+\frac{1}{m^{2}}\left[
-\frac{s^{\mathrm{(2D)}\prime}_{3}}{2x_{\perp}}
-\frac{p^{\mathrm{(2D)}\prime}_{3}}{x_{\perp}}+
\frac{2s^{\mathrm{(2D)}}_{3}}{x^{2}_{\perp}}\right]
\right)\bigg{]} \delta_{\sigma' \sigma} \cr 
&+\bigg{[}
  \left(p^{\mathrm{(2D)}}_{0}+\frac{1}{2}s^{\mathrm{(2D)}}_{0}\right)
  + \frac{2}{3} \left(2p^{\mathrm{(2D)}}_{2}+
  s^{\mathrm{(2D)}}_{2}+\frac{1}{m^{2}}\left[
-\frac{s^{\mathrm{(2D)}\prime}_{3}}{2x_{\perp}}-
\frac{p^{\mathrm{(2D)}\prime}_{3}}{x_{\perp}}+\frac{
2s^{\mathrm{(2D)}}_{3}}{x^{2}_{\perp}}\right]\right)\bigg{]}
  \delta_{\lambda' 3}\delta_{3 \lambda} \cr 
&+ \hat{Q}^{rr}_{\lambda' \lambda} \frac{1}{m^{2}}
\left[-p^{\mathrm{(2D)}\prime \prime}_{3}+
\frac{p^{\mathrm{(2D)}\prime}_{3}}{x_{\perp}}-
\frac{s^{\mathrm{(2D)}\prime \prime}_{3}}{2}+
\frac{s^{\mathrm{(2D)}\prime}_{3}}{2x_{\perp}}-
\frac{2s^{\mathrm{(2D)}}_{3}}{x^{2}_{\perp}}\right], \cr
\frac{dF_{r}}{dS_{\theta}}&=\frac{dF_{\theta}}{dS_{r}}= 
\hat{Q}^{r\theta}_{\lambda' \lambda}  \frac{1}{m^{2}}  
\left[-\frac{2s^{\mathrm{(2D)}\prime}_{3}}{x_{\perp}}+
\frac{2s^{\mathrm{(2D)}}_{3}}{x^{2}_{\perp}}\right],  \cr 
\frac{dF_{\theta}}{dS_{\theta}}&= \bigg{[} \left(
p^{\mathrm{(2D)}}_{0}-\frac{1}{2}s^{\mathrm{(2D)}}_{0}\right)
 - \frac{1}{3} \left(2p^{\mathrm{(2D)}}_{2} +  
s^{\mathrm{(2D)}}_{2} +\frac{1}{m^{2}}\left[
\frac{s^{\mathrm{(2D)}\prime}_{3}}{2x_{\perp}}-
\frac{p^{\mathrm{(2D)}\prime}_{3}}{x_{\perp}}\right]\right)\bigg{]} 
\delta_{\sigma' \sigma} \cr
&+\bigg{[} \left(p^{\mathrm{(2D)}}_{0}-
\frac{1}{2}s^{\mathrm{(2D)}}_{0}\right) + \frac{2}{3} 
\left(2p^{\mathrm{(2D)}}_{2}+ s^{\mathrm{(2D)}}_{2}+
\frac{1}{m^{2}}\left[
\frac{s^{\mathrm{(2D)}\prime}_{3}}{2x_{\perp}}-
\frac{p^{\mathrm{(2D)}\prime}_{3}}{x_{\perp}}\right]
\right)\bigg{]} \delta_{\lambda' 3}\delta_{3 \lambda} \cr
&+ \hat{Q}^{rr}_{\lambda' \lambda} \bigg{(}-2s^{\mathrm{(2D)}}_{2} 
+\frac{1}{m^{2}} \bigg{[}-p^{\mathrm{(2D)}\prime \prime}_{3}
 + \frac{p^{\mathrm{(2D)}\prime}_{3}}{x_{\perp}}+
\frac{s^{\mathrm{(2D)}\prime \prime}_{3}}{2} -
\frac{s^{\mathrm{(2D)}\prime}_{3}}{2x_{\perp}}\bigg{]}\bigg{)} 
+ \hat{Q}^{\theta\theta}_{\lambda' \lambda} \bigg{[}- 
2s^{\mathrm{(2D)}}_{2} - \frac{2}{m^{2}}
\frac{s^{\mathrm{(2D)}}_{3}}{x^{2}_{\perp}}\bigg{]}.
\label{eq:52}
\end{align}
If the target is polarized along the logitudual direction, there is no
$\theta$ dependence in the infinitesimal forces, such as
$\hat{Q}^{rr}_{33}=\hat{Q}^{\theta \theta}_{33}=1/3$ and
$\hat{Q}^{r\theta}_{33}=0$. However, if a target is polarized along
one axis of the transverse plane, the angle dependence on $\theta$
appears (see Appendix~\ref{app:c}).  

\subsection{EMT densities in the 2D infinite momentum frame}
As mentioned previously, the EMT densities in the 2D IMF do not
require any relativistic corrections, so that they have a
probabilistic meaning. In the IMF, we divide the mass and mechanical
densities respectively by the Lorentz factors $P_{0}/m$ and $m/P_{0}$
to remove the kinamatical divergence and suppression in these
densities:
\begin{align}
T^{00}_{\mathrm{IMF}}({\bm
  x}_{\perp},\lambda',\lambda)&:=T^{00}_{\mathrm{EF}}({\bm
  x}_{\perp},P_{z},\lambda',\lambda)
  \frac{m}{P_{0}} \bigg{|}_{P_{z}\to
  \infty}, \cr 
T^{ij}_{\mathrm{IMF}}(\bm{x}_{\perp},\lambda',\lambda)
&:=T^{ij}_{\mathrm{EF}} (\bm{x}_{\perp},P_{z},\lambda',\lambda) 
\frac{P_{0}}{m} \bigg{|}_{P_{z}\to \infty}. 
\end{align}
Thus, the $T^{00}_{\mathrm{IMF}}$ is normalized to be its mass $m$
instead of its momentum $P_{z}$. This normalization is the same as the
$T^{00}_{\mathrm{BF}}$ one. One should keep in mind that this mass
density $\sim m T^{++}/ P^{+}$ is actually connected to the momentum
density $T^{++}$ defined in the light-cone basis. It is
different from the higher-twist mass density that arises from the
\emph{bad} component of the EMT current. In this paper, the \emph{mass 
density in 2D IMF} indicates the momentum density normalized as a
hadron mass. Similarly, the normalization of the
$T^{ij}_{\mathrm{IMF}}$ is consistent with that of
$T^{ij}_{\mathrm{BF}}$.  The temporal component of the EMT current in 
the 2D IMF is given by  
\begin{align}
T^{00}_{\mathrm{IMF}}({\bm x}_{\perp},\lambda',\lambda)&  =
   \delta_{3\lambda}   \delta_{\lambda'3} 
    \varepsilon^{\mathrm{IMF}}_{(0,0)}(x_{\perp})+
    \delta_{\sigma'\sigma}
    \varepsilon^{\mathrm{IMF}}_{(0,1)}(x_{\perp})+ \epsilon^{3jk}
    \hat{S}^{j} X^{k}_{1}(\theta_{x_{\perp}})
    \varepsilon^{\mathrm{IMF}}_{1}(x_{\perp})+ \hat{Q}^{ij}
    X^{ij}_{2}(\theta_{x_{\perp}})
    \varepsilon^{\mathrm{IMF}}_{2}(x_{\perp}), 
\label{eq:j01}
\end{align}
where 2D mass densities in the IMF are defined as
\begin{align}
\varepsilon^{\mathrm{IMF}}_{(0,l)}(x_{\perp}) =  m {\tilde {\cal
  E}}^{\mathrm{IMF}}_{(0,l)}(x_{\perp}),\
  \varepsilon^{\mathrm{IMF}}_{1}(x_{\perp}) = -\frac{1}{2}
  \frac{d}{dx_{\perp}} {\tilde {\cal
  E}}^{\mathrm{IMF}}_{1}(x_{\perp}), \
  \varepsilon^{\mathrm{IMF}}_2(x_{\perp}) = -{1\over 2m}
  x_{\perp}{d\over dx_{\perp}} {1\over x_{\perp}}{d\over dx_{\perp}}
  \,  {\tilde {\mathcal E}}^{\mathrm{IMF}}_2(x_{\perp})
\label{eq:mass_2DIMF}
\end{align}
with
\begin{align}
{\tilde {\mathcal E}}^{\mathrm{IMF}}_{(0,l),\, n}(x_{\perp}) =
  \int{d^2\Delta\over (2\pi)^2} e^{-i{\bm\Delta}\cdot{\bm r}}
  {\mathcal E}^{\mathrm{IMF}}_{(0,l),\, n}(t), \ \ \ (n=1,2 \
  \mathrm{and} \ l=0,1) .
\end{align}

As discussed in the 2D BF, the geometrical difference between 2D
and 3D space brings about the induced monopole term. It
contributes to $\varepsilon^{\mathrm{IMF}}_{(0,l)}$ together with
the 3D BF distribution. In addition to that, 
$\varepsilon^{\mathrm{IMF}}_{(0,l)}$ is subject to the relativistic
effects arising from the Lorentz boost. On the other hand, the 
$\varepsilon^{\mathrm{IMF}}_{2}$ consists of the pure 3D BF
distribution together with the relativistic effects. 

Interestingly, the dipole mass density is induced in the 2D IMF as
shown in Eq.~\eqref{eq:j01}. On the other hand, the 2D BF mass
distributions in Eq.~\eqref{eq:j0} do not contain the dipole
one. It implies that the dipole mass density for a spin-1 particle is
induced by the Lorentz boost. At the zero momentum transfer $t=0$, the
normalizations of the GFFs in the 2D IMF are found to be 
\begin{align}
\mathcal{E}^{\mathrm{IMF}}_{(0,l)}(0)= \frac{1}{m} \int d^{2}
  x_{\perp} \varepsilon^{\mathrm{IMF}}_{(0,l)}(x_{\perp}), \ \ 
  \mathcal{E}^{\mathrm{IMF}}_{2}(0)= -\frac{m}{4}\int d^{2} x_{\perp}
  \, x^{2}_{\perp} \varepsilon^{\mathrm{IMF}}_{2}(x_{\perp}), \ \
  \mathrm{with} \ \ (l=1,2) .
\end{align}
Since the 2D angle integration of the irreducible tensor $X^{i\ldots
  i_{n}}_{n}$, we obtain the normalization of the 2D mass
distributions  
\begin{align}
\int d^{2} x_{\perp} \, T^{00}_{\mathrm{IMF}} 
(\bm{x}_{\perp},\lambda',\lambda)
&= \int d^{2}x_{\perp}  \, \left[
\varepsilon^{\mathrm{IMF}}_{(0,0)}(x_{\perp}) 
\delta_{\lambda' 3} \delta_{3 \lambda}+
\varepsilon^{\mathrm{IMF}}_{(0,1)}(x_{\perp}) 
\delta_{\sigma' \sigma} \right] \cr
&= m\left[\mathcal{E}^{\mathrm{IMF}}_{(0,0)}(0) 
\delta_{\lambda' 3} \delta_{3 \lambda}+
\mathcal{E}^{\mathrm{IMF}}_{(0,1)}(0) \delta_{\sigma' \sigma}
  \right]. 
\end{align}
As shown in Eq.~\eqref{eq:IMF_nor}, we find 
$\mathcal{E}^{\mathrm{IMF}}_{(0,0)}(0) =
\mathcal{E}^{\mathrm{IMF}}_{(0,1)}(0) =\mathcal{E}_{0}(0)=1$, so 
the $T^{00}_{\mathrm{IMF}}$ is properly normalized to its mass  
regardless of it spin polarization. 
We want to mention that the normalizations of the mass
and spin lead to
$\mathcal{E}^{\mathrm{IMF}}_{1}(0)=2\mathcal{J}_{1}(0)-2\mathcal{E}_{0}(0)=0$
in Eq.~\eqref{eq:IMF_nor}, which yields the 
interesting non-trivial constraint on the induced dipole mass
density: 
 \begin{align} 
 \mathcal{E}^{\mathrm{IMF}}_{1}(0)=  \int d^{2}
  x_{\perp} \, x_{\perp} \varepsilon^{\mathrm{IMF}}_{1}(x_{\perp})=0.
\label{eq:e1_constr}
 \end{align}

The spin density is obtained as 
\begin{align}
J^{3}_{\mathrm{IMF}}(\bm{x}_{\perp},\lambda', \lambda)&=\epsilon^{3jk}
  x_{\perp}^{j}T^{0  k}_{\mathrm{EF}}(\bm{x}_{\perp},P_{z}
  ,\lambda', \lambda) \bigg{|}_{P_{z}\to \infty}  \cr 
&= \hat{S}^{3}_{\lambda' \lambda}   \int
  \frac{d^{2}\bm{\Delta}}{(2\pi)^{2}} e^{-i\bm{x}_{\perp} \cdot
  \bm{\Delta}_{\perp} } 
  \left[ \mathcal{J}_{1}(t) + t \frac{d\mathcal{J}_{1}(t)}{dt}\right]
  \cr 
 & +i \epsilon^{3jl}\hat{Q}^{3l}_{\lambda' \lambda} \frac{1}{2m}
    \int   \frac{d^{2}\bm{\Delta}}{(2\pi)^{2}} e^{-i\bm{x}_{\perp}
    \cdot \bm{\Delta}_{\perp} } \Delta^{j} 
  \left[ 3\mathcal{J}_{2}(t) + 2 t
    \frac{d\mathcal{J}_{2}(t)}{dt}\right].  
\label{eq:501}
\end{align} 
Note that in the IMF the induced quadrupole structure yields the
additional contribution to the spin densities. 
The integration of $J^{3}_{\mathrm{IMF}}$ over $\bm{x}_{\perp}$
removes the quadrupole contribution, except for
$\mathcal{J}_{1}$. We thus have same normalization $\int 
d^{2}x_{\perp} \, J^{3}_{\mathrm{IMF}}(\bm{x}_{\perp},\lambda',
\lambda) = \hat{S}^{3}_{\lambda' \lambda} \mathcal{J}_{1}(0)$ with
$\mathcal{J}_{1}(0)=1$ as in the case of the 2D BF spin distribution.  

We can derive the spatial component of the EMT densities in the 2D IMF
as follows 
\begin{align}
T^{ij}_{\mathrm{IMF}}({\bm x}_{\perp},\lambda',\lambda) 
&=\bigg{(}p^{\mathrm{IMF}}_{(0,1)}(x_{\perp}) - 
\frac{2}{3}p^{\mathrm{IMF}}_2(x_{\perp})\bigg{)} 
\delta_{\sigma' \sigma} \delta^{ij}+
s^{\mathrm{IMF}}_{(0,1)}(x_{\perp}) \delta_{\sigma' \sigma} 
X_2^{ij}(\theta_{x_{\perp}})    \cr
&+ \bigg{(}p^{\mathrm{IMF}}_{(0,0)}(x_{\perp}) + 
\frac{4}{3}p^{\mathrm{IMF}}_2(x_{\perp})\bigg{)} 
\delta_{\lambda' 3}\delta_{\lambda 3} 
\delta^{ij}+s^{\mathrm{IMF}}_{(0,0)}(x_{\perp})  
\delta_{\lambda' 3}\delta_{\lambda 3} X_2^{ij}(\theta_{x_{\perp}})  \cr
&+  2s^{\mathrm{IMF}}_2(x_{\perp}) 
 \left[\hat{Q}^{ip}X_{2}^{pj}(\theta_{x_{\perp}})+
\hat{Q}^{jp} X_{2}^{pi}(\theta_{x_{\perp}}) - 
\delta^{ij} \hat{Q}^{pq}X_{2}^{pq}(\theta_{x_{\perp}})  \right] \cr 
& -\frac{1}{m^{2}} \hat{Q}^{pq} \partial^{p}\partial^{q} 
\bigg{(}s^{\mathrm{IMF}}_{3}(x_{\perp}) X^{ij}_{2}(\theta_{x_{\perp}}) 
+ p^{\mathrm{IMF}}_{3}(x_{\perp}) \delta^{ij}\bigg{)} \cr
&-\frac{2}{m}\epsilon^{lm3}\hat{S}^{l} \partial^{m} 
\bigg{(}s^{\mathrm{IMF}}_{1}(x_{\perp}) X^{ij}_{2}(\theta_{x_{\perp}}) 
+ p^{\mathrm{IMF}}_{1}(x_{\perp}) \delta^{ij}\bigg{)}.
\label{eq:61}
\end{align}
We then have five different equilibrium conditions:
\begin{align}
p_{(0,l),n}^{\mathrm{IMF}\prime}(x_{\perp}) + \frac{1}{2}
  s_{(0,l),n}^{\mathrm{IMF}\prime}(x_{\perp}) +
  \frac{1}{x_{\perp}}s_{(0,l),n}^{\mathrm{IMF}}(x_{\perp}) = 0, \;
  \mbox{ ($n=1,2,3$ and $l=0,\,1$)},
\end{align}
where  $p^{\mathrm{IMF}}_{(0,l),n}(x_{\perp})$ and
$s^{\mathrm{IMF}}_{(0,l),n}(x_{\perp})$ are defined as 
\begin{align}
\label{eq:pressure_force_01}
p^{\mathrm{IMF}}_{(0,l),n}(x_{\perp})  =  {1\over 8m} \
   \frac{1}{x_{\perp}}\frac{d}{dx_{\perp}}
   x_{\perp}\frac{d}{dx_{\perp}}\ {\tilde {\cal
   D}}^{\mathrm{IMF}}_{(0,l),n}(x_{\perp}), \ \ \
   s^{\mathrm{IMF}}_{(0,l),n}(x_{\perp}) = -{1\over 4m}
   x_{\perp}{d\over dx_{\perp}} {1\over x_{\perp}}{d\over dx_{\perp}}
   \, {\tilde {\mathcal D}}^{\mathrm{IMF}}_{(0,l),n}(x_{\perp}), 
  \end{align}
with 
  \begin{align}
{\tilde {\mathcal D}}^{\mathrm{IMF}}_{(0,l),n}(x_{\perp}) =
  \int{d^2\Delta\over (2\pi)^2}
  e^{-i\bm{\Delta}_{\perp}\cdot\bm{x}_{\perp}} {\mathcal
  D}^{\mathrm{IMF}}_{(0,l),n}(t),  \;
\mbox{ ($n=1,2,3$ and $l=0,\,1$)}.  
\end{align}

The difference between 2D BF and 2D IMF pressure and shear force
distributions solely results from the Lorentz boost effects. In
addition to that, $p^{\mathrm{IMF}}_{1}$ and $s^{\mathrm{IMF}}_{1}$
purely originate from the Lorentz boost effects. 
All the pressure densities satisfy the von Laue condition: 
\begin{align}
\int d^{2}x_{\perp} \,
   p^{\mathrm{IMF}}_{(0,l),n}(x_{\perp})=0.
\label{eq:65}
\end{align}
They are related to the constant $D$-terms 
\begin{align}
\mathcal{D}^{\mathrm{IMF}}_{(0,l),n}(0) = 2m \int d^{2}x_{\perp} \,
  x_{\perp}^2 p^{\mathrm{IMF}}_{(0,l),n}(x_{\perp}) = -\frac{1}{2}m
  \int d^{2} x_{\perp} \, x_{\perp}^{2}
  s^{\mathrm{IMF}}_{(0,l),n}(x_{\perp}),  \ \ \ (n=1,2,3 \
  \mathrm{and} \ l=0,1)  
\end{align}
The internal local force fields in the IMF are given by  
\begin{align}
  \hat{\bm{x}}_{\perp}^{i}T^{ij}_{\mathrm{IMF}}(\bm{x}_{\perp},
  \lambda',\lambda)=   \frac{dF_{r}}{dS_{r}}\hat{\bm{x}}^{j}_{\perp} + 
  \frac{dF_{\theta}}{dS_{r}}\hat{\bm{\theta}}^{j}_{\perp}, \ \
  \hat{\bm{\theta}}_{\perp}^{i}T^{ij}_{\mathrm{IMF}}(\bm{x}_{\perp},\lambda',\lambda)=
  \frac{dF_{r}}{dS_{\theta}}\hat{\bm{x}}^{j}_{\perp} +
  \frac{dF_{\theta}}{dS_{\theta}}\hat{\bm{\theta}}^{j}_{\perp}, 
\end{align}
where
\begin{align}
\frac{dF_{r}}{dS_{r}}&= \bigg{[}
   \left(p^{\mathrm{IMF}}_{(0,1)}+\frac{1}{2}s^{\mathrm{IMF}}_{(0,1)}\right)
 - \frac{1}{3} \left(2p^{\mathrm{IMF}}_{2}+
   s^{\mathrm{IMF}}_{2}+\frac{1}{m^{2}}\left[
-\frac{s^{\mathrm{IMF}\prime}_{3}}{2x_{\perp}}-
\frac{p^{\mathrm{IMF}\prime}_{3}}{x_{\perp}}+
\frac{2s^{\mathrm{IMF}}_{3}}{x^{2}_{\perp}}
\right]\right)\bigg{]} \delta_{\sigma' \sigma} \cr 
&+\bigg{[} \left(p^{\mathrm{IMF}}_{(0,0)}+
\frac{1}{2}s^{\mathrm{IMF}}_{(0,0)}\right) + \frac{2}{3} 
\left(2p^{\mathrm{IMF}}_{2}+  s^{\mathrm{IMF}}_{2}+
\frac{1}{m^{2}}\left[
-\frac{s^{\mathrm{IMF}\prime}_{3}}{2x_{\perp}}-
\frac{p^{\mathrm{IMF}\prime}_{3}}{x_{\perp}}+
\frac{2s^{\mathrm{IMF}}_{3}}{x^{2}_{\perp}}
\right]\right)\bigg{]} \delta_{\lambda' 3}\delta_{3 \lambda} \cr 
&+ \hat{Q}^{rr}_{\lambda' \lambda} \frac{1}{m^{2}}
\left[-p^{\mathrm{IMF}\prime \prime}_{3}+
\frac{p^{\mathrm{IMF}\prime}_{3}}{x_{\perp}}-
\frac{s^{\mathrm{IMF}\prime \prime}_{3}}{2}+
\frac{s^{\mathrm{IMF}\prime}_{3}}{2x_{\perp}}-
\frac{2s^{\mathrm{IMF}}_{3}}{x^{2}_{\perp}}\right]
-\epsilon^{lm3}\hat{S}^{l} X^{m}_{1}(\theta_{x_{\perp}})
 \frac{1}{m}\left[2p^{\mathrm{IMF}\prime}_{1} +
 s^{\mathrm{IMF}\prime}_{1}\right], \cr
\frac{dF_{r}}{dS_{\theta}}&=\frac{dF_{\theta}}{dS_{r}}
= \hat{Q}^{r\theta}_{\lambda' \lambda}  \frac{1}{m^{2}}  
\left[-\frac{2s^{\mathrm{IMF}\prime}_{3}}{x_{\perp}}+
\frac{2s^{\mathrm{IMF}}_{3}}{x^{2}_{\perp}}\right] - 
\frac{2}{m}\epsilon^{lm3}\hat{S}^{l}\hat{\theta}^{m}_{\perp} 
\frac{s^{\mathrm{IMF}}_{1}}{x_{\perp}}, \cr
\frac{dF_{\theta}}{dS_{\theta}}&= \bigg{[} 
\left(p^{\mathrm{IMF}}_{(0,1)}-\frac{1}{2}
s^{\mathrm{IMF}}_{(0,1)}\right) - \frac{1}{3}
 \left(2p^{\mathrm{IMF}}_{2} +  s^{\mathrm{IMF}}_{2} +
\frac{1}{m^{2}}\left[
\frac{s^{\mathrm{IMF}\prime}_{3}}{2x_{\perp}}-
\frac{p^{\mathrm{IMF}\prime}_{3}}{x_{\perp}}\right]
\right)\bigg{]} \delta_{\sigma' \sigma} \cr
&+\bigg{[} \left(p^{\mathrm{IMF}}_{(0,0)}-
\frac{1}{2}s^{\mathrm{IMF}}_{(0,0)}\right) + 
\frac{2}{3} \left(2p^{\mathrm{IMF}}_{2}+ 
s^{\mathrm{IMF}}_{2}+\frac{1}{m^{2}}\left[
\frac{s^{\mathrm{IMF}\prime}_{3}}{2x_{\perp}}-
\frac{p^{\mathrm{IMF}\prime}_{3}}{x_{\perp}}\right]
\right)\bigg{]} \delta_{\lambda' 3}\delta_{3 \lambda} \cr
&+ \hat{Q}^{rr}_{\lambda' \lambda}  \bigg{(} 
-2s^{\mathrm{IMF}}_{2} + \frac{1}{m^{2}} 
\bigg{[}-p^{\mathrm{IMF}\prime \prime}_{3} + 
\frac{p^{\mathrm{IMF}\prime}_{3}}{x_{\perp}}+
\frac{s^{\mathrm{IMF}\prime \prime}_{3}}{2} -
\frac{s^{\mathrm{IMF}\prime}_{3}}{2x_{\perp}}
\bigg{]}\bigg{)} + \hat{Q}^{\theta\theta}_{\lambda' \lambda}
 \bigg{[}- 2s^{\mathrm{IMF}}_{2} - \frac{2}{m^{2}}
\frac{s^{\mathrm{IMF}}_{3}}{x^{2}_{\perp}}\bigg{]}\cr
&-\epsilon^{lm3}\hat{S}^{l} X^{m}_{1}(\theta_{x_{\perp}}) 
\frac{1}{m}\left[2p^{\mathrm{IMF}\prime}_{1} - 
s^{\mathrm{IMF}\prime}_{1}\right].
\end{align}

\section{Abel transformation for the spin-1 particle\label{sec:4}}
In the case of the nucleon EM form factors~\cite{Kim:2021kum}, 
the 3D BF charge distributions are distinguished from 2D IMF charge
densities by the relativistic factor $\sqrt{1-t/4m^{2}}$ in the
integrand of the Fourier transform of the EM form factors. 
Since the relativistic factor generates the infinite order of the
Laplacians, it is technically difficult to connect the 3D
charge distributions in the BF to the 2D charge densities in IMF.
It requires us to impose the infinite number of boundary conditions on 
the differential equation. As for the Abel transformation for the
nucleon EMT form factors~\cite{Kim:2021jjf, Panteleeva:2021iip}, on
the other hand, we do not encounter such a
complexity~\cite{Kim:2021kum}, since such a Lorentz boost 
factor $m/E$ does not appear fortuitously. Thus, we can directly
connect the 3D BF distributions of the EMT form factors to the 2D IMF
densities without any technical problems for the nucleon. Furthermore, 
if we consider both the EM and EMT form factors for the higher-spin
particle, the relativistic effects raise similar problems in a more
complicated manner~\cite{Alexandrou:2009hs, Kim:2022bia}. In the
present work, we thus carry out the Abel transformation in projecting
the 3D BF distributions to the 2D BF densities instead of the 2D IMF
ones. 

The Abel transformation and its inverse transformation are defined as  
\begin{align}
A[g](x_{\perp}) =\mathcal{G}(x_{\perp}) = \int^{\infty}_{x_{\perp}}
  \frac{dr}{r} \frac{g(r)}{\sqrt{r^{2}-x_{\perp}^{2}}}, \ \ \  
g(r) =  - \frac{2}{\pi} r^{2} \int^{\infty}_{r} d x_{\perp}
  \frac{d\mathcal{G}(x_{\perp})}{dx_{\perp}} 
  \frac{g(r)}{\sqrt{x_{\perp}^{2}-r^{2}}}.   
\label{eq:Able}
\end{align}
$\mathcal{G}(b)$ is called the Abel image of the function $g(r)$. For
example, if there is no higher multipole distribution in the BF, the
Abel image of the monopole mass distribution is found to be 
\begin{align}
\int dx_{z} \langle \hat{T}^{00}(0) \rangle_{-\bm{r},\bm{0}} = \int
  dx_{z} \, \varepsilon_{0}(r) \delta_{\lambda' \lambda}
  =\varepsilon^{\mathrm{(2D)}}_{0}(x_{\perp}) \delta_{\lambda'
  \lambda}, \ \ \
  \varepsilon^{\mathrm{(2D)}}_{0}(x_{\perp})=\int^{\infty}_{x_{\perp}}
  dr \frac{2r \varepsilon_{0}(r)}{\sqrt{r^{2}-x^{2}_{\perp}}}. 
\label{eq:spherical_line}
\end{align}
The Abel transformation can be straightforwardly applied to the
nucleon and the pion, since they do not have any quadrupole
distributions. However, as we pointed out in Ref.~\cite{Kim:2022bia},
mapping the 3D distribution onto the 2D one in the presence of the
quadrupole structure brings about additional contributions. Collecting 
all the angle-dependent Abel images in 3D space, we are able to
reconstruct them in 2D space. It has been already discussed in various
contexts~\cite{Bracewell:1956, Barrett:1984,Leonhardt,
  Natterer:2001}. While the single Abel image is enough for a
spherically symmetric distribution, the angle-dependent distribution
requires more than one Abel image~\cite{Abel}. The  scanning in all
directions are required to project angle-dependent 3D distributions to
2D ones in general. In fact, the number of the scans depends on the
shape of the distributions~\cite{Bracewell:1956}. In our case, we need
two Abel images only, which is a very special case of the anisotropic 
distributions. To generalize them, we should introduce the Radon
transformation~\cite{Radon}. The Abel transformation we keep using is
just a special case of the Radon transformation and is deeply related
to it~\cite{Leonhardt, Natterer:2001}.  In the present case, 
we need to integrate $\varepsilon(r) Y^{ij}(\Omega_{r})$ over the
$z$-axis for each 3D angle. The angle-dependent Abel transformation
can be analytically achieved as follows:
\begin{align}
 \int dx_{z} \, \varepsilon_{2}(r) Y^{ij}(\Omega_{r})
  \hat{Q}^{ij}_{\lambda' \lambda}
  =\varepsilon^{\mathrm{(2D)}}_{2}(x_{\perp}) X^{ij}_{2}
  (\theta_{x_{\perp}}) \hat{Q}^{ij} +\Delta_{\varepsilon}(x_{\perp})
  \left( -\frac{1}{3}\delta_{\sigma'\sigma} +
  \frac{2}{3}\delta_{\lambda' 3}\delta_{3\lambda} \right),  
\label{eq:Abel_BF}
\end{align}
with
\begin{align}
\varepsilon^{\mathrm{(2D)}}_{2}(x_{\perp})=\int^{\infty}_{x_{\perp}}
  dr \frac{2x^{2}_{\perp}
  \varepsilon_{2}(r)}{r\sqrt{r^{2}-x^{2}_{\perp}}}, \; \; \;
  \Delta_{\varepsilon}(x_{\perp})=\int^{\infty}_{x_{\perp}} dr
  \frac{(3x^{2}_{\perp}-2r^{2})
  \varepsilon_{2}(r)}{r\sqrt{r^{2}-x^{2}_{\perp}}}. 
\end{align}
Employing the above expression, we derive the explicit connections
between the 2D and 3D distributions in the BF by the Abel
transformations
\begin{align}
&\varepsilon^{\mathrm{(2D)}}_{(0,0)}(x_{\perp})+
2\varepsilon^{\mathrm{(2D)}}_{(0,1)}(x_{\perp})
  = 6 \int^{\infty}_{x_{\perp}} \frac{r
  dr}{\sqrt{r^{2}-x^{2}_{\perp}}} \varepsilon_{0}(r) =
  3\varepsilon^{\mathrm{(2D)}}_{0}(x_{\perp}),  
\label{eq:2D_E2}
\\  
&\Delta_{\varepsilon}(x_{\perp}):=\varepsilon^{\mathrm{(2D)}}_{(0,0)} 
(x_{\perp})-\varepsilon^{\mathrm{(2D)}}_{(0,1)}(x_{\perp})
  = -\frac{\partial^{2}_{(2D)}}{4m}
  \tilde{\mathcal{E}}_{2}(x_{\perp}), 
\label{eq:Delta_E}
\end{align}
Under this transformation, we observe that the rank-$2$ irreducible
tensor in 3D space is reduced to the rank-$2$ irreducible tensor
in 2D space and a part of the diagonal contributions leaks out to
the rank-$0$ irreducible tensor in the 2D space. This \emph{induced
  monopole}  distribution $\Delta_{\varepsilon}$ is responsible for
the splitting of the mass distributions with the longitudinally and
transversely polarized spins. Note that the 2D spin distributions of
the nucleon were intensively discussed in Refs.~\cite{Lorce:2017wkb,
  Schweitzer:2019kkd, Kim:2021jjf}. 

The nucleon monopole pressure and shear forces were investigated in
Ref.~\cite{Panteleeva:2021iip, Kim:2021jjf} by means of the Abel
transformation. 
When it comes to a non-spherical hadron, all the quadrupole pressure
and shear force distributions are connected to each other via the Abel 
transformation in the same manner: 
\begin{align}
&p^{\mathrm{(2D)}}_{n}(x_{\perp})+\frac{1}{2}s^{\mathrm{(2D)}}_{n}(x_{\perp})
  =  \int^{\infty}_{x_{\perp}} \frac{r dr}{\sqrt{r^{2}-x^{2}_{\perp}}}
  \left(p_{n}(r) + \frac{2}{3}s_{n}(r) \right), \cr 
&s^{\mathrm{(2D)}}_{n}(x_{\perp})  = 2 \int^{\infty}_{x_{\perp}}
  \frac{x^{2}_{\perp} dr}{r\sqrt{r^{2}-x^{2}_{\perp}}} s_{n}(r). 
\label{eq:Abelpshden}
\end{align}
Note that the definitions of the pressure and shear forces are
different from those in Ref.~\cite{Panteleeva:2021iip} by factor
$1/2$. 

\section{Numerical results and discussion\label{sec:5}}
Since we want to investigate the general features of the mechanical
structure of the spin-1 particle, we will not use a specific model but
employ a simplified toy model for the GFFs. We utilize the simple
multipole parametrization for the $t$ dependence. The  
mass and spin form factors are normalized to be 1 at $t=0$. The
quadrupole mass form factor at $t=0$ is also taken to be 1. On the
other hand, the values of the $D$-term form factors at $t=0$ are
unknown. To determine them, we need results from models (see
for example Refs.~\cite{Sun:2020wfo,Freese:2019bhb}). However, we take
arbitrary negative values for the constant $D$-terms at $t=0$,
assuming that their negative values impose the stability conditions on
the spin-1 particle. As observed in Eq.~\eqref{eq:IMF_FF}, the
kinematical factor $\tau$ enters in the expressions for the GFFs in
the IMF. This means that when we carry out the inverse 2D Fourier
transforms to obtain the EMT densities in the IMF, $\tau$ may cause
the divergence. While this may have a physical meaning as in the case
of the pion, we do not know if this is the case also for the spin-1
particle. In the current work, we will assume that the EMT densities
do not have any singular behavior. So, we parametrize the GFFs by
using generically the following quadrupole type of the parametrization: 
\begin{align}
G (t)=\frac{G(0)}{(1-t/\Lambda^{2})^{4}}.
\end{align}
Here, we introduce the cutoff value $\Lambda=2 m_{\rho}$, where
$m_\rho$ is the $\rho$-meson mass. The values of $\mathcal{D}_n(0)$
are taken to be
$\mathcal{D}_0(0)=\mathcal{D}_2(0)=\mathcal{D}_3(0)=-1$. Since the
values of all the $D$-term form factors are taken to be the same, we
have the same values of $p_n(x_\perp)$ and $s_n(x_\perp)$ in the 2D
BF. This choice of $D$-term values yields merit that we can
comprehensively examine the relativistic effects when we move from the
2D BF to the 2D IMF, as will be observed later.

\subsection{Mass distributions in the 2D transverse plane}
The mass distributions in the 2D BF can be obtained by the Abel
transformation of those in the 3D BF, as shown in
Eqs.~\eqref{eq:spherical_line} and~\eqref{eq:Abel_BF}. The Abel
transformation of the quadrupole mass distribution in the 3D BF,
$\varepsilon_2(r)$, yields the induced monopole distribution
$\Delta_\varepsilon(x_\perp)$. Thus, we have three different terms for
the mass distributions in the 2D BF: $\varepsilon_0(x_\perp)$,
$\Delta_\varepsilon (x_\perp)$, and $\varepsilon_2(x_\perp)$. 
\begin{figure}[htp]
\includegraphics[scale=0.27]{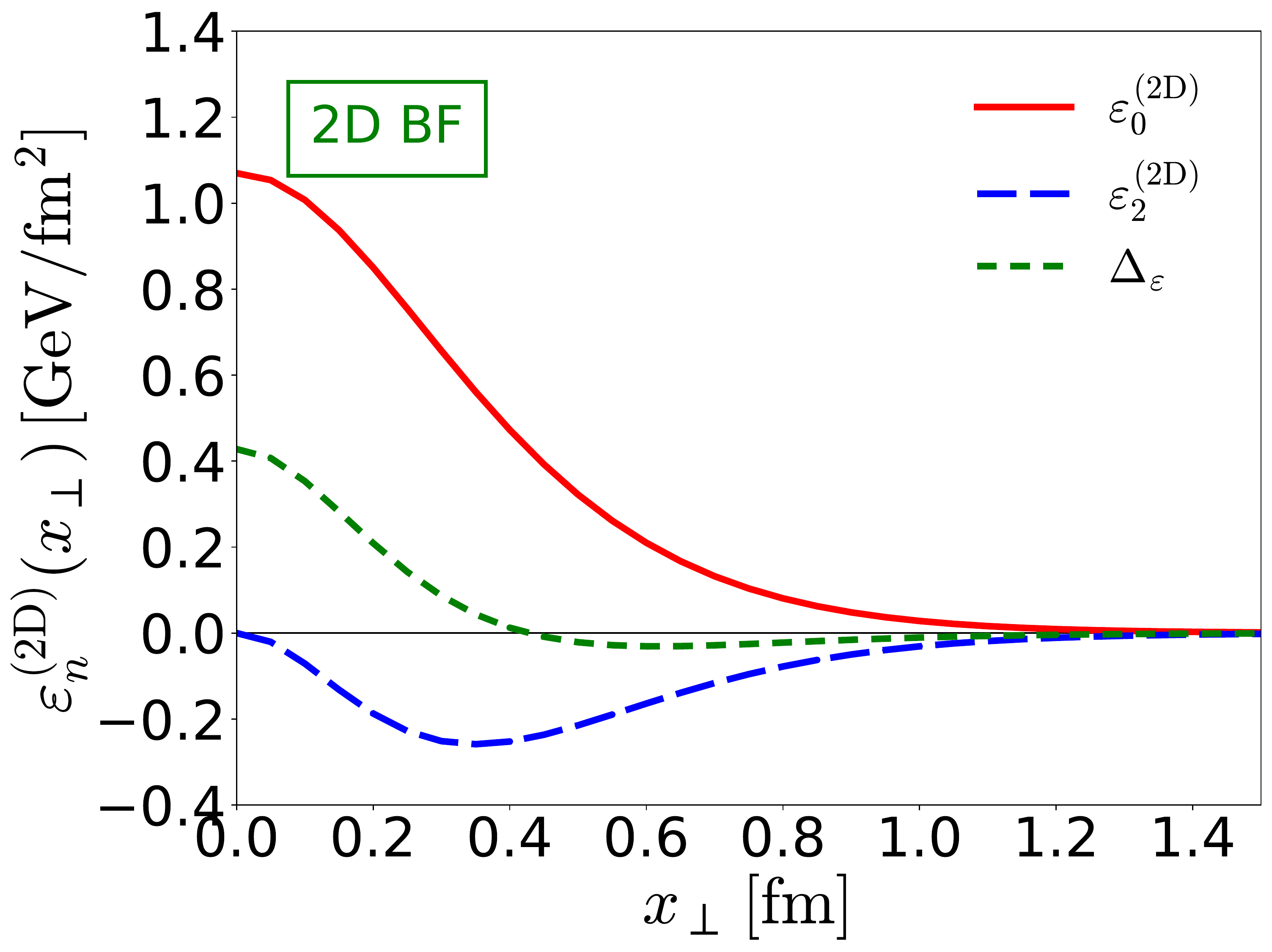} \;\;\;\;\;\;
\includegraphics[scale=0.27]{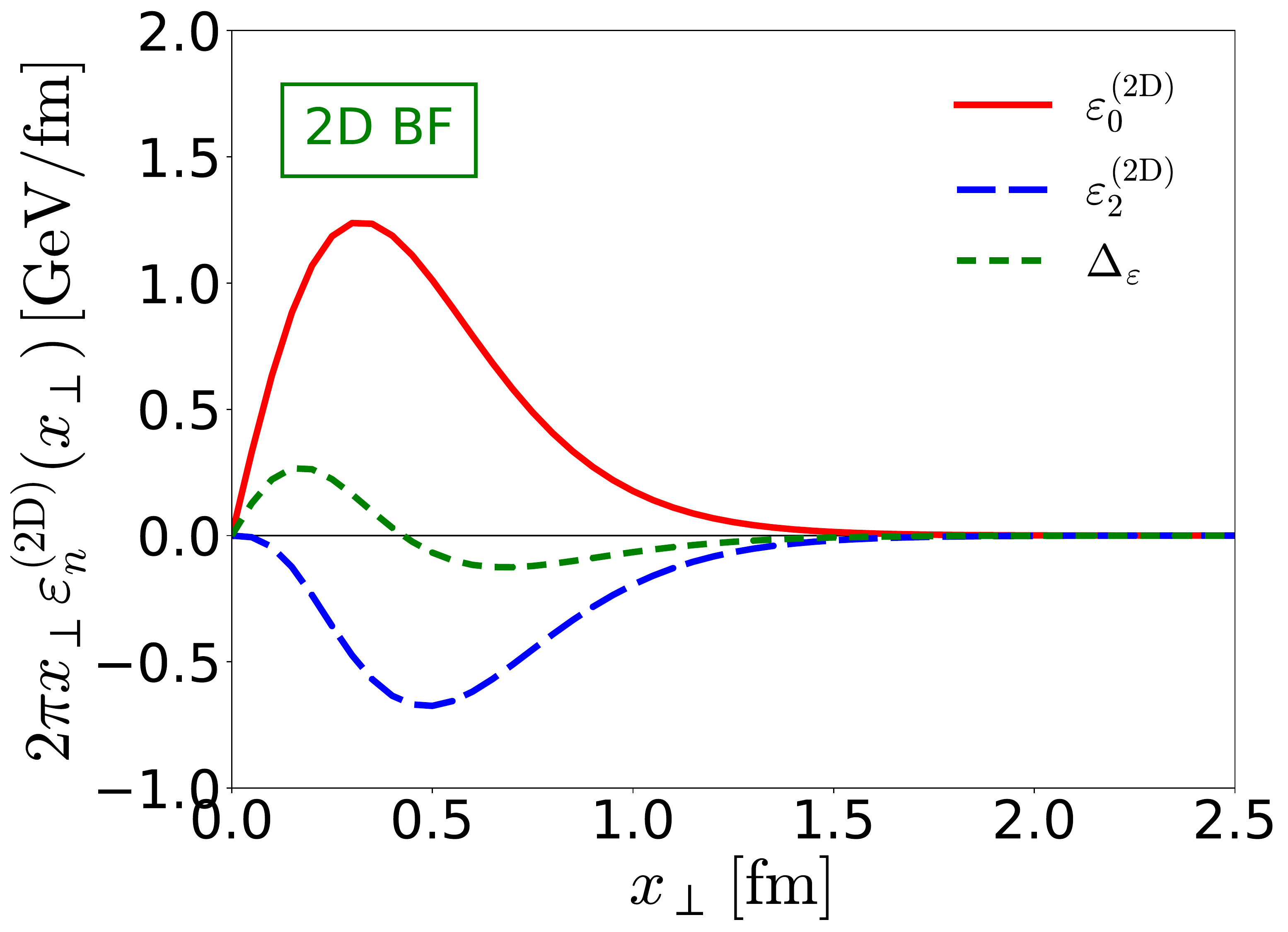}
\caption{Mass distributions of a spin-1 particle in the 2D Breit frame. 
In the left panel, the solid, dashed, and short-dashed curves draw the
numerical results for $\varepsilon_0^{(\mathrm{2D})}$,
$\varepsilon_2^{(\mathrm{2D})}$, and $\Delta_\varepsilon$ defined in 
Eq.~\eqref{eq:spherical_line} and Eq.~\eqref{eq:2D_E2}, respectively.
In the right panel, those weighted by $2\pi
x_\perp$ are exhibited. 
\label{fig:1}
}  
\end{figure}
We draw the numerical results for them in Fig.~\ref{fig:1}. The
monopole mass distribution is dominant over the induced monopole and
quadrupole ones. The quadrupole mass density is negative in the whole
region of $x_\perp$ whereas the induced monopole mass distribution,
$\Delta_\varepsilon(x_\perp)$, is positive till it reaches around 0.4
fm and then turns negative. 
\begin{figure}[htp]
\includegraphics[scale=0.27]{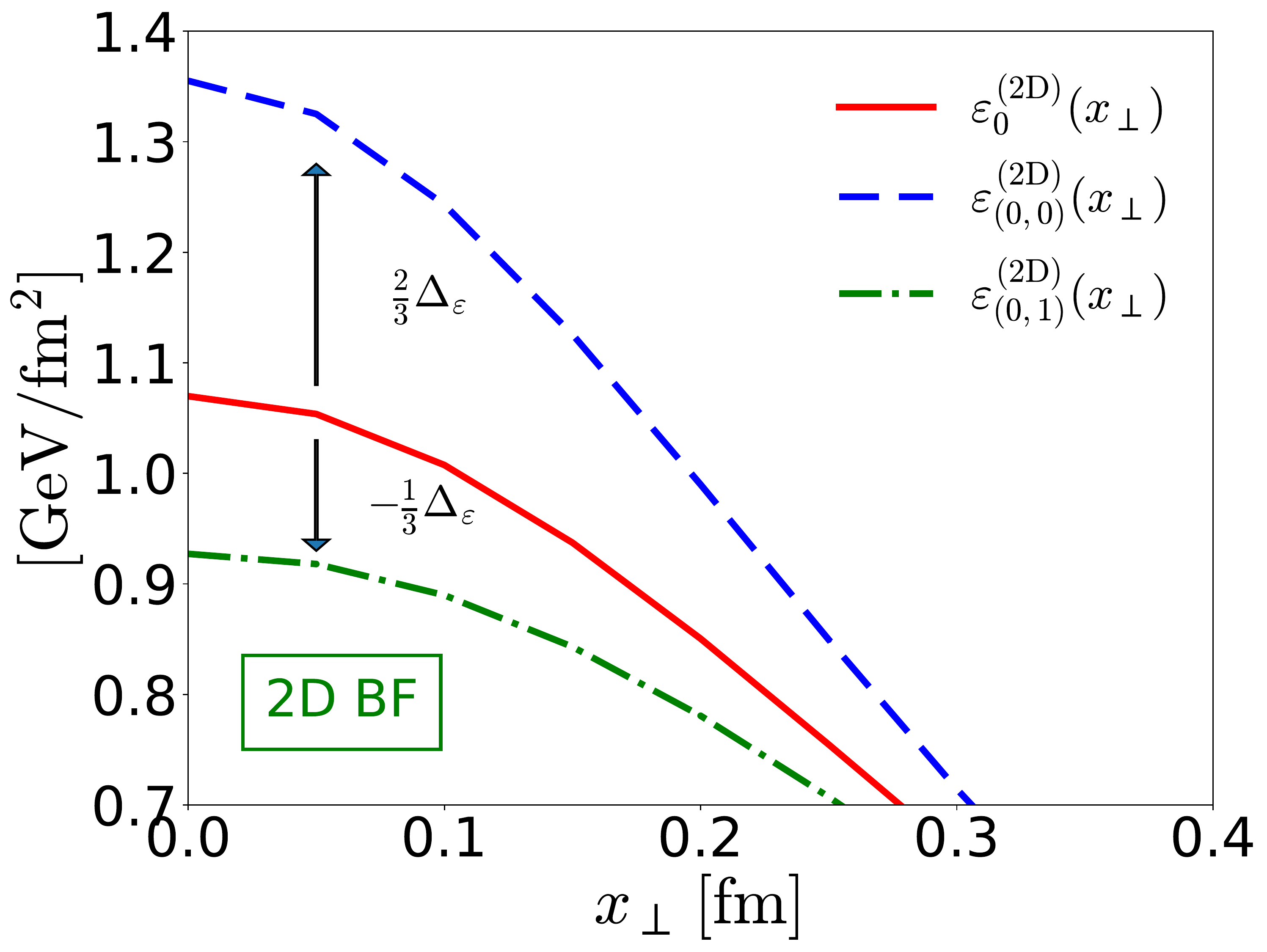}
\caption{Mass densities of a spin-1 particle in the 2D Breit frame. 
The solid, dashed, and dot-dashed curves show the numerical results
for $\varepsilon_0^{(\mathrm{2D})}$, $\varepsilon_{(0,0)}^{(\mathrm{2D})}$, and
$\varepsilon_{(0,1)}^{(\mathrm{2D})}$. $\Delta_\varepsilon$ is 
defined in Eq.~\eqref{eq:Delta_E}.
}  
\label{fig:2}
\end{figure}
As shown in Eqs.~\eqref{eq:2D_E2} and~\eqref{eq:Delta_E},
$\varepsilon_{(0,0)}^{(\mathrm{2D})}$ and
$\varepsilon_{(0,1)}^{(\mathrm{2D})}$ can be expressed in terms of
$\varepsilon_0^{(\mathrm{2D})}$ and $\Delta_\varepsilon$:
\begin{align}
 \varepsilon_{(0,0)}^{(\mathrm{2D})}(x_\perp) =
  \varepsilon_0^{(\mathrm{2D})}(x_\perp) +   \frac23
  \Delta_\varepsilon (x_\perp),\;\;\;
 \varepsilon_{(0,1)}^{(\mathrm{2D})} =
  \varepsilon_0^{(\mathrm{2D})}(x_\perp) -   \frac13
  \Delta_\varepsilon (x_\perp) . 
\end{align}
Thus, $\varepsilon_{(0,0)}^{(\mathrm{2D})}$ and
$\varepsilon_{(0,1)}^{(\mathrm{2D})}$ are split by
$\Delta_\varepsilon$ as shown in Fig.~\ref{fig:2}. It indicates that
the magnitude of $\varepsilon_{(0,0)}^{(\mathrm{2D})}$ is larger than
$\varepsilon_{(0,1)}^{(\mathrm{2D})}$. Since the contribution of the
inner part of the nodal point cancels out that of the outer part,
$\Delta_{\varepsilon}$ does not affect the normalization of the
mass as shown in the following integration 
\begin{align}
\int d^{2} x_{\perp} \, \Delta_{\varepsilon} (x_{\perp}) = -\int d^{2}
  x_{\perp} \,  \frac{\partial^{2}_{\mathrm{(2D)}}
\tilde{\mathcal{E}}_{2}(x_{\perp})}{4m}=0.
\end{align}
The induced monopole mass distribution changes only the shape of the
mass distribution.

\begin{figure}[htp]
\includegraphics[scale=0.27]{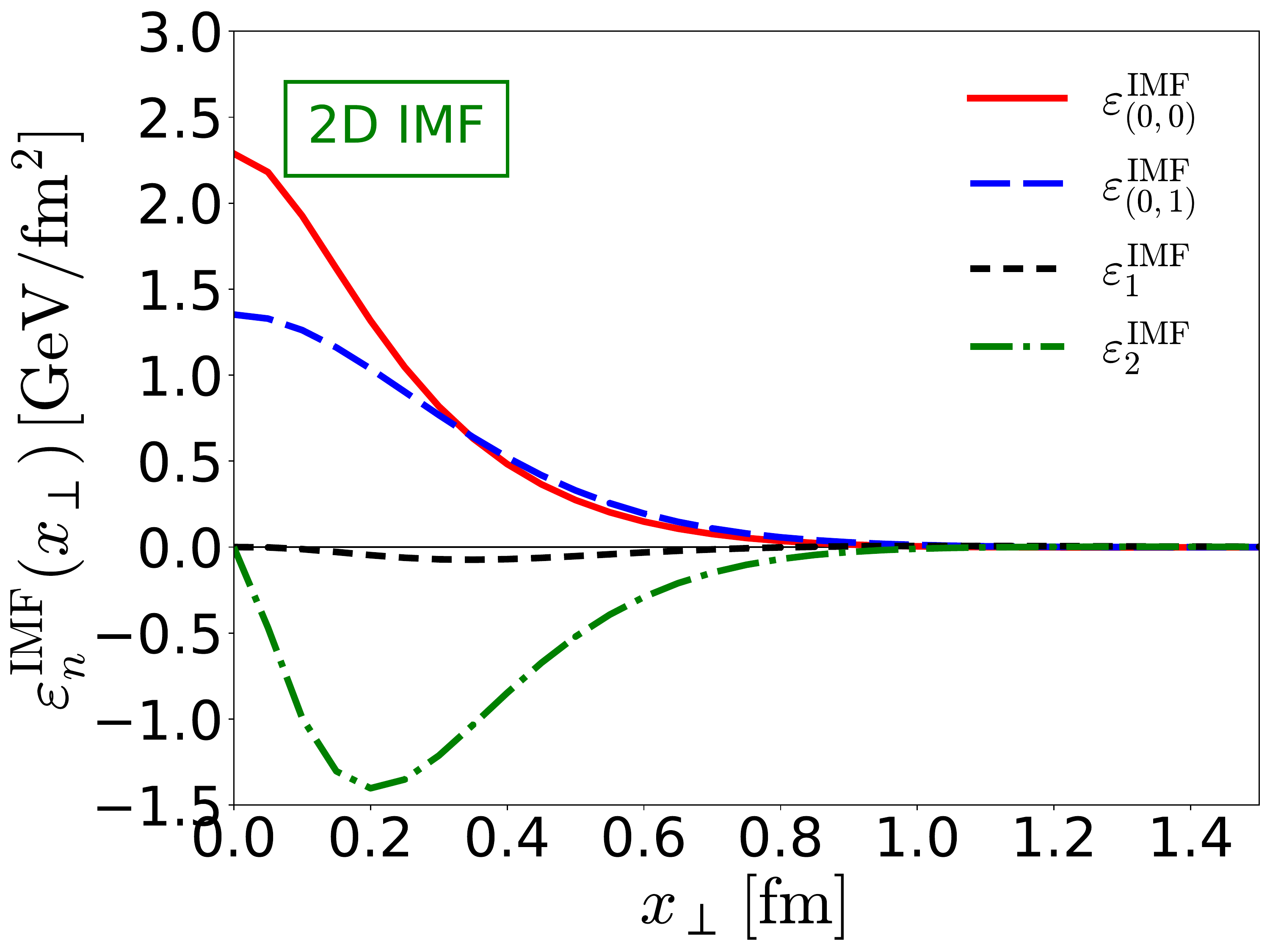}\;\;\;\;\;\;
\includegraphics[scale=0.27]{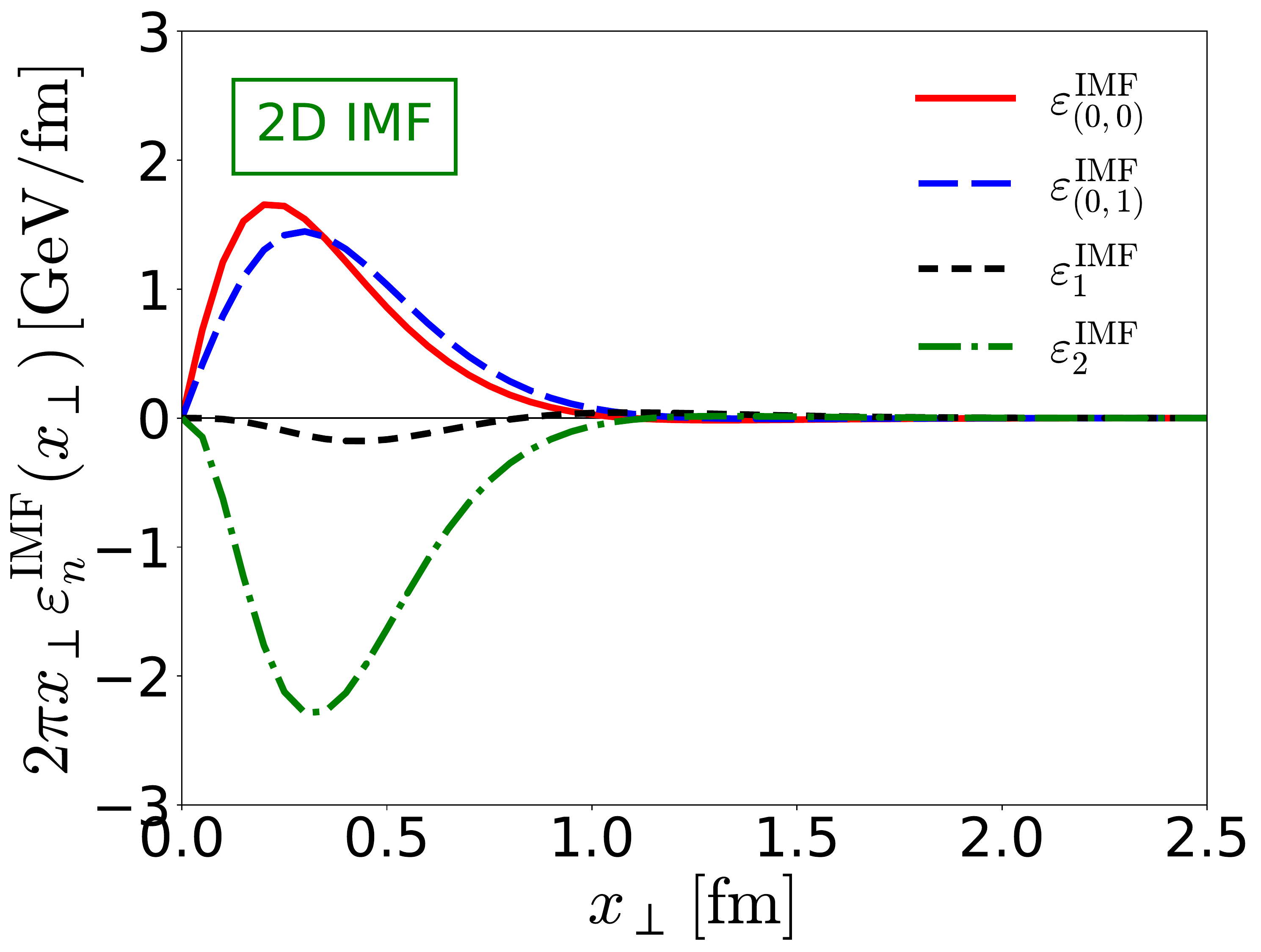}
\caption{Transverse mass densities of a spin-1 particle in the 2D
  infinite-momentum frame. 
  The solid and dashed curves draw the 2D
  mass densities $\varepsilon_{(0,0)}^{\mathrm{IMF}}$ and
  $\varepsilon_{(0,1)}^{\mathrm{IMF}}$, whereas the short-dashed and
  dot-dashed ones depict $\varepsilon_{1}^{\mathrm{IMF}}$ and
  $\varepsilon_{2}^{\mathrm{IMF}}$. The expressions for these mass
  densities are given in Eq.~\eqref{eq:mass_2DIMF}. 
  In the right panel, we draw those weighted by $2\pi x_\perp$.
\label{fig:3}
} 
\end{figure}
Figure~\ref{fig:3} depicts the transverse mass densities in the 2D
IMF. We have the following relations:
\begin{align}
\varepsilon_{(0,0)}^{\mathrm{IMF}}(x_{\perp}) > 0, \;\;\;  
\varepsilon_{(0,1)}^{\mathrm{IMF}}(x_{\perp}) > 0. 
\end{align}
The induced transverse dipole mass density is much smaller than the
other transverse mass densities. Note that
$\varepsilon_{1}^{\mathrm{IMF}}(x_{\perp})$ amd
$\varepsilon_{2}^{\mathrm{IMF}}(x_{\perp})$ vanish at $x_\perp =0$.
As discussed in the previous section, the integral of the
$x_{\perp}$-weighted induced dipole mass density over $x_{\perp}$
vanishes: 
 \begin{align} 
 \int d^{2}
  x_{\perp} \, x_{\perp} \varepsilon^{\mathrm{IMF}}_{1}(x_{\perp})=0.
 \end{align}
Note that this condition is highly nontrivial. The right panel depicts
the mass distributions weighted by $2\pi x_\perp$.

\begin{figure}[htp]
\includegraphics[scale=0.8]{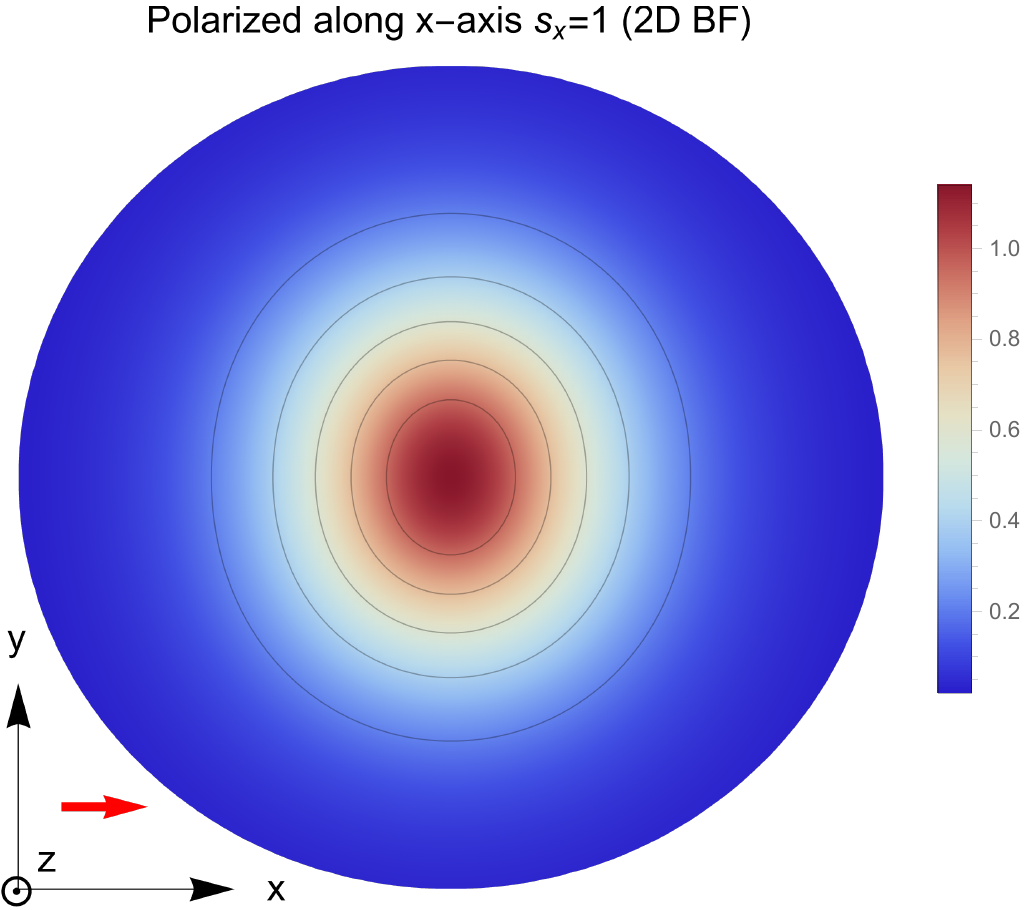}
\includegraphics[scale=0.8]{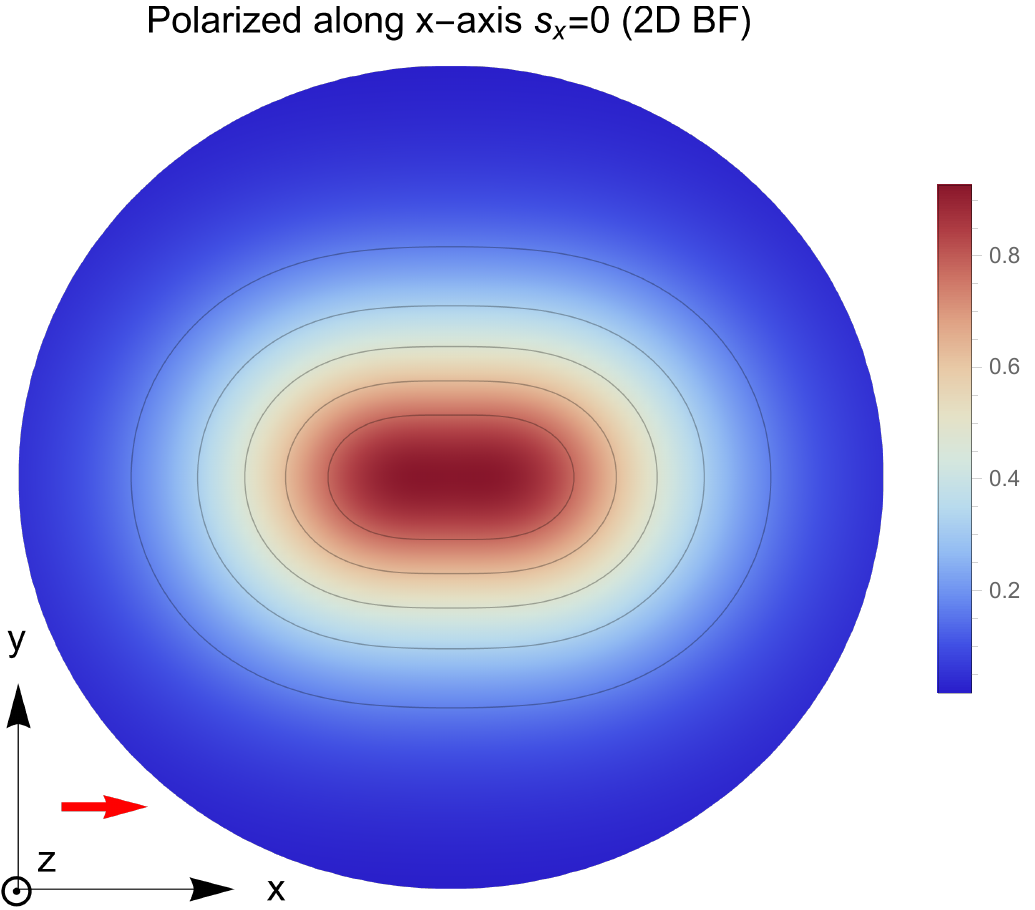}
\includegraphics[scale=0.8]{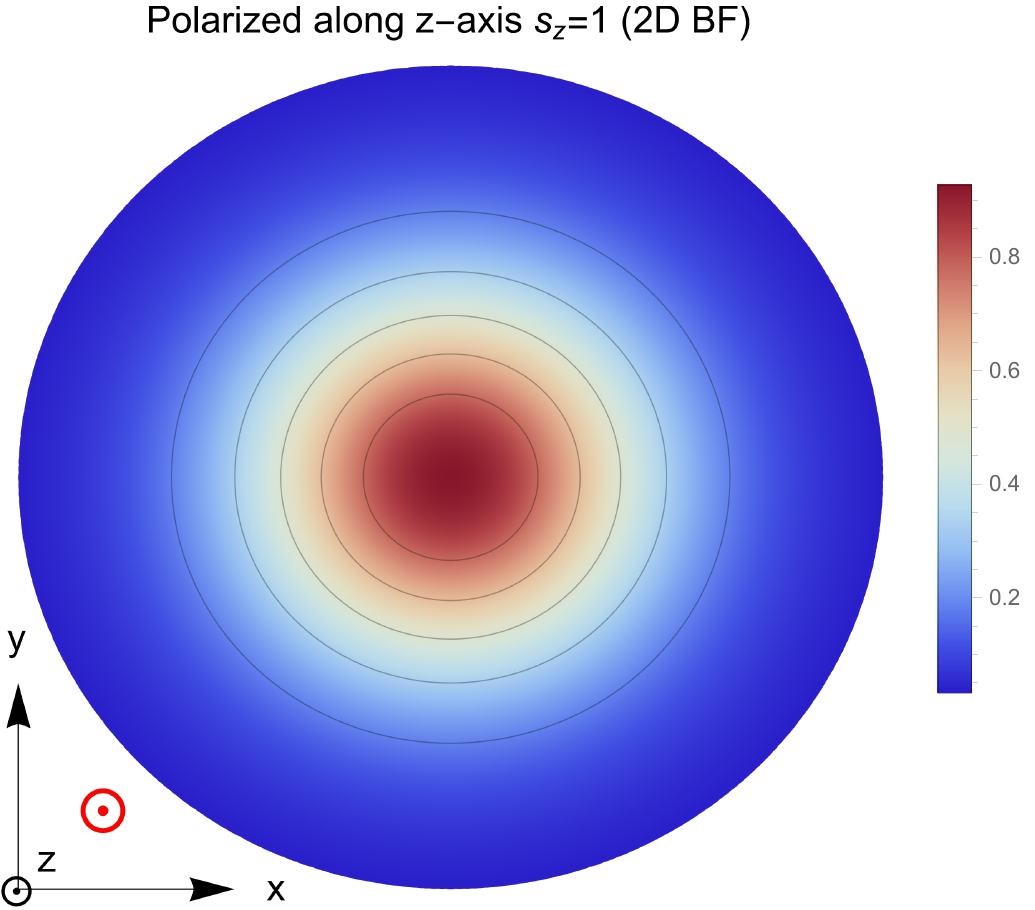}
\includegraphics[scale=0.8]{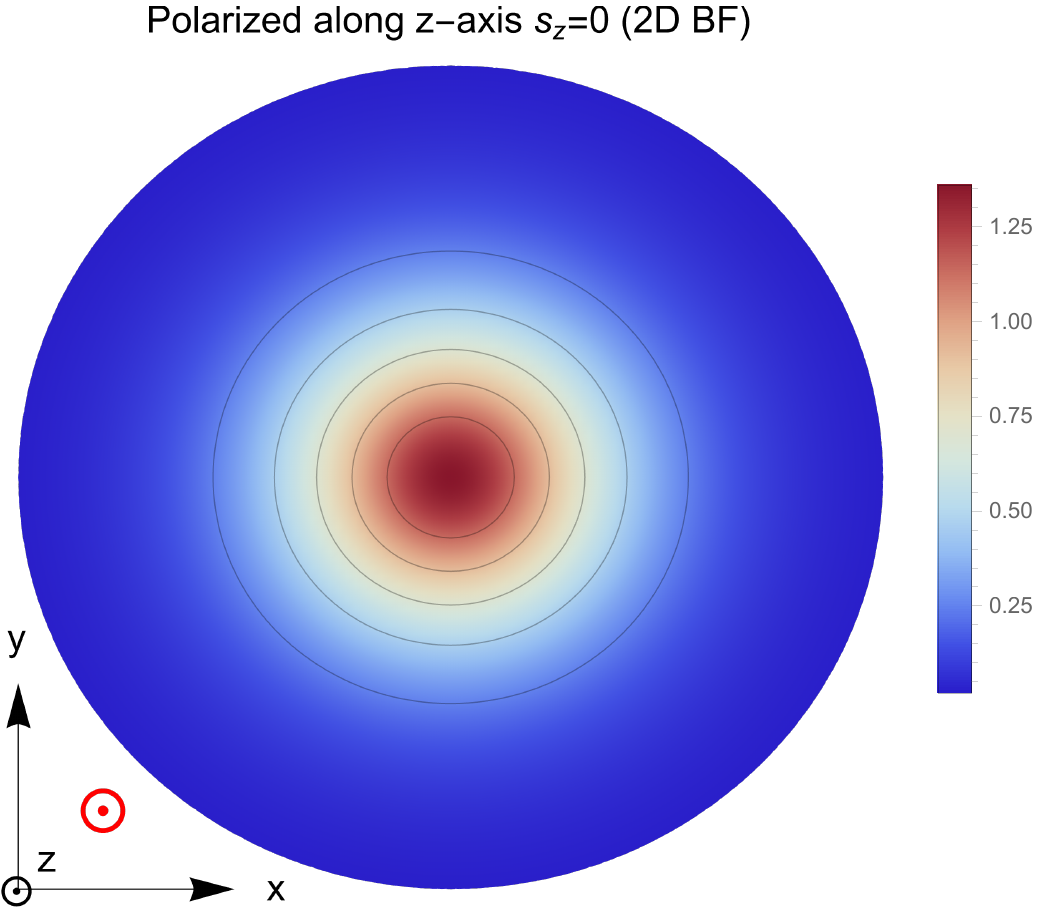}
\caption{$T^{00}(x_{\perp})$ visualized in the 2D BF by choosing
  a specific polarization. In the upper-left (upper-right) panel, we
  draw the mass distribution when the spin-1 particle is polarized
  with $s_x=1$ ($s_x=0$). In the lower-left (lower-right) panel, we
  illustrate that with $s_z=1$ ($s_z=0$).
}  
\label{fig:4}
\end{figure}

It is of great use to visualize the 2D mass distributions for the
polarized spin-1 particle, so that we can see how its mass
is distributed and influenced under the polarization.
As shown explicitly in Appendix~\ref{app:b}, we can derive the
expressions for the mass distribution in the 2D BF by choosing
specifically the polarization. In Fig.~\ref{fig:4}, we draw
$T^{00}_{\mathrm{EF}}(\bm{x}_{\perp},0,\lambda',  \lambda)$  by
choosing four different polarizations. 
The upper-left panel of Fig.~\ref{fig:4} illustrates the mass
distribution of the spin-1 particle with $s_x=1$ when it is polarized
along the $x$ axis. As written in Eq.~\eqref{eq:appb-x1}, the mass
distribution contains all contributions. The quadrupole
term $\cos2\theta \varepsilon_2^{(2\mathrm{D})}(x_{\perp})/4$
introduces the angular dependence of the mass distribution. As shown
in Fig.~\ref{fig:1}, the quadrupole term is negative over the whole region
in the 2D BF. Putting them together, we observe that the mass
distribution in the 2D BF is squeezed into a 2D prolate form. On the
other hand, when the spin-1 particle is polarized along the $x$-axis
with $s_x=0$ chosen, the scalar term
$\varepsilon_{(0,0)}^{(2\mathrm{D})} (x_\perp)$ vanishes, so that we
have only two different contributions, i.e.,
$\varepsilon_{(0,1)}^{(2\mathrm{D})}$ and $-\cos2\theta
\varepsilon_2^{(2\mathrm{D})}/2$ as given in
Eq.~\eqref{eq:appb-x0}. Since we have already shown in 
Fig.~\ref{fig:3} that $\varepsilon_{(0,1)}^{(2\mathrm{D})}$ is smaller
than $\varepsilon_{(0,0)}^{(2\mathrm{D})}$ and we have a larger
positive quadrupole contribution to the mass distribution, compared to   
the previous case ($s_x=1$), we expect that the mass distribution
should be compressed along the $y$-axis. Thus, its shape becomes
oblate as shown in the upper-right panel of Fig.~\ref{fig:4}. When the
spin-1 particle is polarized along the $z$-axis, the scalar
contribution survives only regardless of choosing a specific value of
$s_z$. Thus, the mass distribution of the spin-1 particle polarized
along the $z$-axis is always in a spherical shape, as visualized in the
lower panel of Fig.~\ref{fig:4}. 

\begin{figure}[htp]
\includegraphics[scale=0.8]{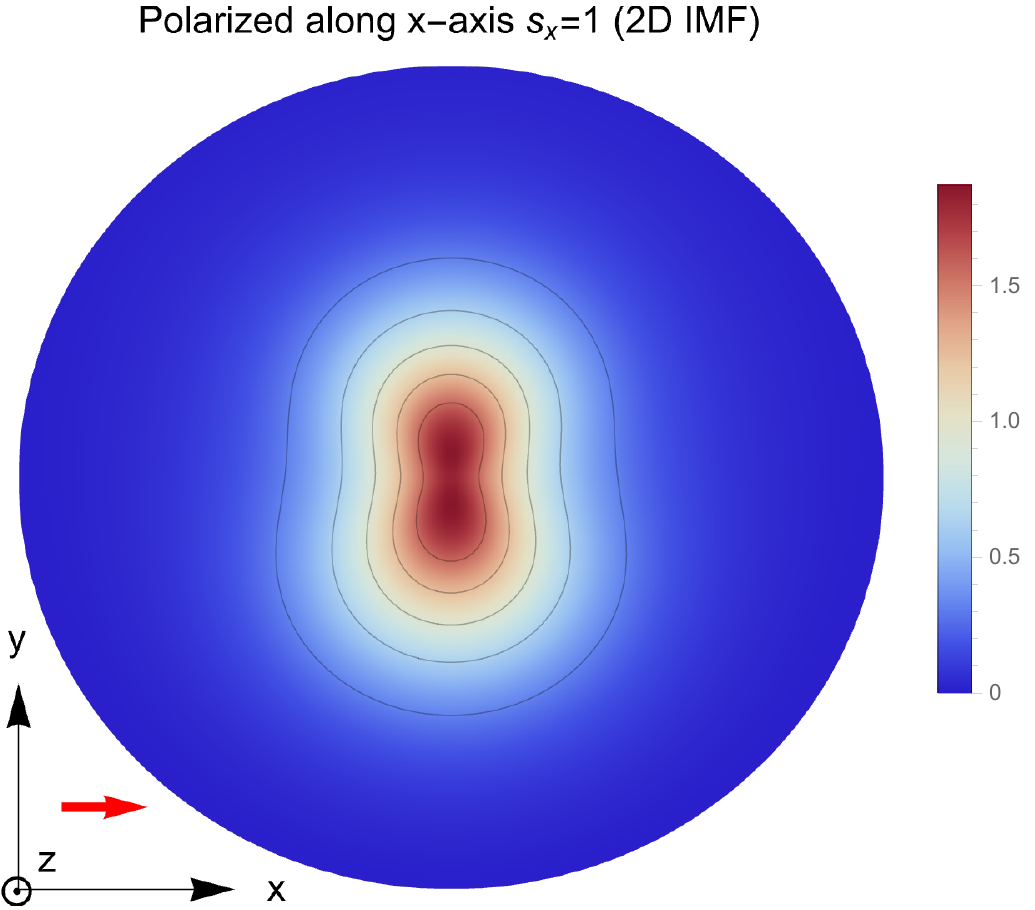}
\includegraphics[scale=0.8]{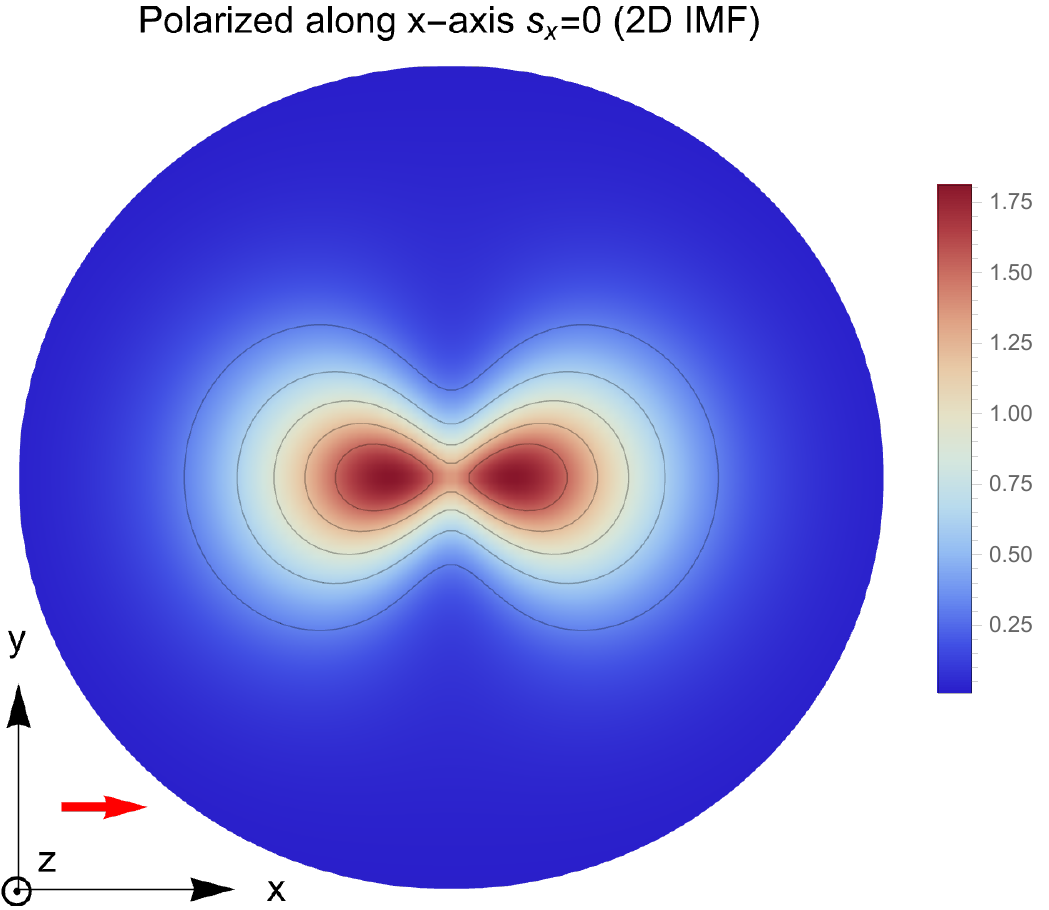}
\includegraphics[scale=0.8]{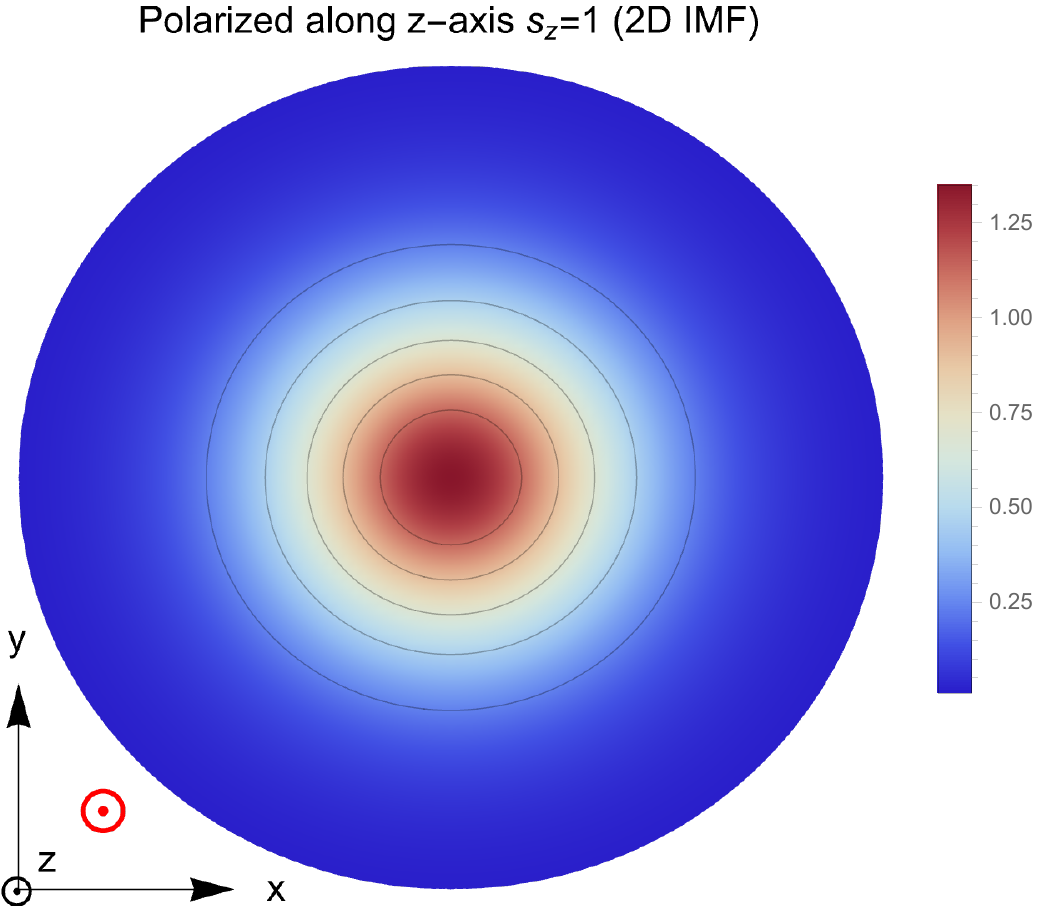}
\includegraphics[scale=0.8]{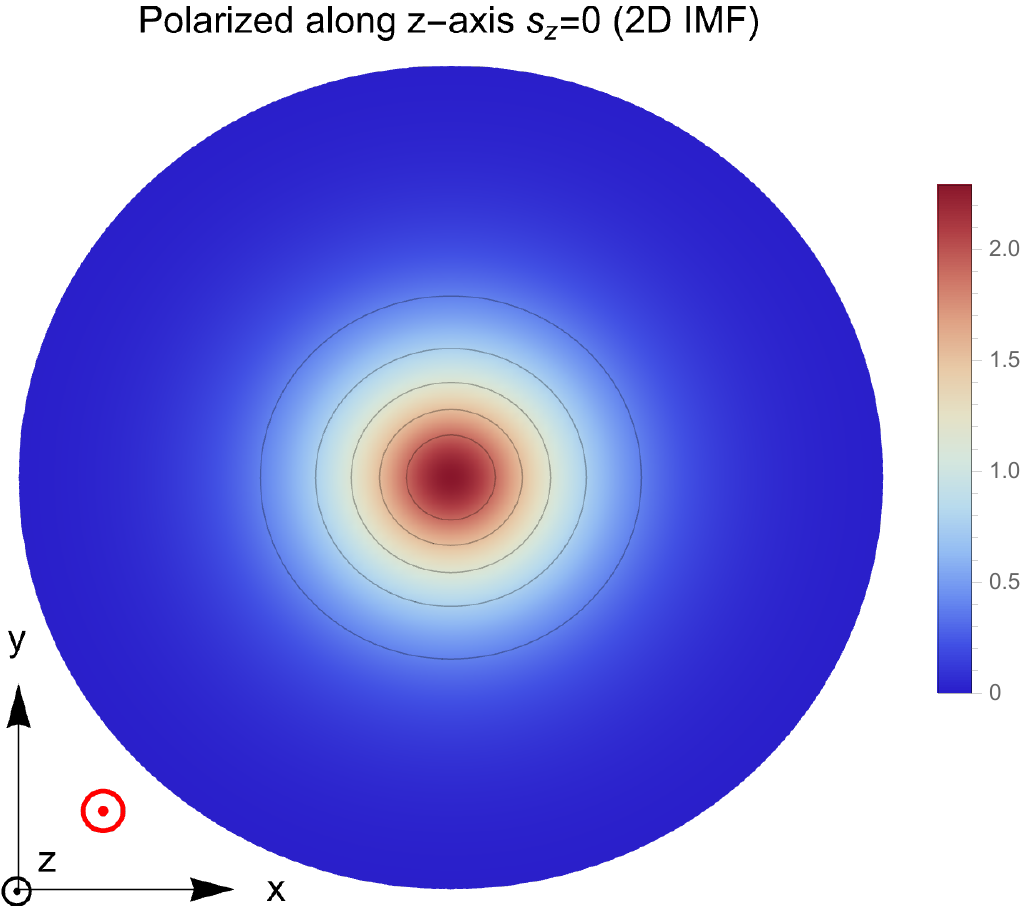}
\caption{
$T^{00}(x_{\perp})$ visualized in the 2D IMF by choosing
  a specific polarization. In the upper-left (upper-right) panel, we
  draw the mass distribution when the spin-1 particle is polarized
  with $s_x=1$ ($s_x=0$). In the lower-left (lower-right) panel, we
  illustrate that with $s_z=1$ ($s_z=0$).
}
  \label{fig:5}
\end{figure}
We now consider the mass distribution of the spin-1 particle in the 2D
IMF. As written in Eq.~\eqref{eq:j01}, it acquires the induced dipole
contribution that arises from the Lorentz boost. We have presented its
explicit expressions in Eq.~\eqref{eq:appb-inddip} given in
Appendix~\ref{app:b} when the polarization of the spin-1 particle is
fixed. In addition to the quadrupole contribution, the induced dipole
one provides an additional angular dependence of the mass
distribution. Thus, when the spin-1 particle is polarized along the
$x$-axis with $s_x=1$, the mass distribution is contracted to be
prolate, which is similar to that in the 2D BF. However, we can
observe a clear signature of the dipole feature when the spin-1
particle is polarized along $s_x=1$, as drawn in the upper-left panel
of Fig.~\ref{fig:5}. The upper-right panel displays the clear
quadrupole structure. It indicates that the Lorentz boost intensifies
the quadrupole pattern. When the spin-1 particle is polarized along
the $z$-axis, only the scalar contribution survives as in the case of 
the mass distribution in the 2D BF.   

\subsection{Mechanical properties of a spin-1 particle}
\begin{figure}[htp]
\includegraphics[scale=0.27]{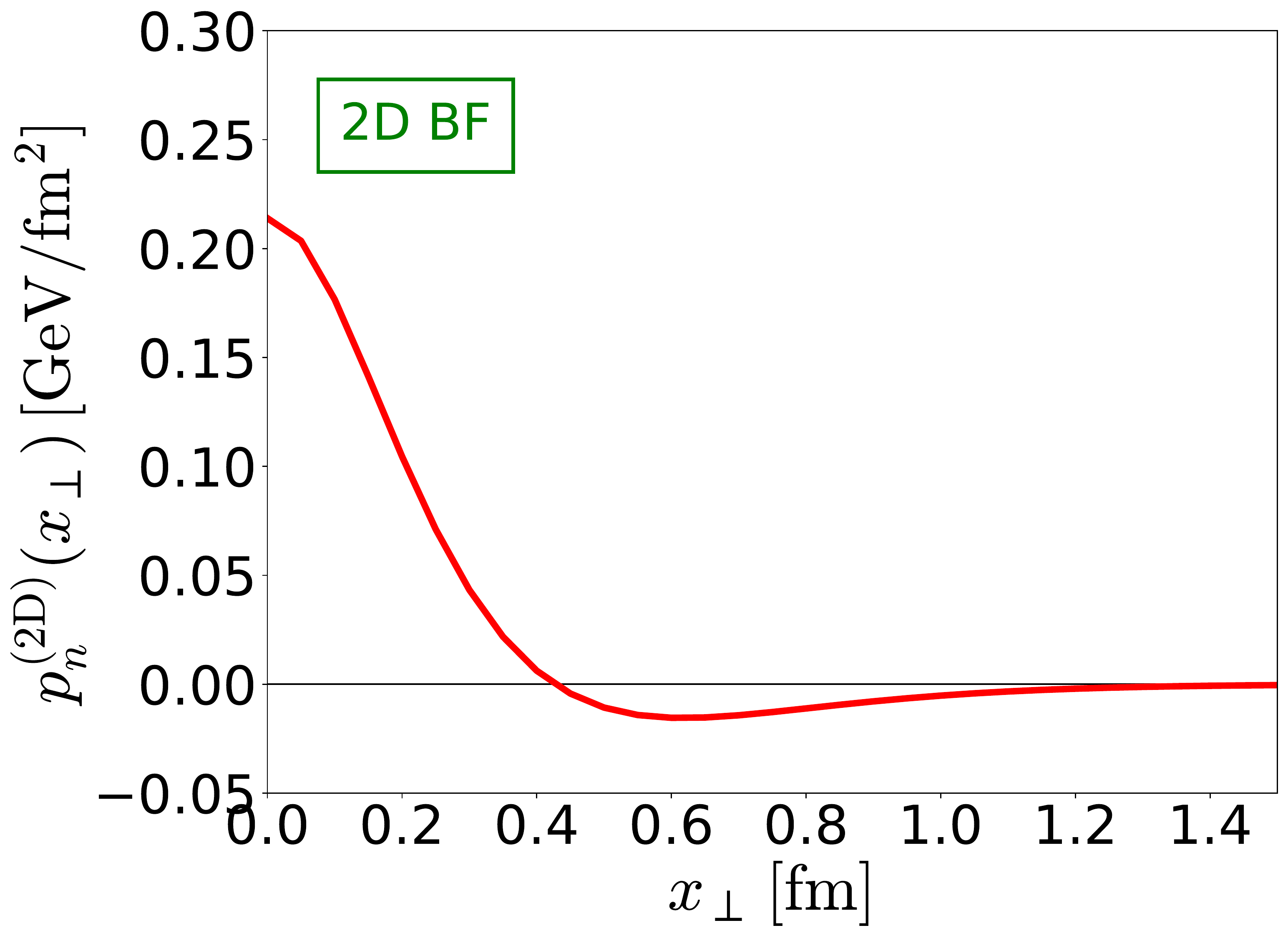}\;\;\;\;\;\;
\includegraphics[scale=0.27]{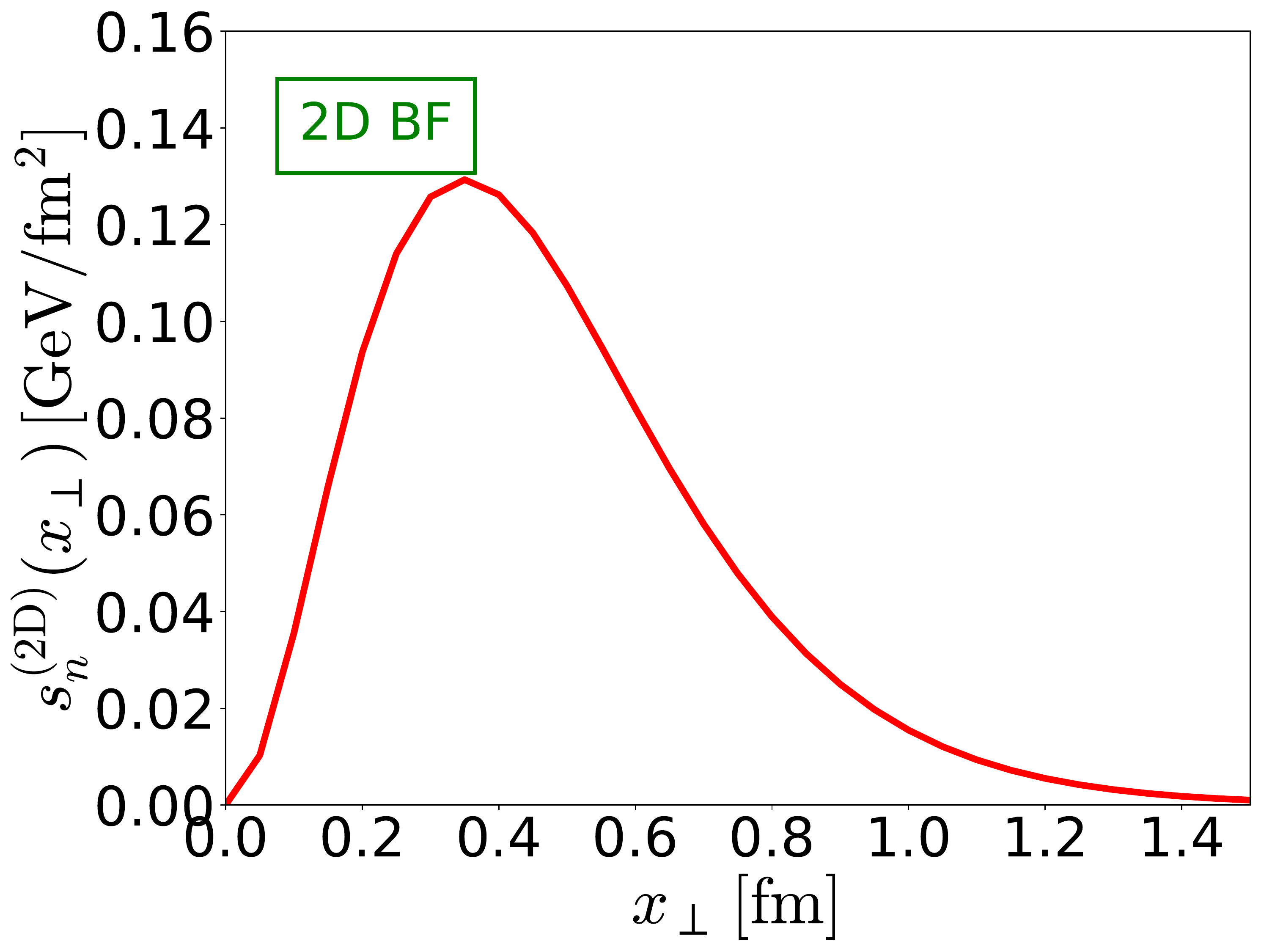}
\includegraphics[scale=0.27]{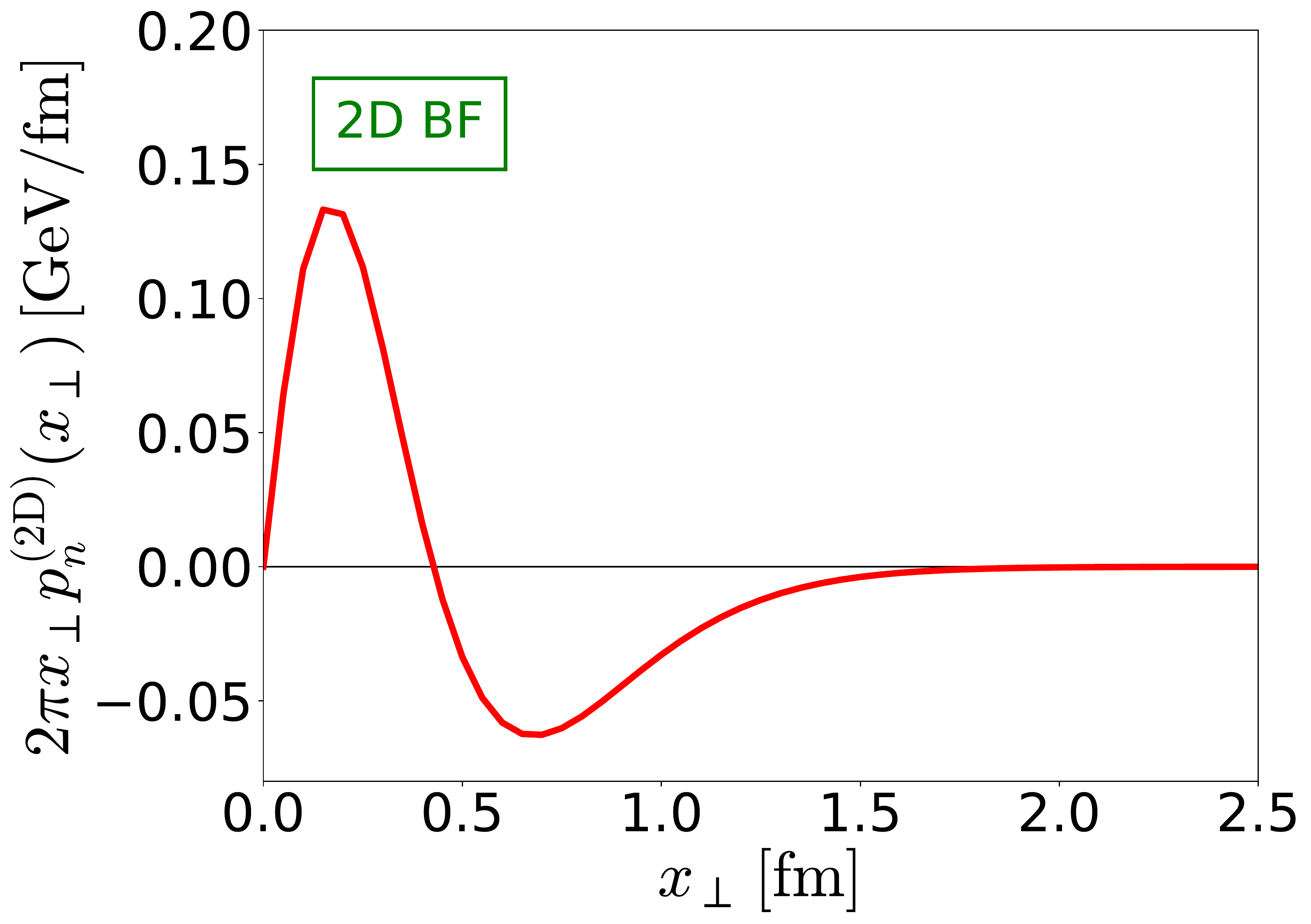}\;\;\;\;\;\;
\includegraphics[scale=0.27]{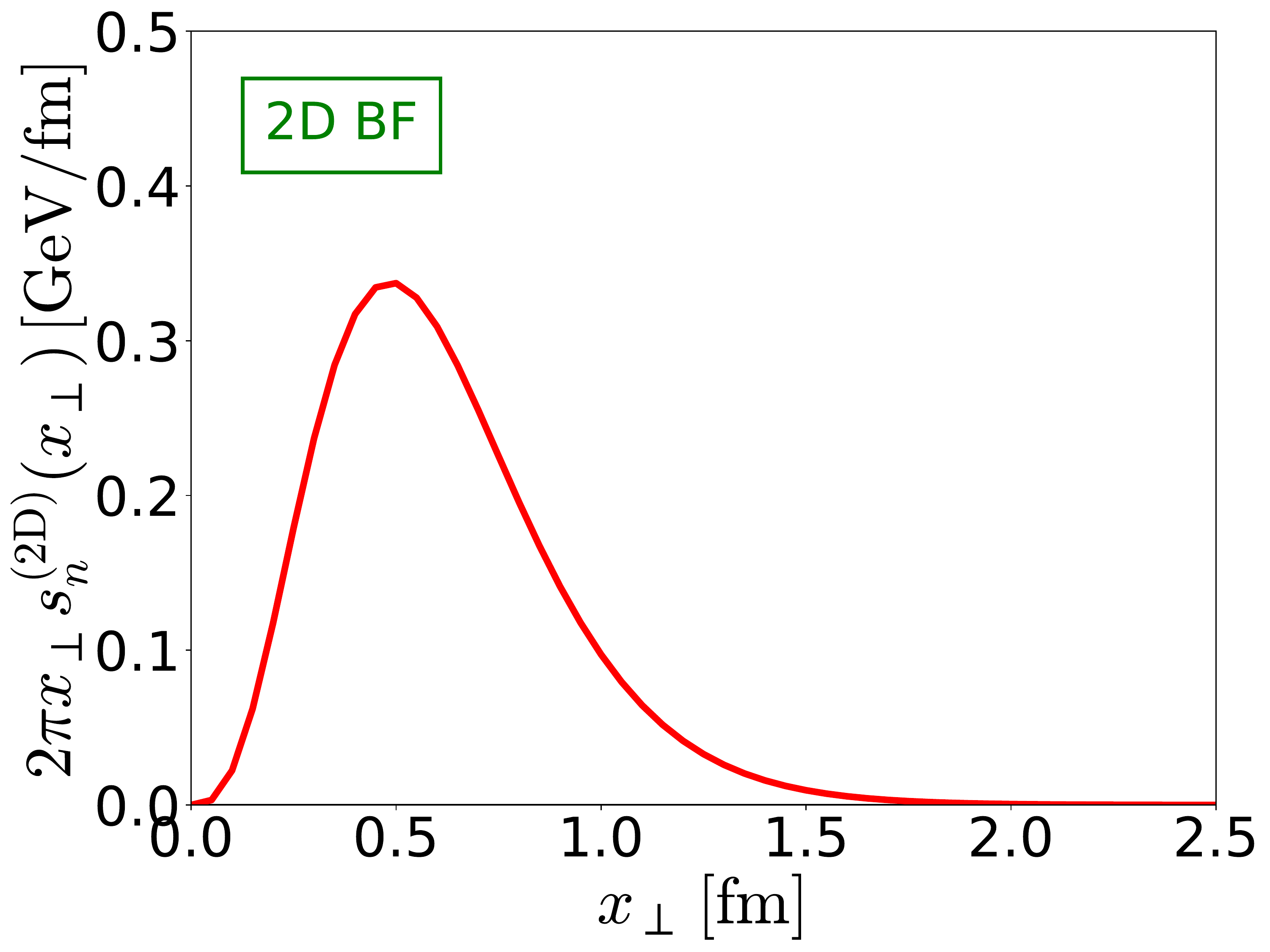}
\caption{Pressure and shear-force distributions ($p_n^{(\mathrm{2D})}$
  and $s_n^{(\mathrm{2D})}$) in the 2D BF. In the lower panel, we draw
  those weighted by $2\pi x_\perp$.} 
\label{fig:6}
\end{figure}
We are now in a position to discuss the mechanical properties of a
spin-1 particle. The 3D pressure and shear-force densities can be
derived by the inverse Fourier transformation as given in
Eqs.~\eqref{eq:D-termden} and \eqref{eq:pressure_3D}. Transforming the
derived 3D BF pressure and sear-force distributions by the Abel
transformation, we obtain those in the 2D BF. Since we employ the
same form for the GFFs of the spin-1 particle, the pressure and
shear-force distributions become degenerate respectively for all $n$,
as shown in Fig.~\ref{fig:6}. This degenerate forms of $p_n$ and $s_n$
have a virtue that the relativistic effects can clearly emerge when the
longitudinal momentum of the spin-1 particle goes to infinity. 

\begin{figure}[htp]
\includegraphics[scale=0.27]{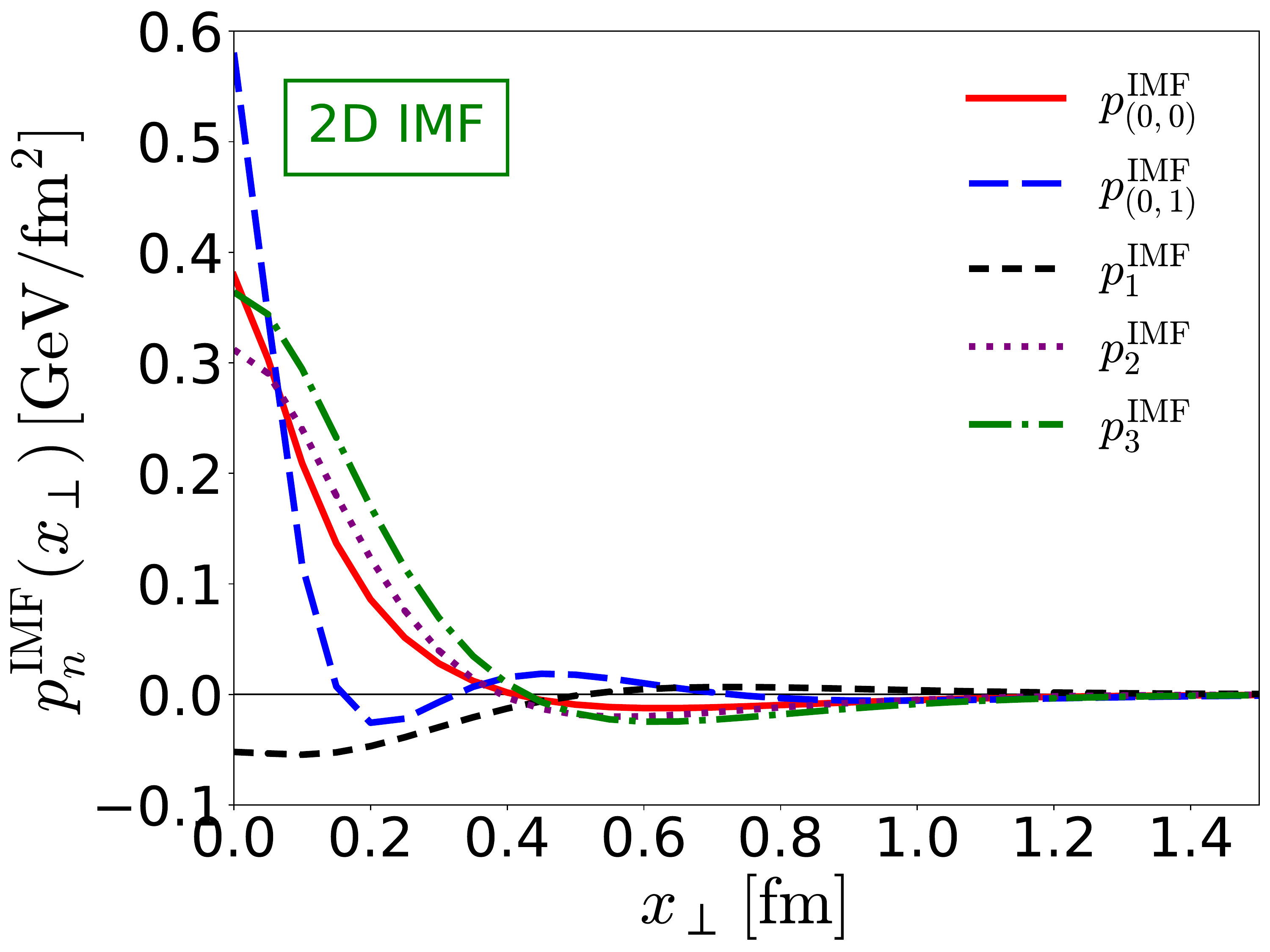}\;\;\;\;\;\;
\includegraphics[scale=0.27]{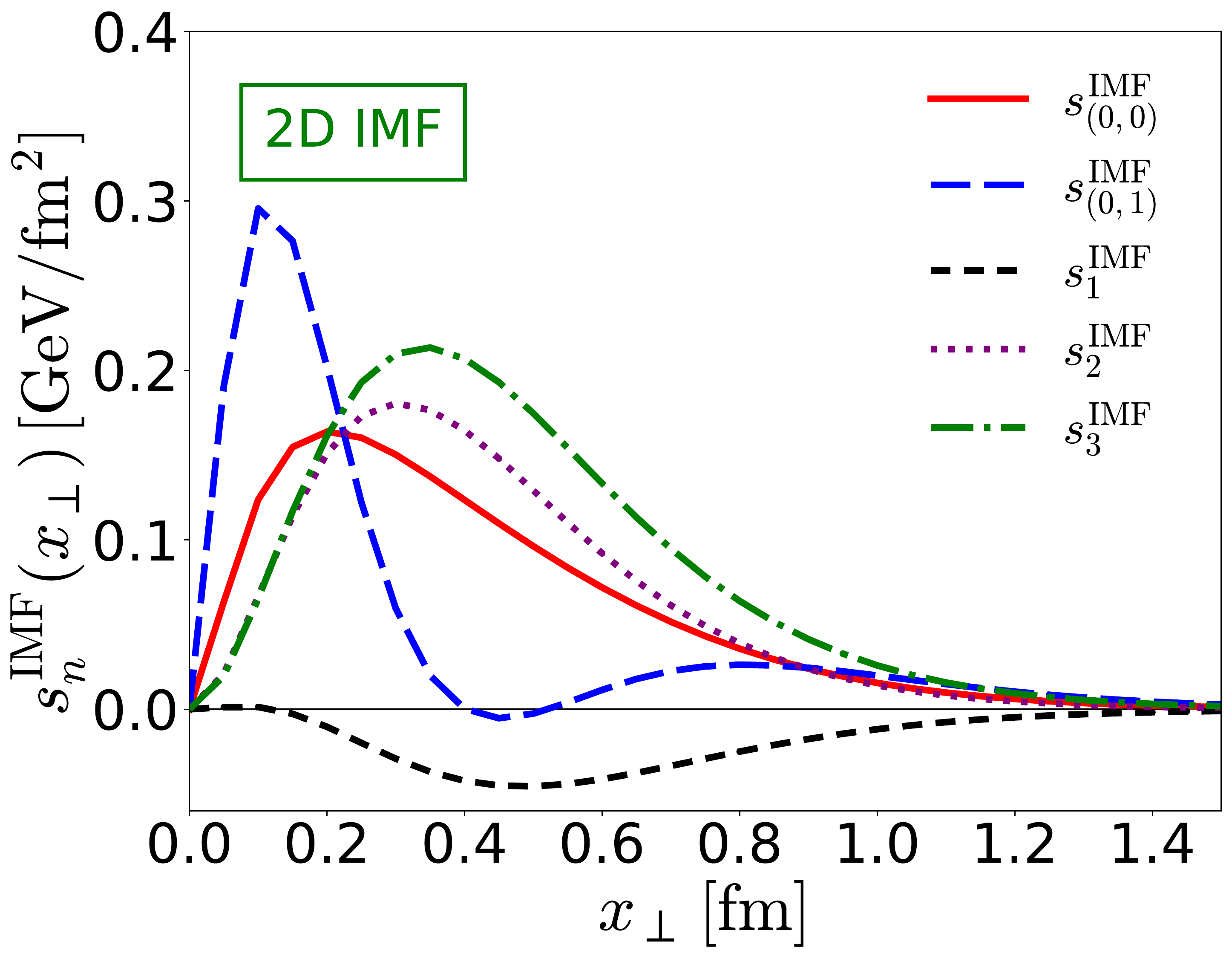}
\includegraphics[scale=0.27]{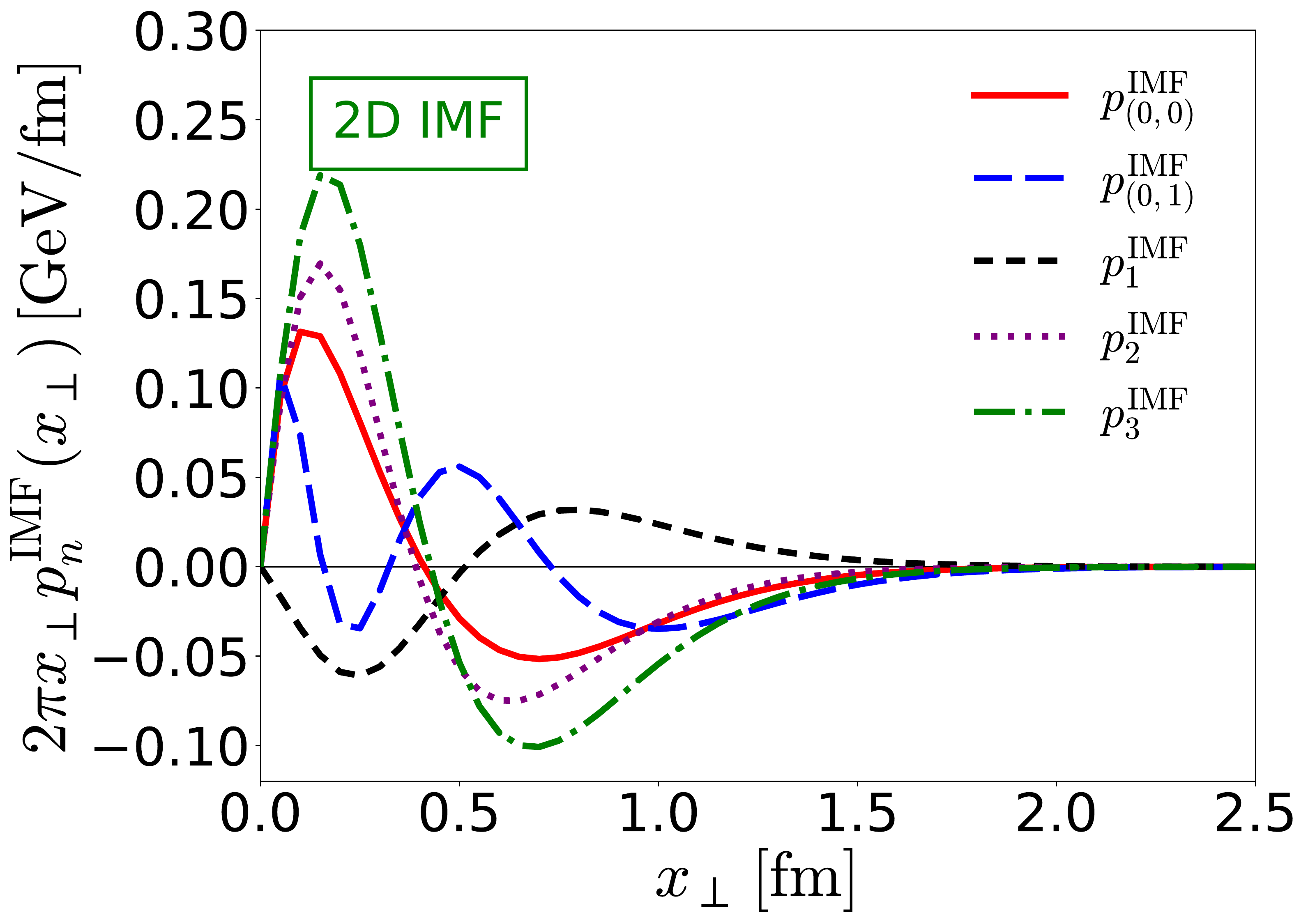}\;\;\;\;\;\;
\includegraphics[scale=0.27]{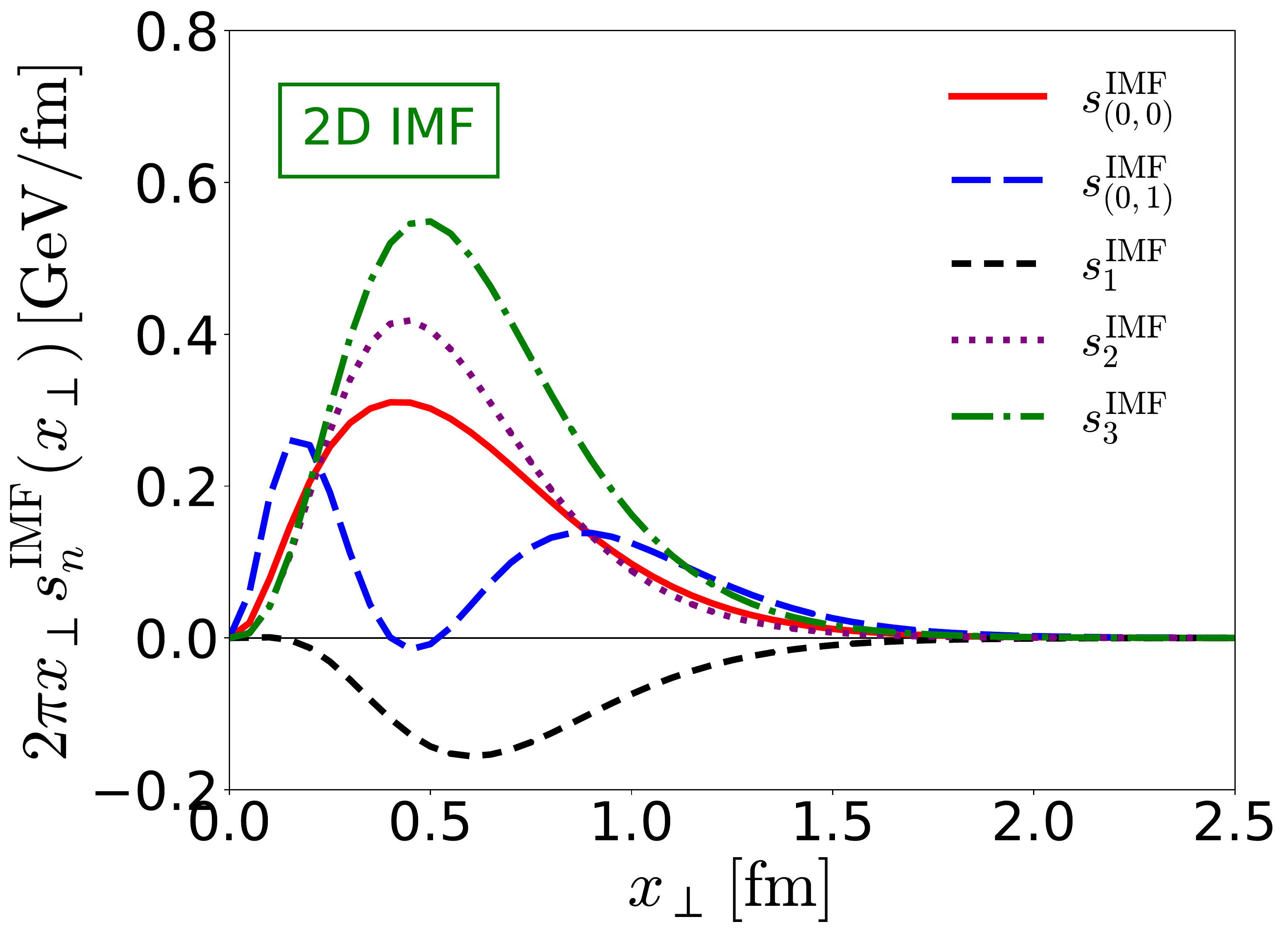}
\caption{Pressure and shear-force densities in the 2D IMF. In the
  lower panel, we draw those weighted by $2\pi x_\perp$. 
The decomposition of the pressure and shear-force densities are
defined in Eq.~\eqref{eq:61}.
}
\label{fig:7}
\end{figure}
When the spin-1 particle is boosted by the Lorentz transformation, the
$p_n$ and $s_n$ distributions undergo drastic changes. So, the
pressure and shear-force densities in the 2D IMF reveal the
relativistic effects as exhibited in Fig.~\ref{fig:7}. We 
observe that the degeneracy imposed on the pressure and shear-force
distributions are removed. As mentioned previously,
$p_1^{\mathrm{IMF}}$ and $s_1^{\mathrm{IMF}}$ arise from the Lorentz
boost. The pressure densities satisfy the stability condition as in
Eq.~\eqref{eq:65}. This indicates that the pressure densities in the
2D IMF should have the odd number of the nodal points. As shown in the
lower-left panel, we depict the pressure densities weighted by $2\pi
x_\perp$. $p_{(0,1)}^{\mathrm{IMF}}$ has three nodal points whereas
all other pressure densities have only one as those in the 2D BF.  
It is interesting to see that the newly emerged induced pressure
$p_1^{\mathrm{IMF}}$ under Lorentz boost is negative at the center in
contrast to other pressures. The shear-force densities in the 2D
IMF also exhibit interesting features. As shown in Fig.~\ref{fig:6}, the
shear-force distributions are all positive and degenerate. When the
momentum of the spin-1 particle becomes infinite, all the components
of the shear-force densities come apart. The sign of
$s_1^{\mathrm{IMF}}(x_\perp)$ becomes even negative. The negative
features of the pressure and shear-force densities are related to the
positive $D$-term in the IMF~\eqref{eq:IMF_nor}, which originates from
the solely relativistic effects. More interestingly,
$s_{(0,1)}^{\mathrm{IMF}}$ has two nodal points. These are unique
features for the spin-1 particle and higher spin states.  

\begin{figure}[htp]
\includegraphics[scale=0.83]{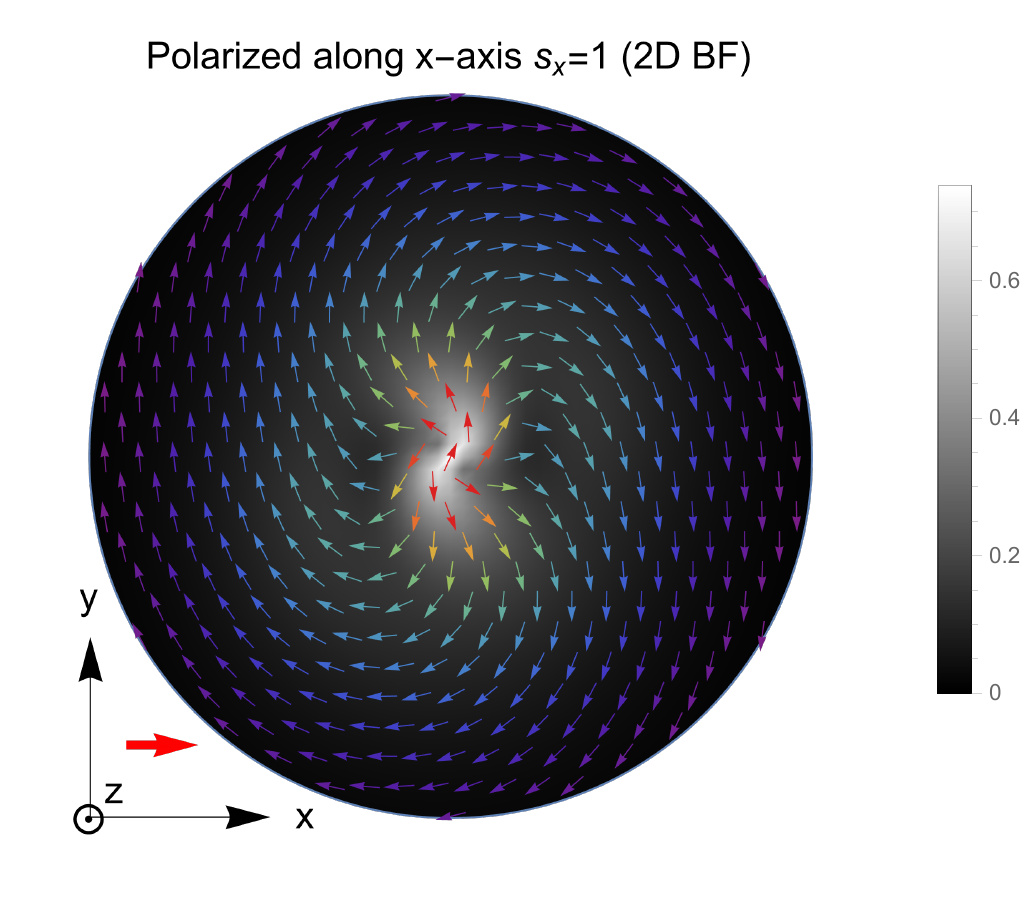}
\includegraphics[scale=0.83]{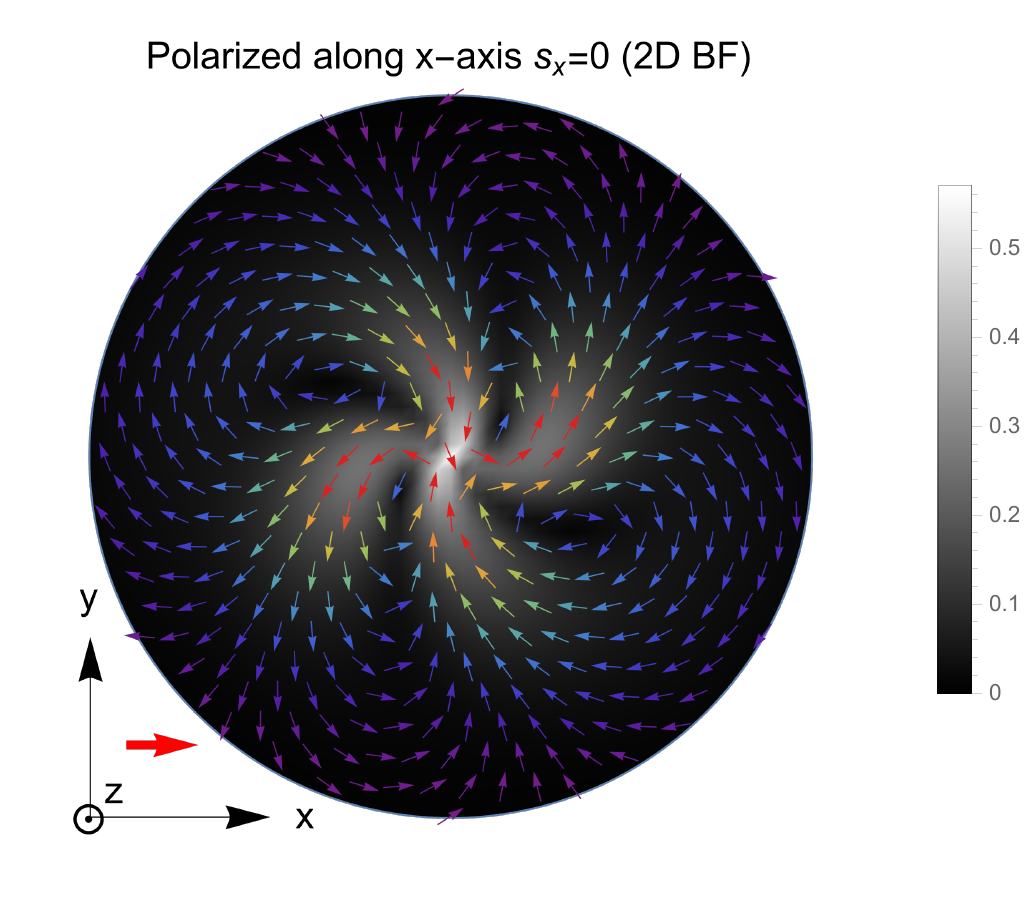}
\includegraphics[scale=0.83]{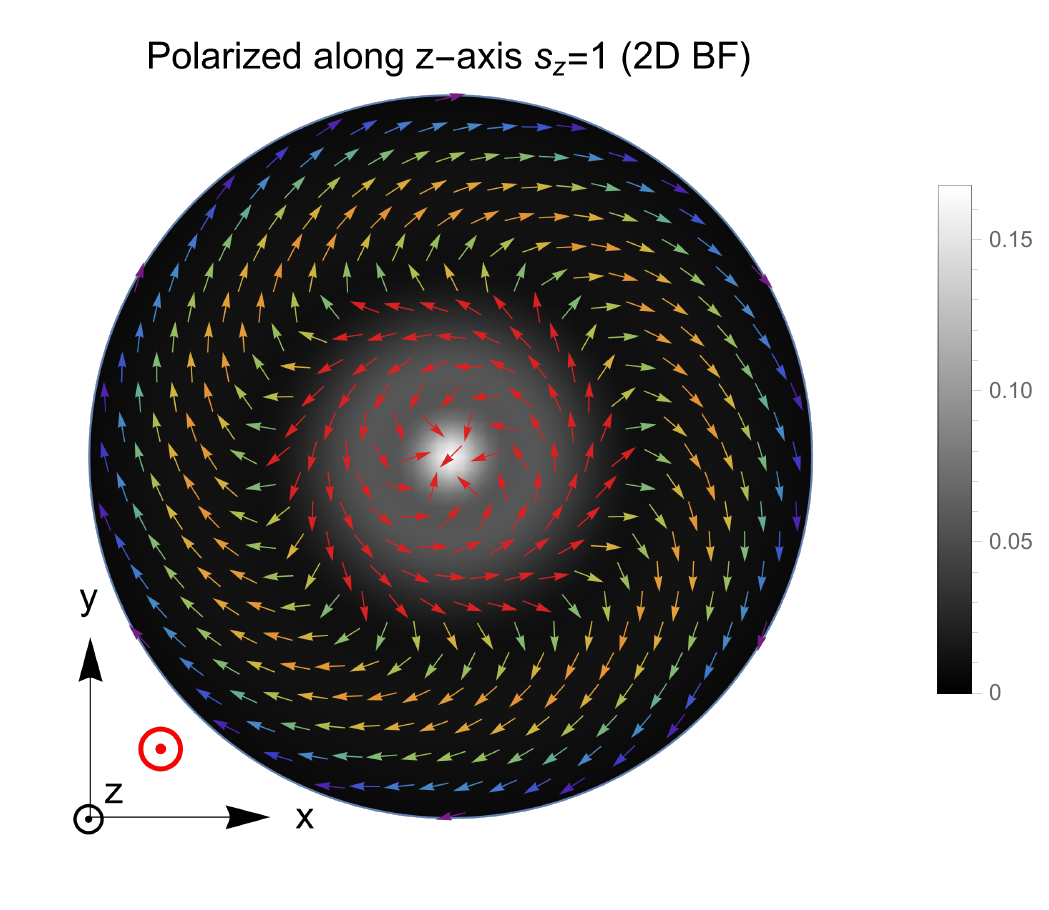}
\includegraphics[scale=0.83]{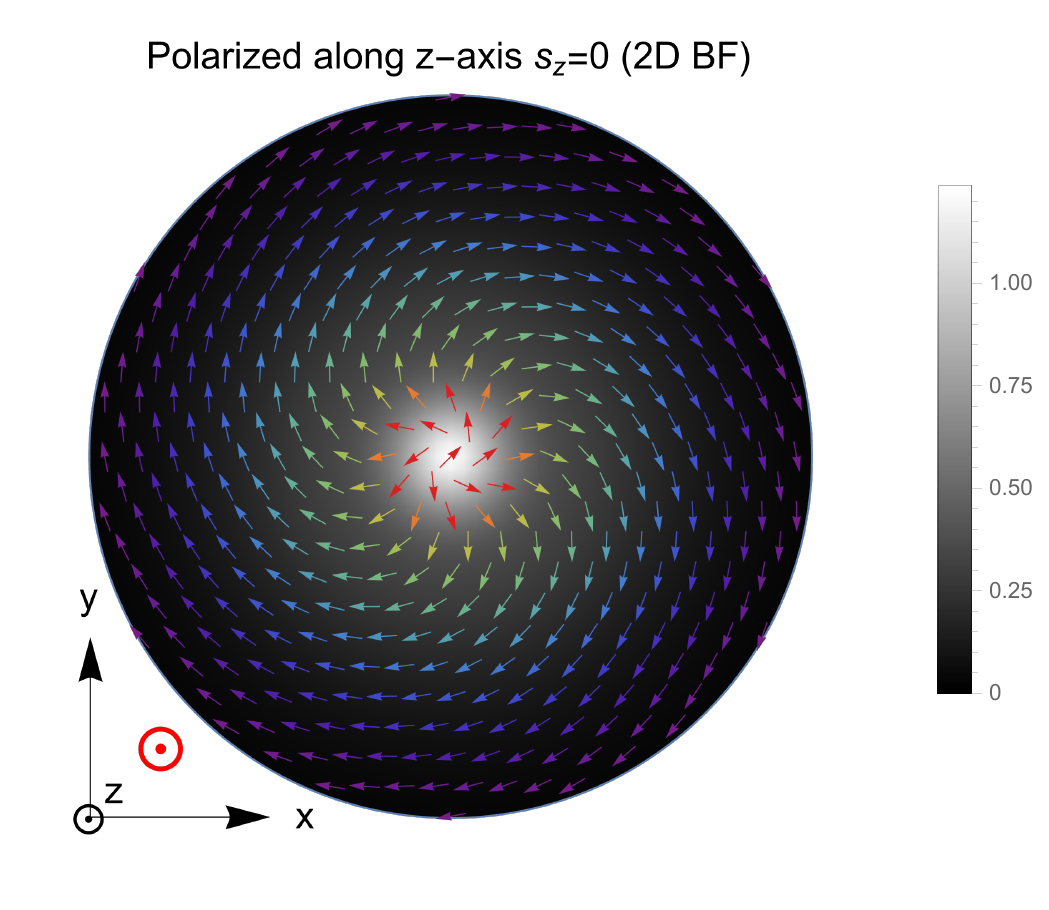}
\caption{Strong force fields inside a spin-1 particle in the rest frame
  ($P_{z}=0$) are visualized in the 2D plane when the target is
  polarized along $x$- 
  and $z$-axis.} 
\label{fig:8}
\end{figure}
\begin{figure}[htp]
\includegraphics[scale=0.83]{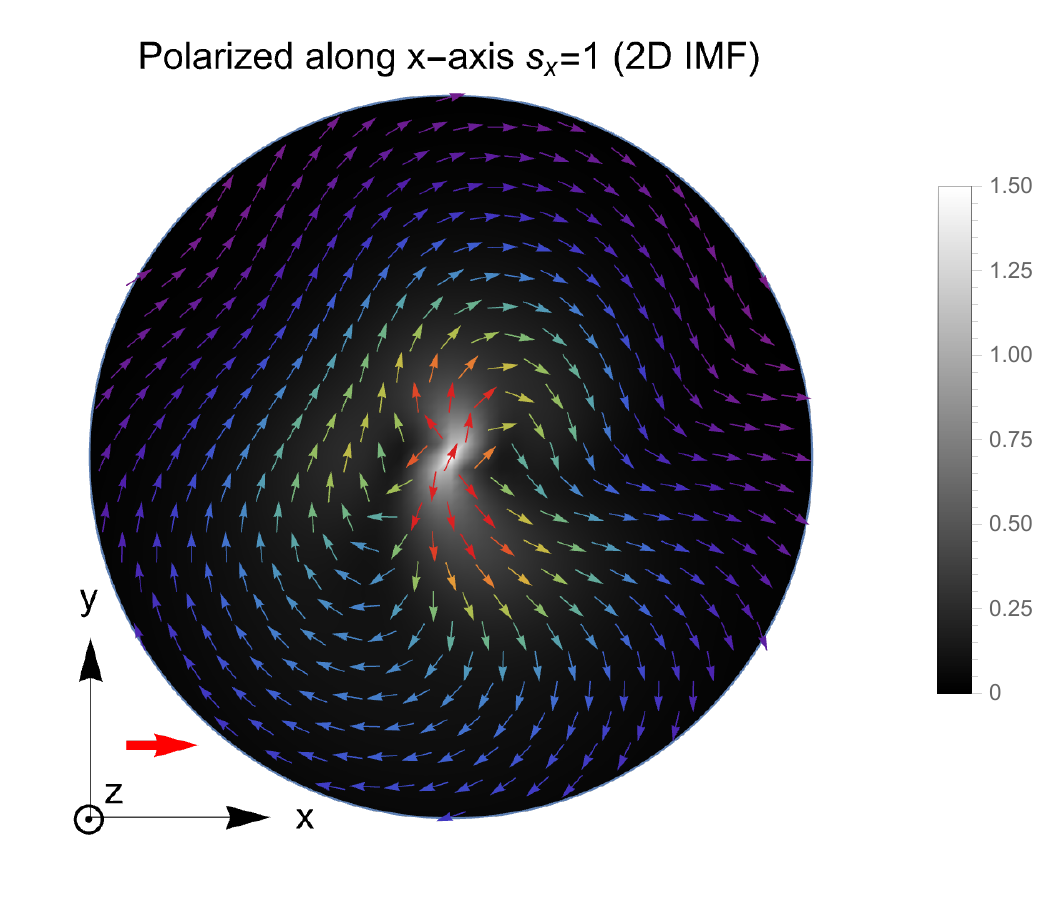}
\includegraphics[scale=0.83]{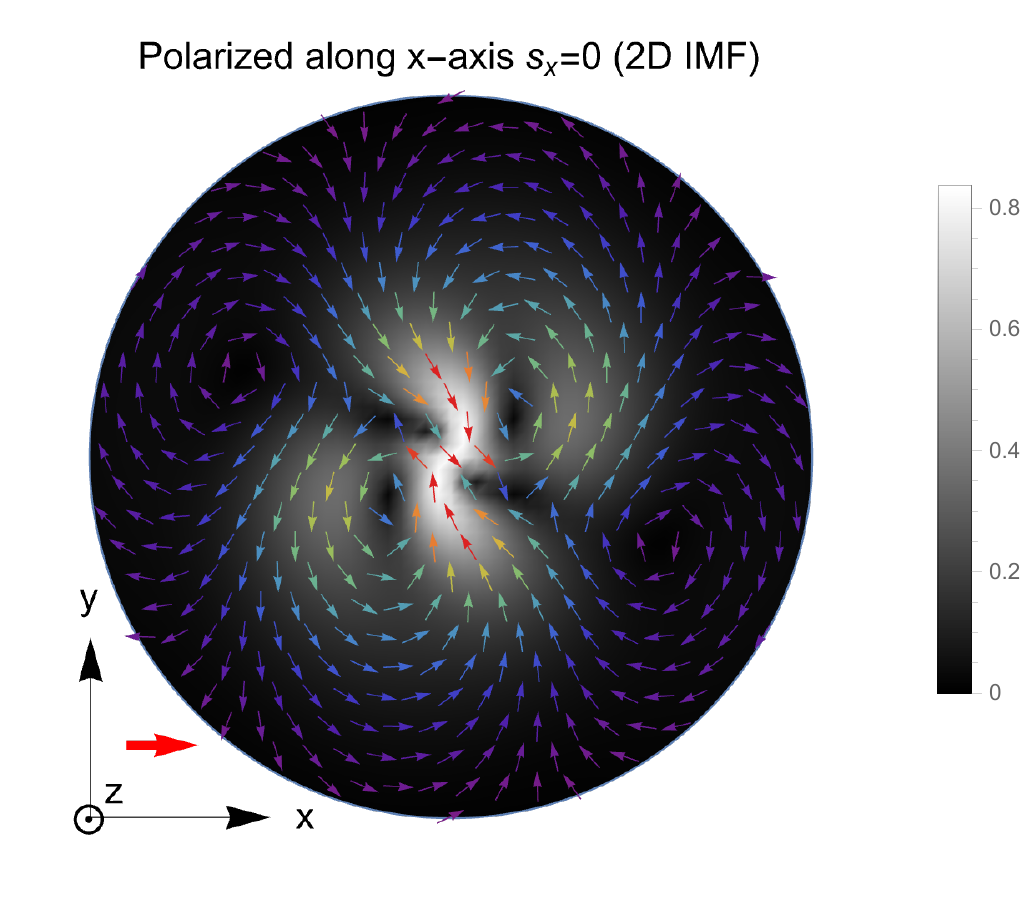}
\includegraphics[scale=0.83]{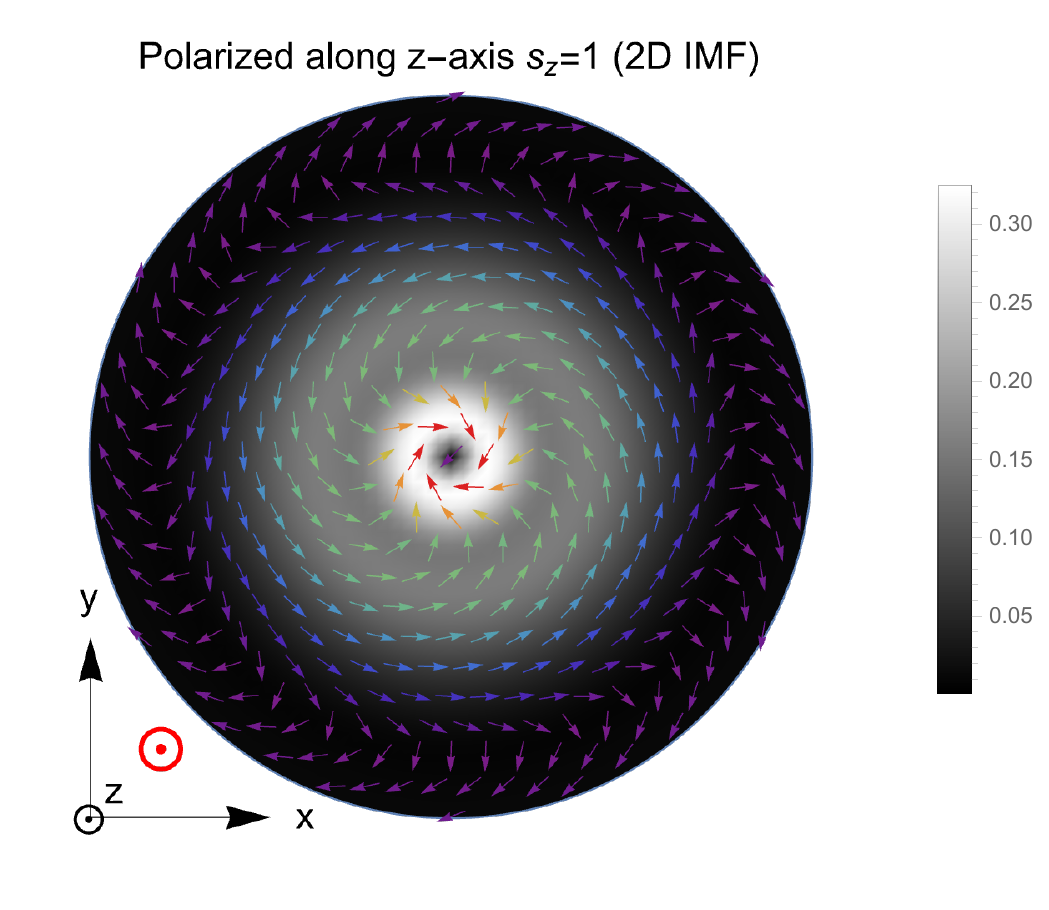}
\includegraphics[scale=0.83]{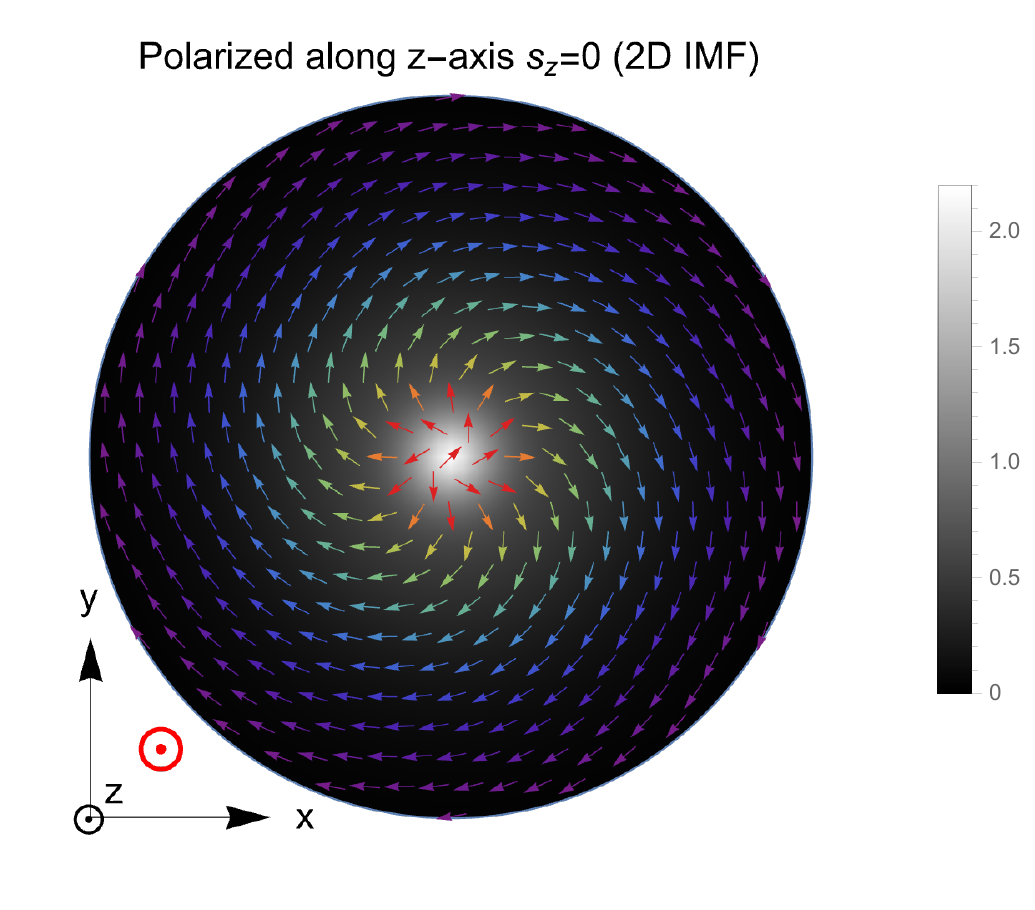}
\caption{Strong force fields inside a fastly moving spi-1 particle
  ($P_{z}=\infty$) are visualized in the 2D plane when the target is
  polarized along $x$- and $z$-axis.} 
\label{fig:9}
\end{figure}
In Figs.~\ref{fig:8} and~\ref{fig:9}, we illustrate the strong force
fields inside a spin-1 at rest and in the fast-moving frame,
respectively, when it is polarized along the $x$-axis (upper panel)
and $z$-axis (lower panel). In Appendix~\ref{app:c}, we have
explicitly written the expressions for the strong force fields with
the direction of the polarization chosen. In the 2D
IMF, we observe the dipole pattern in the upper-left panel and
the quadrupole structure in the upper-right panel as in the mass
distributions. Note that the strong force fields shown in 
Figs.~\ref{fig:8} and~\ref{fig:9} consist of partial ones. If one
considers all the force fields, they vanish at each local
point. In other words, the energy and momentum flow in both
directions along the arrows in Figs.~\ref{fig:8} and~\ref{fig:9}, so
that the net flows vanish. For simplicity, we present only one direction
indicated by the arrows. Nevertheless, the strong force fields
visualized in Figs.~\ref{fig:8} and~\ref{fig:9} exhibit how the
relativistic effects change the behavior of the strong force fields
inside a spin-1 particle. 

As written in Eq.~\eqref{eq:52}, the strong
force fields in the 2D BF have only the monopole and quadrupole
terms. Thus, when the spin-1 particle is polarized along the $x$-axis
with both $s_x=1$ and $s_x=0$, the qudrupole patterns are seen as in
the upper panel of Fig.~\ref{fig:8}. When it is polarized along the
$z$-axis, we find only the monopole pattern. On the other hand, the
Lorentz boost induces the dipole term in the spatial component of the
EMT densities (see Eq.~\eqref{eq:61}). When the spin-1 particle is
polarized along the $x$-axis with $s_x=1$ chosen, all the dipole and
quadrupole terms contribute to the strong force fields as shown in
Eq.~\eqref{eq:c8}. The upper-left panel of Fig.~\ref{fig:9} reveals
clearly this feature. However, if one selects $s_x=0$ with the spin-1
particle polarized along the $x$-axis, all the dipole contributions
vanish. So, the upper-right panel of Fig.~\ref{fig:9} demonstrates the
quadrupole pattern, being similar to that in the 2D BF. When the
spin-1 particle is polarized along the $z$-axis, the strong force
fields in the 2D IMF are not much different from those in the 2D BF. 
These features of the strong force fields resemble the mass
2D distributions.

\section{Summary and conclusions\label{sec:6}}
In the present work, we scrutinized the mechanical structure of the
spin-1 particle. We first formulated the gravitational form factors
and the energy-momentum tensor distributions in three different
frames: the three-dimensional Breit frame, the 
two-dimensional Breit frame, and the two-dimensional infinite-momentum
frame (or equivalently the Drell-Yan frame). By introducing these
three different frames, we were able to distinguish the geometric
effects from the relativistic effects that arise from the Lorentz
boost. A prominent point is that an additional monopole structure
is induced by going from the three-dimensional Breit frame to the
two-dimensional one. This can be achieved by the Abel
transformation. The two-dimensional infinite-momentum frame can be 
reached by taking the limit of the infinite longitudinal momentum of
the spin-1 particle. This Lorentz boost induces the dipole mass
density, of which the integration over the transverse plane vanishes
because of the normalization of the mass and spin of the spin-1
particle. This provides a nontrivial constraint on the induced dipole
mass density. When the spin-1 particle is polarized along the
$x$-axis with $s_x=1$, the mass distribution reveals the 
quadrupole pattern in the two-dimensional Breit frame. When its
momentum goes to infinity, the induced dipole mass is
generated, so that one can clearly observe the dipole pattern in the
mass density. When the spin state of the spin-1 particle is taken to
be $s_x=0$, the quadrupole structure is enhanced by the Lorentz boost.  

The relativistic effects on the pressure and shear-force densities 
are even more prominent. Since we take the same quadratic form of the
gravitational form factors, the pressure and shear-force distributions
become degenerate regardless of the value of the subscript
$n$. However, when we go from the two-dimensional Breit frame to the
two-dimensional infinite momentum frame, the pressure and shear-force
densities undergo drastic changes. While the pressure distributions
in the two-dimensional Breit frame have only one nodal point, which is
essential for them to satisfy the stability conditions, 
$p_{(0,1)}^{\mathrm{IMF}}$ has even three nodal points because of the
Lorentz boost. Interestingly, the sign of $p_1^{\mathrm{IMF}}$ becomes 
negative in the core part in contrast to the any other pressures. The
shear-force densities also exhibit unique features under 
the Lorentz boost. $s_{(0,1)}^{\mathrm{IMF}}$ has two nodal points and
$s_1^{\mathrm{IMF}}$ becomes negative. The negative values of the
pressure and shear-force densities are related to the positive
$D$-term in the IMF, which arises from the solely relativistic
effects. We also visualized the strong force fields inside a spin-1
particle. Though they vanish at each local point, they display the
multipole structure of the spin-1 particle, in particular, when it is
polarized along the $x$-axis. The most interesting behavior of the
strong force fields can be found by choosing $s_x=1$. In this case,
all the dipole and quadrupole terms contribute to the strong force
fields.  

In the current work, we focused on the mechanical structure of a generic
spin-1 particle. However, we want to mention that there are several
spin-1 particles such as the $\rho$ meson, the $\omega$ meson, the 
vector kaon, the $a_1$ meson, and the deuteron. While they share the
general features discussed in the present work, they will possibly
show quantitative differences each other. In particular, the deuteron
can be treated as a non-relativistic particle, whereas all the vector
mesons should be considered as relativistic ones. Apart from the
relativistic effects, there may be certain effects due to the
different quark configuration of the vector mesons. The corresponding
study will appear elsewhere. 

\begin{acknowledgments}
Authors want to express their gratitude to C.Lorcé for invaluable
comments and criticism. J.-Y.K. is supported by the Deutscher
Akademischer Austauschdienst(DAAD) doctoral scholarship.  
The work of H.-Ch.K. was supported by Basic Science Research Program
through the National Research Foundation of Korea funded by the Korean
government (Ministry of Education, Science and Technology, MEST),
Grant-No. 2021R1A2C2093368 and 2018R1A5A1025563. The work of
B.-D.S. is supported by Guangdong Provincial funding with Grant
No. 2019QN01X172, the National Natural Science Foundation of China
with Grant No. 12035007 and No. 11947228, Guangdong Major Project of
Basic and Applied Basic Research No. 2020B0301030008, and the Department of Science and Technology of Guangdong Province with Grant No. 2022A0505030010.
\end{acknowledgments}

\appendix

\section{Gravitational form factors of a spin-1 particle in the
  two-dimensional Drell-Yan frame \label{Appendix:A}}
The results from the light-front formalism should coincide with the
those from the IMF. As a cross-check, we are able to carry out the
same calculation in the light-front formalism. A light-front
four-vector is given by  
\begin{align}
x^{\nu}= \{x^{+},x^{-},\bm{x}_{\perp}\},
\end{align}
with $x^{\pm}= (x^{0}\pm x^{3})/\sqrt{2}$. The Drell-Yan frame is
defined by $\Delta^{+}=0$. The kinamatics are given by 
\begin{align}
P=\frac{1}{2}(p'+p)= (P^{+},P^{-},\bm{0}), \ \ \ \Delta=(p'-p)=
  (0,0,\bm{\Delta}_{\perp}), \ \ \  \mathrm{with} \ \
  P^{-}=\frac{1}{2P^{+}} \left( m^{2} +
  \frac{\bm{\Delta}^{2}_{\perp}}{4} \right). 
\end{align}
The matrix elements of $T^{++}$, $T^{+i}$, and $T^{ij}$ are
respectively decomposed in terms of the gravitational form factors as
follows: 
\begin{align}
\langle p', \lambda' | T^{++} (0) | p, \lambda \rangle &= 2P^{+2}
  \mathcal{E}^{\mathrm{IMF}}_{(0,0)}(t)  \delta_{\lambda'
  3}\delta_{3  \lambda} +  2P^{+2}
    \mathcal{E}^{\mathrm{IMF}}_{(0,1)}(t) 
   \delta_{\sigma'   \sigma} \cr 
 &+  2P^{+2} \sqrt{\tau}\mathcal{E}^{\mathrm{IMF}}_{1}(t) 
i \epsilon^{3jk} \hat{S}^{j}_{\lambda' \lambda} 
X^{k}_{1}(\theta_{\Delta_{\perp}})   +4P^{+2} \tau 
\mathcal{E}^{\mathrm{IMF}}_{2}(t) \hat{Q}^{kl} 
X^{kl}_{2}(\theta_{\Delta_{\perp}}),  \cr
\langle p', \lambda' | T^{+i} (0) | p, \lambda 
\rangle &=  2 m P^{+} \sqrt{\tau} i \epsilon^{3li} 
\hat{S}^{3}_{\lambda' \lambda} X^{l}_{1}(\theta_{\Delta_{\perp}}) 
\mathcal{J}^{\mathrm{IMF}}_{1}(t) + 4m P^{+} 
\tau \left(X^{ik}_{2}(\theta_{\Delta_{\perp}})-
\frac{1}{2}\delta^{ik}\right) 
\mathcal{J}^{\mathrm{IMF}}_{2}(t) \hat{Q}^{3k}, \cr
\langle p', \lambda' | T^{ij} (0) | p, \lambda \rangle 
&= 
 2m^{2}\tau \left[ \left( \frac{1}{3} 
\mathcal{D}^{\mathrm{IMF}}_{2} -\frac{1}{2}
\mathcal{D}^{\mathrm{IMF}}_{(0,1)} \right) 
\delta_{\sigma' \sigma}  + \left( -\frac{2}{3} 
\mathcal{D}^{\mathrm{IMF}}_{2} -\frac{1}{2}
\mathcal{D}^{\mathrm{IMF}}_{(0,0)} \right) 
\delta_{\lambda' 3} \delta_{\lambda 3} \right] \delta^{ij}  \cr
& +  2m^{2}\tau X^{ij}_{2}(\theta_{\Delta_{\perp}})  
\left[\delta_{\sigma' \sigma} 
\mathcal{D}^{\mathrm{IMF}}_{(0,1)}(t) +  
\delta_{\lambda' 3}\delta_{\lambda 3} 
\mathcal{D}^{\mathrm{IMF}}_{(0,0)}(t) \right] \cr
&+ 4m^{2}\tau \bigg{[} \hat{Q}^{ik} 
X^{jk}_{2}(\theta_{\Delta_{\perp}}) + \hat{Q}^{jk} 
X^{ik}_{2}(\theta_{\Delta_{\perp}}) - \hat{Q}^{lm} 
X^{lm}_2(\theta_{\Delta_{\perp}}) \delta^{ij}\bigg{]} 
\mathcal{D}^{\mathrm{IMF}}_{2}(t)   \cr
& +8m^{2} \tau^{3/2} i \epsilon^{lm3} \hat{S}^{l} 
X^{m}_{1}(\theta_{\Delta_{\perp}}) 
\left(X^{ij}_{2}(\theta_{\Delta_{\perp}}) - 
\frac{1}{2}\delta^{ij}\right) 
\mathcal{D}^{\mathrm{IMF}}_{1}(t) \cr
&+  8 m^{2} \tau^{2} \hat{Q}^{lm} 
\left(X^{lm}_{2}(\theta_{\Delta_{\perp}}) 
+ \frac{1}{2}\delta^{lm}\right) 
\left(X^{ij}_{2}(\theta_{\Delta_{\perp}}) - 
\frac{1}{2}\delta^{ij}\right) \mathcal{D}^{\mathrm{IMF}}_{3}(t). 
\label{eq:app1} 
\end{align}
The decompositions in Eq.~\eqref{eq:app1} are identical to those in
the infinite momentum frame. 

\section{Explicit calculation of the mass distributions
\label{app:b}}
Since the basis used in this work is not a standard one, we first
provide the conversion from the current basis to the standard
spherical basis in Table~\ref{tab:1}. 
\begin{table}[htp]
\setlength{\tabcolsep}{5pt}
\renewcommand{\arraystretch}{1.5}
\caption{Conversion from the our basis to the standard spherical basis.} 
\begin{tabular}{c | c c c } 
\hline
\hline{}
\centering
Quantization axis $n$ & $ | s_{n}= 1 \rangle $   & $| s_{n}= 0
  \rangle$ & $| s_{n}= -1 \rangle$   \\ 
\hline
  $x$-axis (transversal) & $\frac{1}{\sqrt{2}}(| z \rangle - i | 
y \rangle) $ & $-| x \rangle$ & $-\frac{1}{\sqrt{2}}(| 
z \rangle + i | y \rangle)$ \\
  $y$-axis (transversal) & $\frac{1}{\sqrt{2}}(| 
x \rangle + i | z \rangle) $ & $-i| y \rangle$ & 
$\frac{1}{\sqrt{2}}(| x \rangle - i | z \rangle)$\\
  $z$-axis (longitudinal) & $-\frac{1}{\sqrt{2}}(| 
x \rangle + i | y \rangle) $ & $| z \rangle$ & 
$\frac{1}{\sqrt{2}}(| x \rangle - i | y \rangle)$ \\
\hline 
\hline
\end{tabular}
\label{tab:1}
\end{table}
In Figs.~\ref{fig:4} and~\ref{fig:5}, we have visualized the mass
distributions in the 2D BF and 2D IMF, respectively, taking a specific
polarization of the spin-1 particle. We show here how to compute the
mass distributions when the polarization is fixed. In Eq.~\eqref{eq:j0},
the mass distribution in the 2D BF is written as 
\begin{align}
 T^{00}_{\mathrm{EF}}(\bm{x}_{\perp},0,\lambda',
  \lambda)&=  \delta_{3\lambda} \delta_{\lambda'3}
  \varepsilon^{\mathrm{(2D)}}_{(0,0)}(x_{\perp})+
  \delta_{\sigma'\sigma}
   \varepsilon^{\mathrm{(2D)}}_{(0,1)}(x_{\perp})
+  \hat{Q}^{ij}_{\lambda' \lambda} X^{ij}_{2}(\theta) 
  \varepsilon^{\mathrm{(2D)}}_{2}(x_{\perp}),
\label{eq:j0}
\end{align}
where $\lambda$ and $\lambda'$ run over $x$, $y$, and $z$ whereas
$\sigma$ and $\sigma'$ lie in the 2D EF, i.e. they take either $x$ or
$y$. We shall use $\theta$ instead of $\theta_{x_{\perp}}$ for
convenience. When the spin-1 particle is polarized along the $z$-axis 
($s_{z}$) and $s_z$ is explicitly given, we have to get the matrix
elements of $T^{00}$ as follows: 
\begin{align}
s_{z}=0 &\to \langle z | \hat{O} | z \rangle \cr
s_{z}=1 &\to \frac{1}{2}  \bigg{[}\langle x | \hat{O} | x \rangle +
            \langle y | \hat{O} | y \rangle + i \langle x | \hat{O} |
            y \rangle -i\langle y | \hat{O} | x \rangle \bigg{]} \cr 
s_{z}=-1 &\to \frac{1}{2}  \bigg{[}\langle x | \hat{O} | x \rangle +
             \langle y | \hat{O} | y \rangle - i \langle x | \hat{O} |
             y \rangle +i\langle y | \hat{O} | x \rangle \bigg{]} .
\end{align}
On the other hand, when it is polarized along the $x$-axis ($s_{x}$),
we find the matrix elements as 
\begin{align}
s_{x}=0 &\to \langle x | \hat{O} | x \rangle \cr
s_{x}=1 &\to \frac{1}{2}  \bigg{[}\langle z | \hat{O} | z \rangle +
            \langle y | \hat{O} | y \rangle - i \langle z | \hat{O} |
            y \rangle +i\langle y | \hat{O} | z \rangle \bigg{]} \cr 
s_{x}=-1 &\to \frac{1}{2}  \bigg{[}\langle z| \hat{O} | z \rangle +
             \langle y | \hat{O} | y \rangle + i \langle z | \hat{O} |
             y \rangle -i\langle y | \hat{O} | z \rangle \bigg{]} .
\end{align}
With the spin-1 particle polarized along the $y$-axis ($s_{y}$),
we get 
\begin{align}
s_{y}=0 &\to \langle y | \hat{O} | y \rangle \cr
s_{y}=1 &\to \frac{1}{2}  \bigg{[}\langle x | \hat{O} | x \rangle +
          \langle z | \hat{O} | z \rangle + i \langle x | \hat{O} | z
          \rangle -i\langle z | \hat{O} | x \rangle \bigg{]} \cr 
s_{y}=-1 &\to \frac{1}{2}  \bigg{[}\langle x| \hat{O} | x \rangle +
           \langle z | \hat{O} | z \rangle - i \langle x | \hat{O} | z
           \rangle +i\langle z | \hat{O} | x \rangle \bigg{]}, 
\end{align}
where $\hat{O}$ is a multipole operator.
In the Cartesian basis, the matrix representations of the quadrupole
operator $\langle \lambda' | \hat{Q}^{ij} | \lambda \rangle 
:=\hat{Q}^{ij}_{\lambda'\lambda}$ are given by 
\begin{align}
\hat{Q}^{xx}_{\lambda'\lambda}= \left(\begin{array}{c c c c}\lambda
                                  \text{\textbackslash} \lambda'&x & y
                                  & z \\ x &-\frac{2}{3} & 0 & 0 \\ y
                                                                & 0 &
  \frac{1}{3} &  0 \\ z & 0 & 0 & \frac{1}{3} \end{array}\right), \ \
  \hat{Q}^{yy}_{\lambda'\lambda}= \left(\begin{array}{c c c c}\lambda
   \text{\textbackslash}
   \lambda'&x & y & z \\ x
           &\frac{1}{3} & 0 & 0
   \\ y & 0 & -\frac{2}{3} &  0
   \\ z & 0 & 0 &
 \frac{1}{3} \end{array}\right), \ \ 
\hat{Q}^{zz}_{\lambda'\lambda}=  \left(\begin{array}{c c c c}
\lambda   \text{\textbackslash}  \lambda'&x
  & y &  z \\ x &\frac{1}{3} & 0 & 0 \\ y & 0 
& \frac{1}{3} &  0 \\ z & 0 & 0
 &  -\frac{2}{3} \end{array}\right)    
\end{align}
\begin{align}
\hat{Q}^{xy}_{\lambda'\lambda}= \left(\begin{array}{c c c c}\lambda
     \text{\textbackslash}
     \lambda'&x & y & z \\ x &
  0 & -\frac{1}{2}  & 0 \\ y & -\frac{1}{2}
                &
  0 &  0 \\ z & 0 & 0 &
      0 \end{array}\right),  \ \  \hat{Q}^{xz}_{\lambda'\lambda}=
       \left(\begin{array}{c c c c}\lambda \text{\textbackslash}
    \lambda'&x  & y  & z \\ x&0 & 0  & -\frac{1}{2}
     \\ y   &   0   & 0    &  0  \\ z
               & -\frac{1}{2} & 0 & 0 \end{array}\right), \ \
\hat{Q}^{yz}_{\lambda'\lambda}= \left(\begin{array}{c c c c}\lambda
\text{\textbackslash} \lambda'&x & y & z \\ x &0 & 0 & 0
  \\ y & 0 & 0 &  -\frac{1}{2} \\ z & 0 & -\frac{1}{2} & 0
  \end{array}\right).      
\end{align}
Having contracted the $i,j=1,2$ indices to the $x^{i}_{\perp}$ and
$x^{j}_{\perp}$, we have the following expressions:  
\begin{align}
\hat{Q}^{ij}_{\lambda' \lambda} \hat{x}^{i}_{\perp}\hat{x}^{j}_{\perp}
  :=\hat{Q}^{rr}_{\lambda' \lambda} &=
 \left(\begin{array}{c c c}
         \frac{1}{6}(-1-3\cos{2\theta})  & -\cos{\theta} \sin{\theta}  &
  0 \\ -\cos{\theta} \sin{\theta} & \frac{1}{6}(-1+3\cos{2\theta}) &
0 \\  0 & 0 & \frac{1}{3} \end{array}\right), \cr
\hat{Q}^{ij}_{\lambda' \lambda} \hat{\theta}^{i} \hat{x}^{j}_{\perp}=
\hat{Q}^{\theta r}_{\lambda' \lambda} &=
   \left(\begin{array}{c c c}
           \cos{\theta} \sin{\theta} & -\frac{1}{2}\cos{2\theta}   &
0 \\ -\frac{1}{2}\cos{2\theta} & -\cos{\theta} \sin{\theta} &                                            0 \\  0 & 0 & \frac{1}{3} \end{array}\right), \cr
\hat{Q}^{ij}_{\lambda' \lambda} \hat{\theta}^{i}\hat{\theta}^{j}=
\hat{Q}^{\theta \theta}_{\lambda' \lambda} &= \left(\begin{array}{c c c}
  \frac{1}{6}(-1+3\cos{2\theta})  & \cos{\theta} \sin{\theta}  &
   0 \\ \cos{\theta} \sin{\theta} & \frac{1}{6}(-1-3\cos{2\theta}) &
   0 \\  0 & 0 & \frac{1}{3} \end{array}\right).
\end{align}

If the spin-1 particle is polarized with $s_{z}=0$, then there is no
$\theta$ dependence. So, the mass distribution contains only the
scalar one $\varepsilon_{(0,0)}^{(2\mathrm{D})} (x_\perp)$: 
\begin{align}
 T^{00}_{\mathrm{EF}}(\bm{x}_{\perp},0,\lambda',
  \lambda)&=  \delta_{zz} \delta_{zz}
  \varepsilon^{\mathrm{(2D)}}_{(0,0)}(x_{\perp})
+  (\hat{Q}^{rr}_{zz}-\frac{1}{2}\hat{Q}^{ii}_{zz}) 
  \varepsilon^{\mathrm{(2D)}}_{2}(x_{\perp})=
  \varepsilon^{\mathrm{(2D)}}_{(0,0)}(x_{\perp}). 
\label{eq:appb-z0}
\end{align}
If, however, it is polarized with $s_{z}=1$, then we obtain $T^{00}$
as 
\begin{align}
 T^{00}_{\mathrm{EF}}(\bm{x}_{\perp},0,\lambda',
  \lambda)&=  \delta_{\sigma'\sigma}
   \varepsilon^{\mathrm{(2D)}}_{(0,1)}(x_{\perp})
+(\hat{Q}^{rr}_{s'_{z}=1,s_{z}=1}-\frac{1}{2}
\hat{Q}^{ii}_{s'_{z}=1,s_{z}=1}) 
  \varepsilon^{\mathrm{(2D)}}_{2}(x_{\perp}) \cr
  &=  \frac{1}{2}(\delta_{xx}+\delta_{yy})
   \varepsilon^{\mathrm{(2D)}}_{(0,1)}(x_{\perp})
+  (\hat{Q}^{rr}_{s'_{z}=1,s_{z}=1}-\frac{1}{2}
\hat{Q}^{ii}_{s'_{z}=1,s_{z}=1}) 
  \varepsilon^{\mathrm{(2D)}}_{2}(x_{\perp})\cr
  & = \varepsilon^{\mathrm{(2D)}}_{(0,1)}(x_{\perp}). 
\label{eq:appb-z1}
\end{align}
Note that we have the following algebra: 
\begin{align}
&\hat{Q}^{rr}_{s'_{z}=1,s_{z}=1}
-\frac{1}{2}\hat{Q}^{ii}_{s'_{z}=1,s_{z}=1}=0,  \cr 
&\hat{Q}^{rr}_{s'_{z}=0,s_{z}=0}
-\frac{1}{2}\hat{Q}^{ii}_{s'_{z}=0,s_{z}=0}=0, \cr
& \hat{Q}^{rr}_{s'_{x}=0,s_{x}=0}
-\frac{1}{2}\hat{Q}^{ii}_{s'_{x}=0,s_{x}=0}=
  -\frac{1}{2}\cos{2\theta}, \cr
&  \hat{Q}^{rr}_{s'_{x}=1,s_{x}=1}
-\frac{1}{2}\hat{Q}^{ii}_{s'_{x}=1,s_{x}=1}=
  \frac{1}{4}\cos{2\theta}. 
\end{align}
Thus, the quadrupole contribution vanishes when the spin-1 particle is
polarized along the $z$-axis. However, if it is transversly
polarized with $s_{x}=0$, we get the contribution from the quarupole
term 
\begin{align}
 T^{00}_{\mathrm{EF}}(\bm{x}_{\perp},0,\lambda',
  \lambda)&= \delta_{x x}
   \varepsilon^{\mathrm{(2D)}}_{(0,1)}(x_{\perp})
+  (\hat{Q}^{rr}_{xx}-\frac{1}{2}\hat{Q}^{ii}_{xx})
  \varepsilon^{\mathrm{(2D)}}_{2}(x_{\perp}) = 
   \varepsilon^{\mathrm{(2D)}}_{(0,1)}(x_{\perp})
 -\frac{1}{2}\cos{2\theta}
  \varepsilon^{\mathrm{(2D)}}_{2}(x_{\perp}). 
    \label{eq:appb-x0}
\end{align}
When the spin-1 particle is polarized with $s_{x}=1$, we obtain the
following mass distribution
\begin{align}
 T^{00}_{\mathrm{EF}}(\bm{x}_{\perp},0,\lambda',
  \lambda)&=  \frac{1}{2} \delta_{zz} \delta_{zz}
  \varepsilon^{\mathrm{(2D)}}_{(0,0)}(x_{\perp})+
  \frac{1}{2}\delta_{y y}
   \varepsilon^{\mathrm{(2D)}}_{(0,1)}(x_{\perp})
+  (\hat{Q}^{rr}_{s'_{x}=1,s_{x}=1}
-\frac{1}{2}\hat{Q}^{ii}_{s'_{x}=1,s_{x}=1})
  \varepsilon^{\mathrm{(2D)}}_{2}(x_{\perp})  \cr
  &=\frac{1}{2} \varepsilon^{\mathrm{(2D)}}_{(0,0)}(x_{\perp})+\frac{1}{2}
   \varepsilon^{\mathrm{(2D)}}_{(0,1)}(x_{\perp})
+  \frac{1}{4}\cos{2\theta}
    \varepsilon^{\mathrm{(2D)}}_{2}(x_{\perp}).
    \label{eq:appb-x1}
\end{align}

We can carry out a similar algebra to obtain the mass distributions in
the 2D IMF. To obtain the force fields in the 2D BF and 2D IMF, we can
perform a similar calculation. 

In the IMF, we have additional dipole contribution, which is induced
by the Lorentz boost. 
\begin{align}
\sim \epsilon^{3jk} \hat{S}^{j}_{\lambda' \lambda} X^{k}_{1}.
\end{align}
In the Cartesian basis, the matrix representations of the spin operator
$\langle \lambda' | \hat{S}^{i} | \lambda \rangle 
:=\hat{S}^{i}_{\lambda'\lambda}$ are given by 
\begin{align}
\hat{S}^{x}_{\lambda'\lambda}= \left(\begin{array}{c c c c}\lambda
  \text{\textbackslash} \lambda'&x & y
  & z \\ x &0 & 0 & 0 \\ y  & 0 &
  0 &  -i \\ z & 0 & i & 0 \end{array}\right), \ \
   \hat{S}^{y}_{\lambda'\lambda}=
   \left(\begin{array}{c c c c}\lambda
           \text{\textbackslash}
           \lambda'&x & y & z \\ x
                   &0 & 0 & i
           \\ y & 0 & 0 &  0
           \\ z & -i & 0 &
 0 \end{array}\right), \ \ \hat{S}^{z}_{\lambda'\lambda}=
  \left(\begin{array}{cccc}\lambda  \text{\textbackslash}
   \lambda'&x
   & y & z \\ x &0 & -i & 0 \\ y & i & 0 &  0 \\ z & 0 & 0
 &  0 \end{array}\right)    
\end{align}
Having contracted the $j,k=1,2$ indices to the $\epsilon^{3jk}
X^{k}_{1}$, we have the following expressions:   
\begin{align}
\epsilon^{3jk} \hat{S}^{j}_{\lambda' \lambda} X^{k}_{1}
=\left(\begin{array}{c c c c}\lambda
  \text{\textbackslash} \lambda'&x & y
  & z \\ x &0 & 0 & -\cos{\theta} \\ y
   & 0 &  0 &  -i\sin{\theta} \\ z & i\cos{\theta} 
& i\sin{\theta} & 0 \end{array}\right), \ \
\end{align}
Note that the diagonal components vanish, and the transverse states
$|x,y\rangle$ should be mixed with the $|z\rangle$, so that we have a finite
contribution to the mass distributions. Since there is no mixture of
the them, the dipole contribution vanishes when the spin-1 particle is
polarized along the $z$-axis. 

Similarly, if it is transversly
polarized with $s_{x}=0$, we still have null contribution from the
induced dipole term. However, if it is polarized with $s_{x}=\pm 1$ we
then have finite contribution: 
\begin{align}
\epsilon^{3jk} \hat{S}^{j}_{\lambda' \lambda} X^{k}_{1} 
&= \sin{\theta} \quad (s_{x}=1) \cr
\epsilon^{3jk} \hat{S}^{j}_{\lambda' \lambda} X^{k}_{1} 
&= 0 \quad (s_{x}=0)\cr
\epsilon^{3jk} \hat{S}^{j}_{\lambda' \lambda} X^{k}_{1} 
&= -\sin{\theta} \quad (s_{x}=-1)\cr
\epsilon^{3jk} \hat{S}^{j}_{\lambda' \lambda} X^{k}_{1} 
&= \cos{\theta} \quad (s_{y}=1)\cr
\epsilon^{3jk} \hat{S}^{j}_{\lambda' \lambda} X^{k}_{1} 
&= 0 \quad (s_{y}=0)\cr
\epsilon^{3jk}\hat{S}^{j}_{\lambda' \lambda} X^{k}_{1} 
&= -\cos{\theta} \quad (s_{y}=-1)
\label{eq:appb-inddip}
\end{align}
One can clearly see in Fig.~\ref{fig:5} that the dipole contributes to
the mass distribution only with $s_{x}=1$. In the other figures, such 
contirbutions vanish. 

\section{Explicit calculation of the mechanical distributions
  \label{app:c}}
In this Appendix, we show explicitly the mechanical distributions for
a spin-1 particle. When we have $s_{z}=0$, the matrix elements of the
dipole and quadrupole operators are obtained as 
\begin{align}
\hat{Q}^{rr}_{s'_{z}=0,s_{z}=0}=\frac{1}{3}, \quad \hat{Q}^{\theta
  \theta}_{s'_{z}=0,s_{z}=0}=\frac{1}{3}, \quad \hat{Q}^{\theta
  r}_{s'_{z}=0,s_{z}=0}=0, \quad \epsilon^{lm3}\hat{S}^{l}_{s'_{z}=0,s_{z}=0}
  X^{m}_{1}(\theta)=0, \quad \epsilon^{lm3}\hat{S}^{l}_{s'_{z}=0,s_{z}=0}
  \hat{\theta}^{m}_{1}(\theta)=0. 
\end{align}
Thus, we can show that the quadrupole pattern of the distribution
vanishes when the spin-1 particle is polarized along the $z$-axis
($s_{z}=0$):  
\begin{align}
\frac{dF_{r}}{dS_{r}}&= \bigg{[} \left(p^{\mathrm{IMF}}_{(0,0)}+
\frac{1}{2}s^{\mathrm{IMF}}_{(0,0)}\right) + \frac{2}{3} 
\left(2p^{\mathrm{IMF}}_{2}+  s^{\mathrm{IMF}}_{2}+
\frac{1}{m^{2}}\left[
-\frac{s^{\mathrm{IMF}\prime}_{3}}{2x_{\perp}}-
\frac{p^{\mathrm{IMF}\prime}_{3}}{x_{\perp}}+
\frac{2s^{\mathrm{IMF}}_{3}}{x^{2}_{\perp}}
\right]\right)\bigg{]} \cr 
&+ \frac{1}{3} \frac{1}{m^{2}}
\left[-p^{\mathrm{IMF}\prime \prime}_{3}+
\frac{p^{\mathrm{IMF}\prime}_{3}}{x_{\perp}}-
\frac{s^{\mathrm{IMF}\prime \prime}_{3}}{2}+
\frac{s^{\mathrm{IMF}\prime}_{3}}{2x_{\perp}}-
\frac{2s^{\mathrm{IMF}}_{3}}{x^{2}_{\perp}}\right], \cr
\frac{dF_{r}}{dS_{\theta}}&=\frac{dF_{\theta}}{dS_{r}}
= 0, \cr
\frac{dF_{\theta}}{dS_{\theta}}&= \bigg{[} \left(p^{\mathrm{IMF}}_{(0,0)}-
\frac{1}{2}s^{\mathrm{IMF}}_{(0,0)}\right) + 
\frac{2}{3} \left(2p^{\mathrm{IMF}}_{2}+ 
s^{\mathrm{IMF}}_{2}+\frac{1}{m^{2}}\left[
\frac{s^{\mathrm{IMF}\prime}_{3}}{2x_{\perp}}-
\frac{p^{\mathrm{IMF}\prime}_{3}}{x_{\perp}}\right]
\right)\bigg{]} \delta_{\lambda' 3}\delta_{3 \lambda} \cr
&+ \frac{1}{3} \bigg{(} 
-2s^{\mathrm{IMF}}_{2} + \frac{1}{m^{2}} 
\bigg{[}-p^{\mathrm{IMF}\prime \prime}_{3} + 
\frac{p^{\mathrm{IMF}\prime}_{3}}{x_{\perp}}+
\frac{s^{\mathrm{IMF}\prime \prime}_{3}}{2} -
\frac{s^{\mathrm{IMF}\prime}_{3}}{2x_{\perp}}
\bigg{]}\bigg{)} + \frac{1}{3}
 \bigg{[}- 2s^{\mathrm{IMF}}_{2} - \frac{2}{m^{2}}
\frac{s^{\mathrm{IMF}}_{3}}{x^{2}_{\perp}}\bigg{]}.
\end{align}
Since there is no $\theta$ dependence on
$dF_{r,\theta}/dS_{r,\theta}$, the visualized forces looks like the
unpolarized nucleon force distributions, except for the strengths of
them. 

When the spin-1 particle is polarized along the $z$-axis with
$s_{z}=1$, we find the matrix elements of the dipole and quadrupole
operators as follows: 
\begin{align}
&\hat{Q}^{rr}_{s'_{z}=1,s_{z}=1}=-\frac{1}{6}, \quad \hat{Q}^{\theta
                \theta}_{s'_{z}=1,s_{z}=1}=-\frac{1}{6}, \quad
                \hat{Q}^{\theta r}_{s'_{z}=1,s_{z}=1}=0, \cr 
&\epsilon^{lm3}\hat{S}^{l}_{s'_{z}=1,s_{z}=1}
   X^{m}_{1}(\theta)=0, \quad
   \epsilon^{lm3}\hat{S}^{l}_{s'_{z}=1,s_{z}=1}
  \hat{\theta}^{m}_{1}(\theta)=0.
\end{align}
So, the quadrupole pattern of the distribution again vanishes when the
spin-1 particle is polarized along the $z$-axis with $s_{z}=1$: 
\begin{align}
\frac{dF_{r}}{dS_{r}}&= \bigg{[}
   \left(p^{\mathrm{IMF}}_{(0,1)}+\frac{1}{2}s^{\mathrm{IMF}}_{(0,1)}\right)
 - \frac{1}{3} \left(2p^{\mathrm{IMF}}_{2}+
   s^{\mathrm{IMF}}_{2}+\frac{1}{m^{2}}\left[
-\frac{s^{\mathrm{IMF}\prime}_{3}}{2x_{\perp}}-
\frac{p^{\mathrm{IMF}\prime}_{3}}{x_{\perp}}+
\frac{2s^{\mathrm{IMF}}_{3}}{x^{2}_{\perp}}
\right]\right)\bigg{]}  \cr 
&- \frac{1}{6} \frac{1}{m^{2}}
\left[-p^{\mathrm{IMF}\prime \prime}_{3}+
\frac{p^{\mathrm{IMF}\prime}_{3}}{x_{\perp}}-
\frac{s^{\mathrm{IMF}\prime \prime}_{3}}{2}+
\frac{s^{\mathrm{IMF}\prime}_{3}}{2x_{\perp}}-
\frac{2s^{\mathrm{IMF}}_{3}}{x^{2}_{\perp}}\right], \cr
\frac{dF_{r}}{dS_{\theta}}&=\frac{dF_{\theta}}{dS_{r}}
= 0, \cr
\frac{dF_{\theta}}{dS_{\theta}}&= \bigg{[} 
\left(p^{\mathrm{IMF}}_{(0,1)}-\frac{1}{2}
s^{\mathrm{IMF}}_{(0,1)}\right) - \frac{1}{3}
 \left(2p^{\mathrm{IMF}}_{2} +  s^{\mathrm{IMF}}_{2} +
\frac{1}{m^{2}}\left[
\frac{s^{\mathrm{IMF}\prime}_{3}}{2x_{\perp}}-
\frac{p^{\mathrm{IMF}\prime}_{3}}{x_{\perp}}\right]
\right)\bigg{]}  \cr
&- \frac{1}{6} \bigg{(} 
-2s^{\mathrm{IMF}}_{2} + \frac{1}{m^{2}} 
\bigg{[}-p^{\mathrm{IMF}\prime \prime}_{3} + 
\frac{p^{\mathrm{IMF}\prime}_{3}}{x_{\perp}}+
\frac{s^{\mathrm{IMF}\prime \prime}_{3}}{2} -
\frac{s^{\mathrm{IMF}\prime}_{3}}{2x_{\perp}}
\bigg{]}\bigg{)}  - \frac{1}{6}
 \bigg{[}- 2s^{\mathrm{IMF}}_{2} - \frac{2}{m^{2}}
\frac{s^{\mathrm{IMF}}_{3}}{x^{2}_{\perp}}\bigg{]}.
\end{align}
Note that for $s_{z}=1$ they are still independent of $\theta$, which
means that they are spherically symmetric.   

When the spin-1 particle is transversly polarized along $x$-axis with
$s_{x}=0$, $\theta$ dependence of $dF_{r,\theta}/dS_{r,\theta}$
emerges.  In this case, the matrix elements of the dipole and
qudrupole operators
are derived as 
\begin{align}
&\hat{Q}^{rr}_{s'_{x}=0,s_{x}=0}=-\frac{1}{6} (1+3\cos{2\theta}),
                \quad \hat{Q}^{\theta
                \theta}_{s'_{x}=0,s_{x}=0}=-\frac{1}{6}(1-3\cos{2\theta}),
                \quad \hat{Q}^{\theta
                r}_{s'_{x}=0,s_{x}=0}=\cos{\theta} \sin{\theta}, \cr 
&\epsilon^{lm3}\hat{S}^{l}_{s'_{x}=0,s_{x}=0}
  X^{m}_{1}(\theta)=0, \quad \epsilon^{lm3}
  \hat{S}^{l}_{s'_{x}=0,s_{x}=0} \hat{\theta}^{m}_{1}(\theta)=0. 
\end{align}
Thus, the strong force fields can be directly obtained as 
\begin{align}
\frac{dF_{r}}{dS_{r}}&= \bigg{[}
   \left(p^{\mathrm{IMF}}_{(0,1)}+\frac{1}{2}s^{\mathrm{IMF}}_{(0,1)}\right)
 - \frac{1}{3} \left(2p^{\mathrm{IMF}}_{2}+
   s^{\mathrm{IMF}}_{2}+\frac{1}{m^{2}}\left[
-\frac{s^{\mathrm{IMF}\prime}_{3}}{2x_{\perp}}-
\frac{p^{\mathrm{IMF}\prime}_{3}}{x_{\perp}}+
\frac{2s^{\mathrm{IMF}}_{3}}{x^{2}_{\perp}}
\right]\right)\bigg{]}  \cr 
&-\frac{1}{6} (1+3\cos{2\theta}) \frac{1}{m^{2}}
\left[-p^{\mathrm{IMF}\prime \prime}_{3}+
\frac{p^{\mathrm{IMF}\prime}_{3}}{x_{\perp}}-
\frac{s^{\mathrm{IMF}\prime \prime}_{3}}{2}+
\frac{s^{\mathrm{IMF}\prime}_{3}}{2x_{\perp}}-
\frac{2s^{\mathrm{IMF}}_{3}}{x^{2}_{\perp}}\right], \cr
\frac{dF_{r}}{dS_{\theta}}&=\frac{dF_{\theta}}{dS_{r}}
= \cos{\theta} \sin{\theta}  \frac{1}{m^{2}}  
\left[-\frac{2s^{\mathrm{IMF}\prime}_{3}}{x_{\perp}}+
\frac{2s^{\mathrm{IMF}}_{3}}{x^{2}_{\perp}}\right], \cr
\frac{dF_{\theta}}{dS_{\theta}}&= \bigg{[} 
\left(p^{\mathrm{IMF}}_{(0,1)}-\frac{1}{2}
s^{\mathrm{IMF}}_{(0,1)}\right) - \frac{1}{3}
 \left(2p^{\mathrm{IMF}}_{2} +  s^{\mathrm{IMF}}_{2} +
\frac{1}{m^{2}}\left[
\frac{s^{\mathrm{IMF}\prime}_{3}}{2x_{\perp}}-
\frac{p^{\mathrm{IMF}\prime}_{3}}{x_{\perp}}\right]
\right)\bigg{]}  \cr
&-\frac{1}{6} (1+3\cos{2\theta}) \bigg{(} 
-2s^{\mathrm{IMF}}_{2} + \frac{1}{m^{2}} 
\bigg{[}-p^{\mathrm{IMF}\prime \prime}_{3} + 
\frac{p^{\mathrm{IMF}\prime}_{3}}{x_{\perp}}+
\frac{s^{\mathrm{IMF}\prime \prime}_{3}}{2} -
\frac{s^{\mathrm{IMF}\prime}_{3}}{2x_{\perp}}
\bigg{]}\bigg{)} \cr
& -\frac{1}{6}(1-3\cos{2\theta})
 \bigg{[}- 2s^{\mathrm{IMF}}_{2} - \frac{2}{m^{2}}
\frac{s^{\mathrm{IMF}}_{3}}{x^{2}_{\perp}}\bigg{]}.
\end{align}
While all the dipole contribution still found to be zero, the
qudrupole structure brings about the deformation of the strong force
distributions. 

The most interesting case arises when the spin-1 particle is polarized
along the $x$-axis with $s_{x}=1$. The matrix elements of
$\hat{Q}_{s'_{x}=1,s_{x}=1}$ and $\hat{S}_{s'_{x}=1,s_{x}=1}$ are given by 
\begin{align}
&\hat{Q}^{rr}_{s'_{x}=1,s_{x}=1}=\frac{1}{12} (1+3\cos{2\theta}),
                \quad \hat{Q}^{\theta
                \theta}_{s'_{x}=1,s_{x}=1}=\frac{1}{12}(1-3\cos{2\theta}),
                \quad \hat{Q}^{\theta
                r}_{s'_{x}=1,s_{x}=1}=-\frac{1}{2}\cos{\theta}
                \sin{\theta}, \cr 
&\epsilon^{lm3}\hat{S}^{l}_{s'_{x}=1,s_{x}=1}
  X^{m}_{1}(\theta)
=\sin{\theta}, \quad \epsilon^{lm3}\hat{S}^{l}_{s'_{x}=1,s_{x}=1}
\hat{\theta}^{m}_{1}(\theta)=\cos{\theta}.
\end{align}
Using these results, we can show that all the dipole and quadrupole
contributions to the strong force fields emerge as follows: 
\begin{align}
\frac{dF_{r}}{dS_{r}}&= \frac{1}{2}\bigg{[}
   \left(p^{\mathrm{IMF}}_{(0,1)}+\frac{1}{2}s^{\mathrm{IMF}}_{(0,1)}\right)
 - \frac{1}{3} \left(2p^{\mathrm{IMF}}_{2}+
   s^{\mathrm{IMF}}_{2}+\frac{1}{m^{2}}\left[
-\frac{s^{\mathrm{IMF}\prime}_{3}}{2x_{\perp}}-
\frac{p^{\mathrm{IMF}\prime}_{3}}{x_{\perp}}+
\frac{2s^{\mathrm{IMF}}_{3}}{x^{2}_{\perp}}
\right]\right)\bigg{]}  \cr 
&+\frac{1}{2}\bigg{[} \left(p^{\mathrm{IMF}}_{(0,0)}+
\frac{1}{2}s^{\mathrm{IMF}}_{(0,0)}\right) + \frac{2}{3} 
\left(2p^{\mathrm{IMF}}_{2}+  s^{\mathrm{IMF}}_{2}+
\frac{1}{m^{2}}\left[
-\frac{s^{\mathrm{IMF}\prime}_{3}}{2x_{\perp}}-
\frac{p^{\mathrm{IMF}\prime}_{3}}{x_{\perp}}+
\frac{2s^{\mathrm{IMF}}_{3}}{x^{2}_{\perp}}
\right]\right)\bigg{]}  \cr 
&+ \frac{1}{12} (1+3\cos{2\theta}) \frac{1}{m^{2}}
\left[-p^{\mathrm{IMF}\prime \prime}_{3}+
\frac{p^{\mathrm{IMF}\prime}_{3}}{x_{\perp}}-
\frac{s^{\mathrm{IMF}\prime \prime}_{3}}{2}+
\frac{s^{\mathrm{IMF}\prime}_{3}}{2x_{\perp}}-
\frac{2s^{\mathrm{IMF}}_{3}}{x^{2}_{\perp}}\right]
-\sin{\theta}
 \frac{1}{m}\left[2p^{\mathrm{IMF}\prime}_{1} +
 s^{\mathrm{IMF}\prime}_{1}\right], \cr
\frac{dF_{r}}{dS_{\theta}}&=\frac{dF_{\theta}}{dS_{r}}
= -\frac{1}{2}\cos{\theta} \sin{\theta}  \frac{1}{m^{2}}  
\left[-\frac{2s^{\mathrm{IMF}\prime}_{3}}{x_{\perp}}+
\frac{2s^{\mathrm{IMF}}_{3}}{x^{2}_{\perp}}\right] - 
\frac{2}{m}\cos{\theta} 
\frac{s^{\mathrm{IMF}}_{1}}{x_{\perp}}, \cr
\frac{dF_{\theta}}{dS_{\theta}}&= \frac{1}{2} \bigg{[} 
\left(p^{\mathrm{IMF}}_{(0,1)}-\frac{1}{2}
s^{\mathrm{IMF}}_{(0,1)}\right) - \frac{1}{3}
 \left(2p^{\mathrm{IMF}}_{2} +  s^{\mathrm{IMF}}_{2} +
\frac{1}{m^{2}}\left[
\frac{s^{\mathrm{IMF}\prime}_{3}}{2x_{\perp}}-
\frac{p^{\mathrm{IMF}\prime}_{3}}{x_{\perp}}\right]
\right)\bigg{]}  \cr
&+\frac{1}{2}\bigg{[} \left(p^{\mathrm{IMF}}_{(0,0)}-
\frac{1}{2}s^{\mathrm{IMF}}_{(0,0)}\right) + 
\frac{2}{3} \left(2p^{\mathrm{IMF}}_{2}+ 
s^{\mathrm{IMF}}_{2}+\frac{1}{m^{2}}\left[
\frac{s^{\mathrm{IMF}\prime}_{3}}{2x_{\perp}}-
\frac{p^{\mathrm{IMF}\prime}_{3}}{x_{\perp}}\right]
\right)\bigg{]} \cr
&+ \frac{1}{12} (1+3\cos{2\theta})  \bigg{(} 
-2s^{\mathrm{IMF}}_{2} + \frac{1}{m^{2}} 
\bigg{[}-p^{\mathrm{IMF}\prime \prime}_{3} + 
\frac{p^{\mathrm{IMF}\prime}_{3}}{x_{\perp}}+
\frac{s^{\mathrm{IMF}\prime \prime}_{3}}{2} -
\frac{s^{\mathrm{IMF}\prime}_{3}}{2x_{\perp}}
\bigg{]}\bigg{)} \cr
& + \frac{1}{12}(1-3\cos{2\theta})
 \bigg{[}- 2s^{\mathrm{IMF}}_{2} - \frac{2}{m^{2}}
\frac{s^{\mathrm{IMF}}_{3}}{x^{2}_{\perp}}\bigg{]}-\sin{\theta}
\frac{1}{m}\left[2p^{\mathrm{IMF}\prime}_{1} - 
s^{\mathrm{IMF}\prime}_{1}\right].
\label{eq:c8}
\end{align}

\end{document}